\documentclass[10pt,english]{book}
\usepackage[T1]{fontenc}
\usepackage{geometry}
\usepackage{textcomp}
\usepackage{graphicx}
\usepackage{multicol}
\usepackage{t1enc}
\usepackage[latin1,utf8]{inputenc}
\usepackage{amsmath}
\usepackage{amsfonts}
\usepackage{mathrsfs}
\usepackage{tipa}
\usepackage{amssymb}
\usepackage{cancel}
\usepackage{amsthm}
\usepackage{slashed}
\usepackage{quotchap}
\usepackage{fancyhdr}
\makeatletter
\usepackage{babel}
\makeatother
\newcommand{\ud}{\mathrm{d}}
\newcommand{\ui}{\mathrm{i}}
\newcommand{\blambda}{\text{\textcrlambda}}

\begin{document}

\pagenumbering{roman}

\begin{titlepage}

\begin{center}

{\Large UNIVERSIDADE DE SÃO PAULO \\ INSTITUTO DE FÍSICA\\}

\vspace{3.8cm}

{\LARGE {\bf Superradiance: Classical, Relativistic and Quantum Aspects}}

\vspace{1.5cm}
\textsf{\Large  Bruno Arderucio Costa}\\
\end{center}
\vspace{2.1cm}

\begin{flushright}
\textsf{Advisor: Prof. Dr. Alberto Saa}\\
\end{flushright}

\vspace{1.0cm}
\begin{flushright}
\begin{minipage}{9cm}
Updated version of dissertation submitted to the Institute of Physics of the University of São Paulo to obtain the title of Master of Science.
\end{minipage}
\end{flushright}
\vspace{2.5cm} Examining Committee:\\

\small{Prof. Dr. Alberto Saa (Advisor, IF-USP)

Prof. Dr. Diego Trancanelli (IF-USP)

Prof. Dr. George Emanuel Avraam Matsas (IFT-Unesp)}

\vfill
\begin{center}
\vspace{0.1cm} São Paulo\\ 2014\\

\end{center}

\end{titlepage}


\newpage
\mbox{}\thispagestyle{empty}
\newpage
\chapter*{Acknowledgements}

The author wishes to thank everyone who followed him through his master's studies with great interest. Every related conversation I had somehow contributed to this work.

In particular, I mention my admired advisor, of course, Prof. Alberto Saa, who promptly accepted me as his student, originally proposed the topic to be studied and also has frequently exposed some of his ideas to guide me in this and future works. He has always been confident on my competence as a physicist. I appreciate all the time he devoted to me along these two years.

Also my mother deserves to be mentioned separately for encouraging me and for all her help in a wide variety of senses, some of which were crucial to provide me time enough to allow me to dedicate to my studies. She also played an important role on helping me, even invoking huge sacrifices when needed, since my school days until the application process for my future Ph.D. studies.

I have to express my sincere thankfulness to Prof. George Matsas for his several discussions, both about Physics and regarding decisions about my academic future, that provided me guidance for my study, insight on several issues and above all, for being a source of inspiration as a model to be followed in various senses.

Finally, I am grateful to Fundação de Amparo à Pesquisa do Estado de São Paulo (FAPESP) for financial support under the process 2011/15660-0.

I would like to thank the examining committee for pointing out some errors in the original version of this text.

\chapter*{Abstract/Resumo}

Several physical systems can be treated as a scattering process, and, for these processes, a natural observed quantity arises: the ratio between the reflected and incident intensities, known as the \emph{reflection coefficient}. This dissertation is concerned with the phenomenon known as superradiance, that is, when this coefficient is larger than unity.

We shall explore many examples of such systems, and, more importantly, we shall also see how, apart from the interest in its own right, superradiance is related to a number of important current research physical issues. We begin with a small survey of important results on chapter one. On chapter two, we establish a general criteria to decide whether or not superradiant scattering is observed based on the linear, second order, homogeneous ordinary differential equation (ODE) or linear, first order homogeneous systems of ODEs which describes the process and we shall give an example of system in which superradiance is observed. On chapter three, we focus on spinning black hole superradiance, we shall describe how one can compute explicitly the reflection coefficient for different spin waves. 
Chapter four is dedicated to the relations with thermodynamics. We develop what is meant by black hole thermodynamics, particularly the so-called first and second law of black hole thermodynamics, and apply them in the context of superradiance, so we can generalise some of the results from chapter three to more general black holes. Finally, on chapter five, we explore many of the quantum aspects of superradiance, including the relation with the Klein paradox, and the quantum version of black hole superradiance, for the later we will explain briefly how one usually quantise fields in curved space-time. A further connection with thermodynamics is explored. Thorough all this text we analyse the connection between superradiance and spin and statistics.

\vspace{.5cm}
\begin{center}
***
\end{center}
\vspace{.5cm}

Vários sistemas físicos podem ser tratados como problemas de espalhamento e, para esses problemas, uma quantidade observada surge naturalmente: a razão entre as intencidades refletidas e incidentes, a que damos o nome de \emph{coeficiente de reflexão}. Essa dissertação está preocupada com o fenômeno conhecido como superradiância, isto é, quando esse coeficiente é maior que a unidade.

Exploraremos vários exemplos de tais sistemas e, mais importante, veremos como, além do interesse por si só, superradiância está relacionada com um número de questões importantes de pesquisa atual. Começamos com um pequeno resumo de resultados impotantes no capítulo um. No capítulo dois estabeleceremos um critério geral para decidir se espalhamento superradiante é observado baseando-se nas equações diferenciais ordinárias lineares homogêneas de segunda ordem (EDO) ou sistema linear de EDO homgêneas de primeira ordem que descrevem o processo e daremos um exemplo de sistema em que superradiância é observada. No capítulo três, focaremos em superradiância em buracos negros em rotação, e mostraremos como podemos calcular explicitamente o coeficiente de reflexão para ondas incidentes de diferentes spins. 
O capítulo quatro é dedicado à relação com a termodinâmica. Desenvolveremos a chamada termodinâmica de buracos negros, particularmente às assim chamadas primeira e segunda leis da termodinâmica de buracos negros, e aplicaremo-nas no contexto da superradiância de forma a generalizar alguns dos resultados do capítulo dois a buracos negros mais genéricos. Finalmente no capítulo cinco, exploraremos muitos dos aspectos quânticos da superradiância, incluindo a relação com o paradoxo de Klein e a versão quântica da superradiância de buracos negros. Para a última, explicaremos brevemente como normalmente se quantiza campos em espaço-tempo curvo. Exploraremos uma outra conexão com termodinâmica. Ao longo de todo o texto, analisamos a conexão entre superradiância, spin  e estatística.

\tableofcontents

\mainmatter
\pagenumbering{arabic}
\setcounter{page}{9}

\chapter{Introduction}

\section{Notation and Conventions}

We use the metric signature $(-+++)$, apart from the section about spinors, where there is a good reason for the change, which will become clear on that section.

Unless explicit appearances, used where intended to clarify the presence of quantum, relativistic and gravitational arguments, we use natural units such $\hslash=c=G=1$.

$R$ can denote either the curvature scalar or the reflection coefficient. No ambiguity arises, since the appearance of these objects is alway clear and no expression share both quantities. Ricci tensor is $R_{ab}=R^{c}_{\ acb}$, where $R^a_{bcd}$ denotes the Riemann curvature tensor. Weyl tensor is denoted by $C_{abcd}$.

$\mathbf1$ denotes the identity matrix. Its size is not specified by this notation, even though, it will be obvious accordingly to the context. The same symbol is used to denote the identity operator in arbitrary vector space.

Penrose's abstract index notation is used throughout the text by the use of Latin alphabet letters. Greek letters denotes components on a specific basis. And we make use of the following abbreviations: $A_{[ab]}=\frac{1}{2!}(A_{ab}-A_{ba})$, and $A_{(ab)}=\frac{1}{2!}(A_{ab}+A_{ba})$ and similar expression for more than two indices. $\square=\nabla^a\nabla_a$.

The boundary in topological sense $\dot{\mathcal{A}}=\bar{\mathcal{A}}\setminus\mathrm{int}(\mathcal{A})$, where the bar over a letter denotes the closure and $\mathrm{int}$ the interior of a set. The boundary in manifold sense $\partial M$ is the subset of $M$ whose image under a local chart's map lies on the boundary of $\mathbb{H}^n$ in $\mathbb{R}^n$.

Chronological and causal future (past) are denoted respectively by $I^\pm$ and $J^\pm$.

Future (past) null infinities ($\mathscr{I}^\pm$) and related objects are defined according to Hawking \& Ellis \cite{HE}. Their notation is preserved.

Complex conjugation is denoted by *. Where it might cause confusion, we denote it by a bar over the symbol.

Tangent space through a point $P$ of a manifold $M$ is symbolised by $T_PM$. Cotangent space, as any dual space, by $T^*_PM$. The set of vector fields on $M$ is denoted by $\mathscr X(M)$.

Heaviside step function is denoted by $\theta:\mathbb R\to\mathbb R$, $\theta(x)=1$ for $x\geqslant0$ and 0 otherwise. The sign function $\mathrm{sgn}:\mathbb R\to\mathbb R$ is defined by $\mathrm{sgn}(x)=\theta(x)-\theta(-x)$.

The notation $x\to p+0$ means the lateral limit $\lim_{x\to p^+}$.

\newpage

\section{Motivation and Brief Survey}

Reflection and transmission problems appear persistently in several branches of Physics. Common basic examples follow below.

Low amplitude perturbation $y$ on a rope is well known to obey wave equation $\frac{\partial^2y}{\partial x^2}-\frac{1}{v^2}\frac{\partial^2y}{\partial t^2}=0$, where $v=\sqrt{\frac{F}{\lambda}}$, where $\lambda$ and $F$ denote the mass per unit of length and the tension on the rope respectively. Defining the new variables $\xi\equiv x-vt$ and $\eta\equiv x+vt$, the wave equation reduces to $\frac{\partial^2 y}{\partial\eta\partial\xi}=0$, showing the general solution of the one-dimensional wave equation is $y(x,t)=y_+(\xi)+y_-(\eta)$ for two arbitrary one-variable functions $y_+$ and $y_-$. The interpretation of this solution is straightforward: it just represents the superposition of two pulses, one shaped as $y_+$ propagating towards the positive direction of the $x$ axis and a pulse $y_-$ propagating backwards. If two ropes made of different materials, that is, ropes possessing different values for the parameter $\lambda$, two wave equations will hold at the two half-lines and when the continuity of the composed rope is used as a boundary condition, and the possibility of just forward propagation in one of the rope's pieces, we can interpret the problem as having an incident wave which has been partially reflected backward along the first piece and a transmitted part to the second piece. Reflection (transmission) coefficient $R$ ($T$) is defined as the ratio between the energy of reflected (transmitted) and incident part. In this example, these ratios can be computed by means of time averaged power, which for an harmonic wave with frequency $\omega$ and amplitude $A$ reads $\frac{1}{2}\lambda\omega^2A^2v$. They obey the relation $R+T=1$ and $R\to 0$ when $\lambda_1\to\lambda_2$ and $R\to1$ as $\lambda_2\to\infty$. Extensions to non harmonic waves can be dealt with Fourier analysis. 

Other examples constitutes as solving time-independent (one-dimensional, for simplicity) Schrödinger's equation with different potentials, say, the step potential. The solution satisfying the boundary conditions of continuity of the wave and the presence of propagation in only one direction on a half-line can again be interpreted as incident, reflected and transmitted parts. Reflection and transmission coefficients can be computed with aid of conserved probability current of Schrödinger's equation and $R+T=1$ also holds. Combining two step potentials, one can find the possibility of tunneling, for instance. These wave phenomena, including tunneling, have analogous ones in optics, where the reflection and transmission coefficients can be computed with aid of Poyinting vectors. In optics it is possible to obtain vanishing reflection coefficient during refraction, if incident wave polarisation is parallel to incident plane and the incident angle is the Brewster's angle.

All these former examples share an important common feature: all energy balance, or probability balance in case of Schrödinger's equation, is solely dependent on the waves themselves, no energy exchange is allowed between the waves and the media itself. This fact ultimately leads to the conservation relation $R+T=1$\footnote{This fact is manifested mathematically in equation (\ref{condicao}) below, for instance. If its second term on the right-hand side is evaluated at $\xi_0\to-\infty$, where typical solutions approach $T^{\frac{1}{2}}e^{-\ui\omega\xi}$ provided $\Gamma\equiv0$ and $V\to\omega^2$ as $\xi\to-\infty$, expressing the fact that there is no dissipation and we are dealing with the same media at the two asymptotic regions, respectively, then (\ref{condicao}) can be rewritten as $R+T=1$. In fact this result is broader then its hypothesis, for if the media were different, this result would also be valid, but no longer the reflection and transmission would be simply the square modulus of $R^\frac{1}{2}$ and $T^\frac{1}{2}$, instead they would also depend on the different wave numbers on the two asymptotic regions and (\ref{condicao}) would again lead to the desired result. Clearly, one could also have derived this result directly from the conservation relation which applies to the specific problem and in terms of which reflection coefficient is defined.}. As we are going to see during this dissertation, there are physical systems where this is simply not the case. In Zel'dovich's cylinder, for example, where electromagnetic radiation is incident upon a \emph{conducting} cylinder, part of the energy of radiation can be converted in kinetic energy and vice-versa. This may eventually lead to $R>1$, characterising superradiance. Still, on section 5.1 we are going to see another way to produce superradiance whose origin is not understood simply by means of this conservation relation, namely transmission coefficient may be negative, since the direction of propagation of transmitted wave may be constrained to be reversed. We shall explore the details of this subtle case on that section.

Bearing this in mind, and remembering there is a known process (the Penrose process) to extract energy from a rotating black hole, no wonder may they be subjected to superradiance. Penrose process consists on allowing a particle to disintegrate in two parts. The original particle four-momentum is a timelike vector $p_0^a$, and the four-momenta of the parts are the timelike vectors $p_1^a$ and $p_2^a$, such as $p_0^a=p_1^a+p_2^a$. For an observer at infinity, the energy of a particle whose four-momentum is $p^a$ can be calculated as $-p^a\xi_a$, where $\xi_a$ is a timelike Killing field at infinity. This is the common notion of energy of a particle whose world-line does not necessarily intercept the world-line of observer. In Kerr space-time, there is a region (called the ergosphere) where the Killing vector $\xi^a$ becomes spacelike, allowing the energy to be negative within (and only within) this region. Therefore, one can arrange the disintegration such as one of the fragments possesses negative energy and enters the black hole, leaving the second fragment with energy greater than the original particle, extracting energy from the black hole. From certain point of view, superradiance is the wave analogue for Penrose process, and, in fact the presence of an horizon is not crucial for the phenomenon in Kerr space-time, but the ergosphere is \cite{friedman,mauricioalberto}.

In view of this possibility of extracting energy from a black hole, it has been proposed \cite{hod,rosa} a mechanism to extract all possible energy (we are going to establish this limit in the context of so-called black hole thermodynamics) in short period of time (a `black hole bomb'). One possibility is to surround a black hole and electromagnetic radiation at infinity with a mirror that does not alters neither the frequency nor the magnitude of angular momentum of radiation, so that the amplified radiation emitted from the black hole is directed once more towards it, until the black hole rotation is too faint for this process to go on. Another possibility consists on \emph{massive} bosonic incident radiation, the effect of mass could be intuitively imagined similar to the mirror, being attracted again toward the black hole. Consensus has still not been reached nowadays about the quantitative results, particularly because of cumbersome approximations of special functions \cite{hod,rosa}.

Unfortunately, the experimental detection of black hole superradiance, as well as several other black hole phenomena, is far from being a simple task. There is, on the other hand, a consolation: the existence of non-gravitational system that can be described by equations analogous to those of fields around a black hole. Rich examples are found in hydrodynamics, either for sound waves, in the pioneering work \cite{unruh81}, or in gravity waves \cite{unruh02}. There is one of these systems that can potentially be used to detect analogous black hole superradiance \cite{unruh02}, despite of course of being one further physical example of system that exhibits superradiance. We shall not describe it theoretically, since it would be redundant with our black hole description. These analogue systems are much closer to direct experimental verification. Namely, stimulated Hawking emission has already been successfully detected \cite{unruh11} in one of these systems!

It is important to bear in mind that these analogies are formal only, so the limitations are considerable, for example, the system we shall see are analogous only to the kinematics of General Relativity, since the background metric is so chosen in order to the analogy to take place, not because of the solution of some dynamical equation, like Einstein's, and the analogy is lost to that extent. From a certain point of view, these systems can be better interpreted as analogical computers. They are useful for us to see some properties we already expect, but might be too difficult to calculate, however, never could be used to discover novel Physics. Personally, I believe they are a little more than that: they suggest how some properties are changed when some features of the known physics are allowed to vary (the photon dispersion relation constitutes an excellent example), possibly giving us insight for what to expect of a quantum theory of gravity, that is still missing nowadays.

In what follows, I attempt to give a comprehensive even though not exhaustive account of the main results about superradiance, giving emphasis on black holes and its surprisingly rich relation with black hole thermodynamics.

\chapter{General Treatment}

\section{Criteria}

We dedicate this chapter to establish tools that can be used to compute reflection coefficients given appropriate boundary conditions. \cite{Mauricio} described such a procedure for systems governed by a second order linear homogeneous differential equation, to decide whether or not superradiance is present. Such equation can be written in the form
\[h''(\eta)+b(\eta)h(\eta)+c(\eta)h(\eta)=0.\]
It is important for future interpretation to note that when $b$ vanishes, so the equations has already the form to be considered further, the system is Hamiltonian\footnote{Physically, it is customary to say that no dissipation occurs in such system. Mathematically, that there is a function $H(p,h)$ such $h'=\partial H/\partial p$ and $p'=-\partial H/\partial h$ for $p=h'$.}; when it does not, one can either employ the change of independent variables satisfying
\begin{equation}
\frac{\ud^2\xi}{\ud\eta^2}=-b(\eta),
\label{mudanca}
\end{equation}
or one can use the integrating factor to change the dependent variable, $\displaystyle f=\exp\left(\int\ud\eta\ \frac{b}{2}\right)h$, so that the new equation contains no term with first derivatives.

We consider further equations in the form
\begin{equation}
\frac{\ud^2 f}{\ud \xi ^2} + [V(\xi) + \ui\Gamma(\xi)] f = 0.
\label{EDO1}
\end{equation}

Under the assumptions that for large values of the independent variable $\xi$, $V(\xi)$ approaches to a constant $\omega^2$ and $\xi\Gamma(\xi)$ approaches to zero, still on this limit we seek solutions of the form
\begin{equation}
f(\xi)=e^{-\ui\omega\xi}+R^{\frac{1}{2}}e^{\ui\omega\xi},
\label{asform}
\end{equation}
where the meaning of $R=|R^\frac{1}{2}|^2\geqslant0$ is immediate: it represents the fraction of the square of the incident amplitude which is travelling against the $0x$ direction, i.e., the reflection coefficient (as introduced in chapter one). Let us consider the variation of the Wronskian $W(f,f^*)$. We know from the theory of ODE that if the coefficients of (\ref{EDO1}) were real, this Wronskian should be a constant, since both $f$ and $f^*$ would independently satisfy (\ref{EDO1}). This constancy can also be seen directly from Abel's identity with vanishing coefficient for $f'$. This is not the case when $\Gamma\neq0$.
\[\frac{\ud}{\ud\xi}[\ui W(f,f^*)]=\ui\frac{\ud}{\ud\xi}\left[f\frac{\ud f^*}{\ud\xi}-f^*\frac{\ud f}{\ud\xi}\right]=2\Gamma|f|^2,\]
where we have used (\ref{EDO1}) on the last step. Integrating this equation with aid of the assumptions above for the behaviour of all quantities at large values of $\xi$
\[\ui W(f,f^*)|_{\infty}-\ui W(f,f^*)|_{\xi_0}=2\int_{\xi_0}^\infty\Gamma|f|^2 \ud\xi.\]
Substituting (\ref{asform}) we get
\begin{equation}
R=1+\frac{\ui}{2\omega}W(f,f^*)|_{\xi_0}-\frac{1}{\omega}\int_{\xi_0}^\infty\ud\xi\Gamma|f|^2.
\label{condicao}
\end{equation}

Superradiance is then present when
\begin{equation*}
\ui W(f,f^*)|_{\xi_0}-2\int_{\xi_0}^\infty\Gamma(\xi)|f(\xi)|^2\ud\xi>0.
\end{equation*}

On a general physical problem, reflection and transmission coefficient are usually defined as the ratios between the value of a quantity associated with the reflected wave and the same quantity associated with the incident wave. This quantity is usually a conserved quantity $\mathscr{G}$ such as energy, as in the case of the rope mentioned on introduction or in the case of black holes as we are going explore in details on next chapters; or number currents (see sections 3.4 and 5.1), or probability density current in Schrödinger's equation. The later happens to be proportional to the Wronskian, therefore this is the most straightforward way to interpret the results above. In this case, we have shown superradiance may be present when this current is \emph{not} conserved anymore thanks to the presence of an imaginary part on the potential. This may appear as a rather artificial form of producing superradiance, but complex potentials are frequently used as effective theory for describing scattering processes in nuclear physics, for instance.

As mentioned on introduction, a possible way to produce superradince is letting $\mathscr{G}$ not be conserved on the subsystem of interest. For example, if we seek superradiance on electromagnetic waves, $\mathscr{G}$ can be taken as the flux of Poynting vector $\vec{S}$, which obeys $\mathrm{div}\ \vec S+\frac{\partial u_\text{em}}{\partial t}=-\frac{\partial u_\text{mec}}{\partial t}\neq0$, superradiance can only occur provided the right-hand side in different from zero, representing non conservation of electromagnetic energy alone.

Finally, it is important to be cautious if we change variables of an equation in order to cast it in the above criteria (e.g., by using (\ref{mudanca}) or the integrating factor above), and if the asymptotic modes are meaningful in the original variables, which means the amplitudes of reflected and transmitted waves relates with $\mathscr{G}$, because the change of variables will probably spoil the asymptotic behaviour, i.e., the asymptotic behaviour of new variables will no longer relate with $\mathscr{G}$ in the same manner, of course.

\cite{Mauricio} considered only cases for which the reflection and transmission coefficients were given by the expressions above. When studying different systems one must decide the precise form of these coefficients on physical grounds and, together with this criteria decide whether or not superradiance is present by means of the Wronskian relations above. The case just presented represents directly Schrödinger's probability density currents, because this current is proportional to the Wronskian between a solution representing a wave function and its complex conjugate and reflection coefficient is defined as the ratio between reflected and incident currents. When studying a rope, as we mentioned on introduction, the same relations between the amplitudes will be respected, of course, but reflection coefficient is now defined as the ratio of \emph{powers}, not Schrödinger's current. This means we must express how power relates to the wave amplitudes before interpreting Wronskian relations in terms of superradiance.

The case we just studied is useful for the example in the next section, for chapter three, when we manage to separate variables for a system of equations among other cases. However, we shall generalise a partial result from \cite{Mauricio} for a system of linear ODEs. We consider only first-order linear systems, but this means no loss in generality since we can always transform a higher order linear ODE into a system of first order ODEs. Such system can be written as
\begin{equation}
\frac{\ud}{\ud t}X=A(t)X+B(t),
\label{SEDO1}
\end{equation}
where $X$ and $B$ are $n\times1$ matrices and $A$ is an $n\times n$ matrix of complex numbers. Now, consider the $n\times n$ matrix $S$ whose columns are solutions $X_i$ of the system. We denote by $W(t)$ its determinant called the \emph{Wronskian}, that is $W(t)=\det S(t)$ (nomenclature also adopted by \cite{kwa}, we follow here).

Consider, as we did before, the equations to be homogeneous, that is $B(t)\equiv0$. Taking the derivative of the formula for determinants $W=\sum_\sigma\mathrm{sgn}(\sigma)S_{1\sigma(1)}S_{2\sigma(2)}\ldots S_{n\sigma(n)}$, where the sum runs over all possible permutations $\sigma$ of $(1,2,\ldots,n)$, and $\mathrm{sgn}(\sigma)$ is $-1$ for odd permutations and $+1$ for even permutations,
\[\frac{\ud}{\ud t}W=\sum_{i=1}^n\sum_\sigma\mathrm{sgn}(\sigma)S_{1\sigma(1)}\ldots S_{i-1\sigma(i-1)}\frac{\ud S_{i\sigma(i)}}{\ud t}S_{i+1\sigma(i+1)}\ldots S_{n\sigma(n)}.\]
Substituting (\ref{SEDO1}) for the derivative on the right-hand side,
\[\frac{\ud}{\ud t}W=\sum_{i=1}^n\sum_\sigma\sum_{k=1}^n\mathrm{sgn}(\sigma)A_{ik}S_{1\sigma(1)}\ldots S_{k\sigma(i)}\ldots S_{n\sigma(n)}.\]
Let $\tau$ be another permutation obtained when $i\neq k$ by a single permutation over $\sigma$ that interchanges positions $i$ and $k$. Of course $\mathrm{sgn}(\tau)=-\mathrm{sgn}(\sigma)$. This means that only the term $k=i$ survives after the sum over $\sigma$.
\[\frac{\ud}{\ud t}W=\sum_{i=1}^nA_{ii}\sum_\sigma\mathrm{sgn}(\sigma)S_{1\sigma(1)}\ldots S_{i\sigma(i)}\ldots S_{n\sigma(n)}=\mathrm{Tr}A\ W(t),\]
where the formula for the determinant has been used once more. Integrating this equation  we find what we call henceforth as \emph{generalised Abel's identity}:
\begin{equation}
W(t)=W(t_0)\int_{t_0}^t\ud t'\exp\mathrm{Tr}A(t')
\label{GAI}
\end{equation}

To recover the case studied in \cite{Mauricio}, consider the linear first order system associated with (\ref{EDO1}): $\frac{\ud f}{\ud\xi}=g(\xi)$ and $\frac{\ud g}{\ud\xi}=(V+\ui\Gamma)f$. Clearly the Wronskian in the sense above coincides with the Wronskian for second order ODE. Furthermore, because the absence of a term containing first derivatives of $f$ in (\ref{EDO1}), the matrix $A$ is traceless and the Wronskian \emph{between solutions} is conserved by virtue of (\ref{GAI}).

If we impose asymptotic behaviour for the solutions, we can encounter a formula for reflection and transmission coefficients. An example of application of this generalisation will be found on chapter five, below.

\section{Example: Zel'dovich Cylinder}

As an example, we consider (a particular case of) the Zel'dovich's cylinder, that is a conducting rigid cylinder with radius $a$ and conductivity $\sigma\geqslant0$ rotating around its symmetry axis with angular velocity $\Omega$ ($\Omega a<1$) in Minkowski space-time. By Ohm's law, $j^a=\sigma F^{ab}u_b+\rho u^a$ inside it and zero outside, where $\rho$ is the charge density as measured by an observer such that the charges are at rest. For such a rotation, the components of the four-velocity for a point inside the cylinder is given, in cylindrical coordinates such that $\ud s^2=-\ud t^2+\ud r^2+r^2\ud\phi^2+\ud z^2$ by $u_\mu(r)=\frac{1}{\sqrt{1-\Omega^2 r^2}}(-1,0,\Omega r^2,0)$. For consistency, $H^{ab}u_b=\epsilon(\omega)F^{ab}u_b$ and $\varepsilon^{abcd}F_{cd}u_b=\mu(\omega)\varepsilon^{abcd}H_{cd}u_b$, where $\omega$ is measured in cylinder's frame, must satisfy Maxwell's equations: 
\begin{equation}
\partial_{[c}F_{ab]}=0\quad \text{and} \quad\partial_bH^{ab}=4\pi j^a.
\label{max}
\end{equation}
Although there are different modes which experience superradiance (\cite{beke}), for sake of simplicity, we shall consider $\rho=0$ and modes with
\begin{equation}
\begin{array}{c}
\displaystyle\vec{E}=\frac{\omega-m\Omega}{\omega\sqrt{1-\Omega^2r^2}}\frac{f(r)}{\sqrt{r}}e^{\ui(m\phi-\omega t)}\vec{e_z}\\
\displaystyle\vec{B}=\left[\frac{m-\omega\Omega r^2}{\omega r\sqrt{1-\Omega^2r^2}}\vec{e_r}+\frac{\ui}{\omega}\vec{e_\phi}\frac{\ud}{\ud r}\right]\frac{f(r)}{\sqrt{r}}e^{\ui(m\phi-\omega t)},
\label{eb}
\end{array}
\end{equation}
representing electric and magnetic filed measured in the rotating frame. Because $\phi$ is periodic, $m$ must be an integer, say positive.

It is straightforward to write down Maxwell's equations (\ref{max}) in components in this particular coordinate system. The equation not satisfied trivially gives

\begin{equation}
\frac{\ud^2f}{\ud r^2}+\omega^2f+\frac{(1-\epsilon\mu)(\omega-m\Omega)^2}{1-\Omega^2r^2}f-\frac{4m^2-1}{4r^2}f+4\pi\ui\mu\sigma\frac{\omega-m\Omega}{\sqrt{1-\Omega^2r^2}}f=0
\label{dentro}
\end{equation}
for $r<a$ and
\[\frac{\ud^2f}{\ud r^2}+\omega^2f-\frac{4m^2-1}{4r^2}f=0\]
for $r>a$, because we settled $\epsilon=\mu=0$ for $r>a$, which means we are considering vacuum outside the cylinder. The homogeneous equations are consistent with the ansatz choice.

Although it is far from being simple to compute the exact value for the reflection coefficient, (\ref{condicao}) may indicate when superradiance is possible. We must apply an appropriate boundary condition (at the point $r=0$). The third and fifth term in (\ref{dentro}) may be approximated with aid of $(1-\Omega r^2)^{-1}\approx1+\Omega r^2$ and $(1-\Omega r^2)^{-1/2}\approx1+\frac{1}{2}\Omega r^2$ near the axis. Within this approximation (\ref{dentro}) can be written as 
\[\frac{\ud^2f}{\ud r^2}+C_0f+C_2r^2f-\frac{4m^2-1}{4r^2}f=0,\]
where $C_0$ and $C_2$ are constants. Since this equation possesses a singular point on the axis, we may search solution in series form
\begin{equation}
f(r)=\sum_{n=0}^\infty A_nr^{n+\lambda}.
\label{serie}
\end{equation}
The indicial equation\footnote{for discussion of this method, see, for example, chapter nine of \cite{arfken}}, i.e., the algebraic equation for $n=0$ after (\ref{serie}) is substituted back into the differential equation, is then $\lambda(\lambda-1)-m^2-\frac{1}{4}=0$, with roots $\frac{1}{2}\pm m$. As we are interested only in regular solutions, we consider only the upper sign. The indicial equation gives the lowest power in $r$ appearing in the solution (\ref{serie}). So, near the axis, this term will give the most important contribution. So $f(r)\approx A_0r^{m+1/2}$ near $r=0$. Now, substituting on (\ref{condicao}) and bearing in mind that the Wronskian term trivially vanishes, we obtain

\[R=1-\frac{4\pi\mu\sigma}{\omega}\int_0^a\ud r\ \frac{(\omega-m\Omega)|f(r)|^2}{r\sqrt{1-\Omega^2r^2}},\]
from which we see that superradiance is present whenever $\omega<m\Omega$. We also see that in case of an insulator cylinder, the reflection coefficient will be unitary.

To investigate energy conservation, we compute the complex Poynting vector (\cite{jackson}) for the modes with aid of (\ref{eb}): $\vec{S}=EB^*_r\vec{e_\phi}-EB^*_\phi\vec{e_r}$. It is clear that the energy carried away by the electromagnetic field per unit of length and per unit of time is obtained by the flux of the real part of the Poynting vector across the cylinder's surface.

From the well-known result from Electrodynamics $\nabla\cdot\vec{S}+\frac{\partial}{\partial t}(u_{\text{mec}}+u_{\text{em}})=0$, we learn that in the superradiant regime all gain in radiation energy must be compensated by a mechanical energy lost, mechanical energy from the cylinder. Similarly, from the continuity equation associated with angular momentum conservation in Electrodynamics \cite{jackson}, we learn that the cylinder must lose angular momentum in favour of the field. The time averaged angular momentum of the electromagnetic field is $\vec{L}=\int\ud^3x\ \vec{r}\times\vec{S}$, giving contribution of $2\pi\int\ud r\ rE(r)B^*_r(r)\vec{e_z}$ per unit length, directed lengthwise, as we might expect\footnote{The easiest way to convince oneself is by writing the cross product between the position vector and the complex Poynting vector in rectangular coordinates. The components directed along $\vec{e_x}$ and $\vec{e_y}$ vanish when integrated over the angular variable.}.

\chapter{Kerr Black Holes}

\section{Spinors}
The purpose of this section is merely to fix notation and to give readers some operational techniques. We do not intend to construct (when possible) a formal definition of spinors in curved space-time, which is quite extensive. We refer to \cite{penrose} for an account of this task.

A spinor space $(W,\epsilon_{AB})$ is a two-dimensional vector space $W$ over $\mathbb{C}$, whose elements are denoted with superscripts ($\xi^A\in W$) whilst elements of its dual with subscripts ($\xi_A\in W^*$), and an antisymmetric tensor $\epsilon_{AB}:W\otimes W\rightarrow\mathbb{C}$ usually called \emph{skew-metric}.

From $W$ we can construct the complex conjugate dual space $\bar W^*$, corresponding to antilinear maps from $W$ to $\mathbb{C}$. Elements of this space are denoted by a primed subscript ($\xi_{A'}\in\bar W^*$). $\epsilon_{AB}$ can be used to form a correspondence between vectors of $W$ and $W^*$ by $\xi_B=\epsilon_{AB}\xi^A$, and the reciprocal by defining $\epsilon^{AB}$ by $\epsilon^{AB}\epsilon_{BC}=-\delta^A_{\ C}$. Similar relations hold for primed indices, by substituting $\epsilon_{AB}$ by $\bar\epsilon_{A'B'}$, which is just the complex conjugate of $\epsilon_{AB}$.

Consider a spinorial basis $(o^A,\iota^A)$ of $W$ satisfying $o_A\iota^A=1$. Then
\begin{equation}
\begin{array}{c}
\displaystyle t^{AA'}=\frac{1}{\sqrt{2}}(o^A\bar o^{A'}+\iota^A\bar\iota^{A'}),\\
\displaystyle x^{AA'}=\frac{1}{\sqrt{2}}(o^A\bar\iota^{A'}+\iota^A\bar o^{A'}),\\
\displaystyle y^{AA'}=\frac{\ui}{\sqrt{2}}(o^A\bar\iota^{A'}-\iota^A\bar o^{A'}),\\
\displaystyle z^{AA'}=\frac{1}{\sqrt{2}}(o^A\bar o^{A'}-\iota^A\bar\iota^{A'}),
\end{array}
\label{usualbasis}
\end{equation}
comprises a basis of the four complex dimensional vector space $Y$ of tensors ($W^*\otimes\bar W^*\rightarrow\mathbb{C}$). Each element above, say $t^{AA'}$, has the property $\bar t^{A'A}=t^{AA'}$, so they are referred as real and an element of $Y$ is real if and only if it is expanded on the basis above with real coefficients. By linearity, the subset $V\subset Y$ of real elements is a four real dimensional vector space over $\mathbb{R}$. In order to compare with results easily found in the literature, written in terms of this basis $\epsilon_{AB}=o_A\iota_B-\iota_Ao_B$, we may represent the components $\epsilon_{\Xi\Theta}$ of $\epsilon_{AB}$ on a matrix,
\[\epsilon_{\Xi\Theta}=
\begin{bmatrix}
0 & 1\\
-1 & 0
\end{bmatrix},\]
then $(x^\mu)^{\Xi\Xi'}=\frac{1}{\sqrt{2}} ^{\mu}\sigma^{\Xi\Xi'}$, where $(x^\mu)$ denotes any one of the elements ($x,y,z$) of the basis of $V$ and $ ^{\mu}\sigma$ denotes the corresponding Pauli matrix and $t^{\Xi\Xi'}=\frac{1}{\sqrt{2}}\mathbf{1}$.

Consider
\[g_{AA'BB'}=\epsilon_{AB}\bar\epsilon_{A'B'},\]
multi-linear map $V\times V\rightarrow\mathbb{R}$ non-degenerate, symmetric, which defines a Lorentzian metric on $V$ with signature $(+ - - -)$. This metric is flat, and, in fact in terms of the above basis of $V$, using the letter without indices to symbolise the coordinate corresponding to that vector of the basis, $\ud s^2=\ud t^2-\ud x^2-\ud y^2-\ud z^2$. We can make the correspondence between spinors and four-vector (elements of $T_PM$), that is, we can associate each element of $V$ to an element of $T_PM$. It is important to stress that for each $P$ there will be a different basis $\{\xi_\Sigma, \Sigma\in\{0,1\}\}=(o^A,\iota^A)$. 
A $T_PM$ vector can be written on the basis $(t^a,\mathbf{x}^a)$, constructed by means of a $V$ vector:
\begin{equation}
v^a=\underbrace{(t^at_{AA'}-\mathbf{x}^a\mathbf{x}_{AA'})}_{\equiv \sigma^a_{AA'}}v^{AA'}.
\label{mM}
\end{equation}
It is straightaway 
to generalise this construction for tensor of space-time, that is, correspondences between $V\otimes\ldots\otimes V\otimes V^*\otimes\ldots\otimes V^*$ and $T_PM\otimes\ldots\otimes T_PM\otimes T^*_PM\otimes\ldots\otimes T^*_PM$. In this section (not on the entire chapter) we changed the signature of Lorentzian metrics in order to avoid confusion, that is, when we write an object in the form $v^{AA'}$ no matter if we view it as an element of $Y$ (and therefore raise and lower indices with $\epsilon_{AB}$) or of $T_PM$ associated with $V$ as above (and therefore raise and lower pair of indices with $g_{AA'BB'}$).

The object $\sigma_{AA'}^a$ defined above is the responsible for the connection between the spinor space and space-time. It has the suggestive symbol $\sigma$ which can be understood to remind Pauli matrices. If we choose them in a coordinate system such $\sigma^{\Xi\Xi'}_0=2^{-1/2}\mathbf1$ and $\sigma^{\Xi\Xi'}_\alpha=2^{-1/2}\sigma_\alpha$, where $\alpha\in\{1,2,3\}$ and $\sigma_\alpha$ represents the corresponding Pauli matrix, the correspondence shown in (\ref{mM}) is nothing but the connection between four-vectors and Hermitian $2\times2$ matrices used as a representation of the homogeneous Lorentz group, found in most textbooks, for instance \cite{streater}. However, we shall not employ these representations here. Instead, we proceed as follows.

It is easy to check that the (null, because of the antisymmetry property of $\epsilon_{AB}$) vectors
\begin{equation}
\begin{array}{c}
\displaystyle\mathbf{l}^{AA'}=o^A\bar{o}^{A'}\\
\displaystyle\mathbf{n}^{AA'}=\iota^A\bar{\iota}^{A'}\\
\displaystyle\mathbf{m}^{AA'}=\iota^A\bar{o}^{A'}\\
\displaystyle\mathbf{\bar{m}}^{AA'}=o^A\bar{\iota}^{A'}.
\end{array}
\label{spinorNP}
\end{equation}
form a Newman-Penrose basis (see next section for a definition), i.e., when we understand $g_{AA'BB'}$ obtained from the skew-metric as the metric for the space-time, they obey the orthogonality relations imposed on a Newman-Penrose basis. If we interpret the vectors of the left hand side of (\ref{spinorNP}) as vector of space-time, we conclude that, on this basis, it is required
\begin{equation}
\sigma^{\Xi\Xi'}_\mu=\frac{1}{\sqrt{2}}
\begin{bmatrix}
n_\mu & -\bar m_\mu\\
-m_\mu & l_\mu
\end{bmatrix}\quad\text{and}\quad\sigma^\mu_{\Xi\Xi'}=\frac{1}{\sqrt{2}}
\begin{bmatrix}
l^\mu & m^\mu\\
\bar m^\mu & n^\mu
\end{bmatrix}.
\label{NPPauli}
\end{equation}
Then, it suffices to find a Newman-Penrose basis to employ this representation.

Also, from (\ref{spinorNP}) we find a recipe to calculate the metric tensor
\[g_{ab}\leadsto g_{AA'BB'}=\epsilon_{AB}\bar\epsilon_{AB'}=(o_A\iota_B-\iota_A o_B)(\bar o_{A'}\bar\iota_{B'}-\bar\iota_{A'}\bar o_{B'})\leadsto l_an_b+n_al_b-m_a\bar m_b-\bar m_am_b.\]
Here the symbol $\leadsto$ must be understood as using the correspondence (\ref{mM}) between squares of spinorial objects and objects from space-time to link the passages.

Covariant derivatives $\nabla_{AA'}$ are defined to satisfy the all properties $\nabla_a$ do and also we impose it to be real, i.e., $\overline{\nabla\psi}=\nabla\bar\psi$ and compatible with the skew-metric. They may be computed first by the following identity by J. Friedman, easily checked:
\begin{equation}
\gamma_{AA'\Sigma\Lambda}\equiv(\xi_\Sigma)_B\nabla_{AA'}(\xi_\Lambda)^B=\frac{1}{2}\sum_{\Gamma',\Delta'=0}^1 \bar{\epsilon}^{\Gamma'\Delta'}(\bar{\xi}_{\Gamma'})_{B'}(\xi_\Sigma)_B\nabla_{AA'}[(\xi_\Lambda)^B(\bar{\xi}_{\Delta'})^{B'}],
\label{spinorderivative}
\end{equation}
where the symbol $\nabla_{AA'}[(\xi_\Lambda)^B(\bar{\xi}_{\Delta'})^{B'}]$ is well defined with aid of (\ref{mM}) twice. Second, once the spin coefficients are known, we may contract (\ref{spinorderivative}) with $\xi^{\Sigma C}$ and use the orthogonality relations on the basis $\xi_\Sigma$, $\xi_{\Sigma A}\xi^{\Xi A}=\delta_\Sigma^\Xi$ and $\xi_{\Sigma A}\xi_\Xi^A=\epsilon_{\Sigma\Xi}$, to obtain
\begin{equation}
\nabla_{AA'}\xi_{\Sigma B}=\xi_{\Xi B}\gamma^{\ \ \ \Xi}_{AA'\ \Sigma}.
\label{friedmancollorary}
\end{equation}

For more complicated objects, we recall any spinorial tensor can be expressed as a sum of scalar basis components and elements of the basis $\xi_\Sigma$. Using Leibniz's rule, it suffices to calculate the ordinary derivative of these scalar quantities and the spin connections $\gamma_{AA'\Sigma\Lambda}$ above.

Dirac equations for spin-\textonehalf\ field in curved space-times are
\begin{subequations}
\begin{align}
\nabla_{AA'}\phi^A=\frac{m}{\sqrt{2}}\varsigma_{A'}\\
\nabla^{AA'}\varsigma_{A'}=-\frac{m}{\sqrt{2}}\phi^A.
\end{align}
\label{DiracSpinor}
\end{subequations}

If we apply $\nabla^{A'}_{\ B}$ on equation (a) and substitute equation (b) and apply a number of identities involving the commutator of two spin derivatives (analogous to identities of Riemann curvature tensor) we shall not list, we conclude $\left(\square+m^2+\frac{R}{4}\right)\phi^A=0$. This conclusion is indispensable in order for the theory to be compatible with relativity. In Kerr geometry $R=0$ and this reduces to Klein-Gordon equation with any coupling with curvature, of course.

It is indeed a generalisation of usual Dirac equation found in most textbooks. 
To cast it in the usual form, consider the Dirac spinor $\psi=\phi^A\oplus\varsigma_{A'}$ (here the abstract index notation fails to accomplish the vector space structure of this object), and the condense the matrices appearing in (\ref{NPPauli}) as four $4\times4$ matrices
\begin{equation*}
\gamma^\mu=2\ui
\begin{bmatrix}
0 & -\sigma^{\mu\Xi\Xi'}\\
+\sigma^\mu_{\Xi\Xi'} & 0
\end{bmatrix},
\end{equation*}
so that equations (\ref{DiracSpinor}) reads
\begin{equation}
\ui\gamma^\mu\nabla_\mu\psi- m\psi=0,
\label{DiracUsual}
\end{equation}
which \emph{formally} reassembles its usual form, but one must bear in mind when computing the derivatives appearing above $\psi\in W\oplus\bar W^*$ and must be computed with aid of algebraic manipulations from (\ref{spinorderivative}). In fact, because there is a term proportional to $[(\xi_\Lambda)^B(\bar{\xi}_{\Delta'})^{B'}]$ in the right hand side of (\ref{spinorderivative}), corresponding to the $\Gamma^\mu_{\nu\rho}$ coefficients of Levi-Civita connection of space-time, the covariant derivative operator acting on a Dirac spinor $\nabla_\mu\psi$ will be understood as $(\partial_\mu+\Gamma_\mu)\psi$ for some $4\times4$ matrices $\Gamma_\mu$. The precise form of those are known when a particular basis of space-time is used explicitly. The most trivial case occurs in flat space-time in coordinates for which the connection coefficients vanish identically, leading to $\Gamma_\mu=0$, that is, the usual Dirac equation as derived by himself. Because this form of writing Dirac equations mingles objects with coordinates, it will be avoided here whenever possible, for sake of clarity.

Moreover, from the fact we have already proved that any `component' of spinor field which satisfies (\ref{DiracSpinor}) also satisfies Klein-Gordon equation, so will a component of Dirac spinor. Applying the operator $(\ui\gamma^\nu\nabla_\nu+ m)$ on (\ref{DiracUsual}) we conclude $\gamma^{(a}\gamma^{b)}=g^{ab}$ is satisfied identically.

There is a (covariantly) conserved current by virtue of the dynamical equations (\ref{DiracSpinor}): 
\begin{equation}
j^{AA'}=\bar\phi^{A'}\phi^A+\bar\varsigma^A\varsigma^{A'},
\label{DiracCurrentSpinor}
\end{equation}
since $\nabla_{AA'}j^{AA'}=0$, can be readily verified. From (\ref{DiracCurrentSpinor}), we see $j^{AA'}$ is a sum of two future directed null vectors, and hence it is a future directed timelike vector\footnote{To prove this assertion, simply choose an orthonormal basis and write in components $g(j,j)=2g(n_1,n_2)$ and $g(n_1,n_1)=g(n_2,n_2)=0$. Apply Cauchy-Schwartz inequality on the first expression and substitute the second and third equalities.}. If the field were massless, equation (\ref{DiracSpinor}b) would be redundant and the dynamical equation would be simply $\nabla_{AA'}\phi^A=0$ and the term containing $\varsigma$ in (\ref{DiracCurrentSpinor}) would be suppressed. In this case, $j^{AA'}$ is clearly a future directed null vector.

It is possible to write this conserved current in terms of Dirac spinors, which will be useful when we quantise spinor field below. The current can be written as
\[\psi^\dagger\begin{bmatrix}
\delta_\Xi^{\ \Theta}\delta_{\Xi'}^{\ \Theta} & 0\\
0 & \bar\epsilon^{\Xi'\Theta'}\epsilon^{\Xi\Theta}
\end{bmatrix}\psi,\]
directly checked to coincide with $j^{\Theta\Theta'}$ from (\ref{DiracCurrentSpinor}). The association with vectors in space-time can be made clear my writing as $\bar\psi\gamma^\mu\psi$, where the Dirac adjoint $\bar\psi\equiv\psi^\dagger\alpha$, where $\alpha\gamma^\mu$ has to coincide with the $4\times4$ matrix above. So we can take
\[\alpha=\begin{bmatrix}
0 &\mathbf1\\
-\mathbf1 & 0
\end{bmatrix},\]
but it is important to bear in mind this representation is conditioned to our choice of representation a Dirac spinor as $\phi^A\oplus\varsigma_{A'}$ and our representation for Dirac matrices $\gamma^\mu$ above. It is possible to find in the literature, other representations which may differ form ours. Our choice is made primarily to construct a smooth transition between the two formalisms for spinors in curved space-time.

Before the end of this section, we present a construction we are going to make use on chapter five, an inner product between solutions $\phi^A$ and $\psi^A$ of the massless Dirac equation for spin-\textonehalf. Let $\Sigma$ be a Cauchy surface with future-directed normal $n^a$, from the rules above we construct $n_{AA'}$. Then
\begin{equation}
(\phi,\psi)_D\equiv\int_\Sigma\ud\sigma\ n_{AA'}\bar\phi^{A'}\psi^A.
\label{DiracProductCurved}
\end{equation}
This product does not depend on the choice of the Cauchy surface, since $\nabla_{AA'}\bar\phi^{A'}\phi^A=0$ by virtue of the dynamical equation, which means $\bar\phi^{A'}\phi^A$ may be understood as a conserved current in this case. In Dirac spinor notation, it is written as $\bar\psi\gamma^\mu\psi$ which leads to $(\psi_1,\psi_2)_D\equiv\int_\Sigma\ud\sigma\ n_\mu\bar\psi_1\gamma^\mu\psi_2$. This form is going to be used on chapter five. Derivatives acting on the Dirac adjoint, must be again be understood as we mentioned to Dirac spinors, the sign of $\Gamma_\mu$ is reversed, (\ref{spinorderivative}).

In principle, it would be possible to generalise equations (\ref{DiracSpinor}), the conserved current and the inner product for higher spins by adding more indices on spinors. It is possible to show \cite{WaldGR}, however, that for spins higher than one, these equations do not have a well posed initial value formulation in curved space-time.

\section{Newman-Penrose Formalism}

\subsection{Generalities and Application in Kerr Metric}

The main reference for this topic is S. Chandrasekhar, in \cite{Survey} or in \cite{chandrasekhar}. Newman-Penrose formalism is similar to the tetrads, except for the choice of basis: it is chosen a vector basis $(\mathbf{l},\mathbf{n},\mathbf{m},\mathbf{\bar{m}})$ such that the former two are real null vectors and the latter two are a null pair of complex-complex conjugate vectors, instead of the usual orthonormal one, denoted by $\mathbf{e}_{(a)}$. It is imposed the following orthonormality conditions\footnote{If we had chosen to keep the signature convention from previous section, we would have to change the sign of this normalisation criteria, of course.}
\begin{equation}
\begin{array}{c}
\displaystyle \mathbf{l}\cdot\mathbf{m}=\mathbf{l}\cdot\bar{\mathbf{m}}=\mathbf{n}\cdot\mathbf{m}=\mathbf{n}\cdot\bar{\mathbf{m}}=0,\\
\mathbf{l}\cdot\mathbf{n}=-\mathbf{m}\cdot\bar{\mathbf{m}}=-1.
\end{array}
\label{onNP}
\end{equation}

For instance,  we may always obtain a Newman-Penrose basis from an orthonormal basis $\{\mathbf e_i\}$ by choosing
\[\mathbf{l,n}=\frac{1}{\sqrt{2}}\left(\mathbf{e}_t\pm\mathbf{e}_z\right), \quad \mathbf{m,\bar{m}}=\frac{1}{\sqrt{2}}\left(\mathbf{e}_x\pm\ui\mathbf{e}_y\right),\]
for which,
\begin{equation*}
\eta_{\mu\nu}\equiv\mathbf{e}^i_\mu\mathbf{e}_{i\nu}=
\begin{bmatrix}
0 & -1 & 0 & 0\\
-1 & 0 & 0 & 0\\
0 & 0 & 0 & 1\\
0 & 0 & 1 & 0
\end{bmatrix}
\end{equation*}

Once chosen a Newman-Penrose basis, we compute the \emph{spin coefficients} similarly as one would compute the rotation coefficients with tetrads:
\[\gamma_{abc}=\mathbf{e}^k_c\mathbf{e}_{ak;i}\mathbf{e}_b^i\]

Derivatives with respect to a basis component is defined simply as the projection of the usual derivative onto the basis vector. The \emph{intrinsic derivative} is defined as $A_{a|b}=A_{i;j}\mathbf{e}_a^i\mathbf{e}_b^j$.

After projecting Ricci identity $\nabla_l\nabla_k\mathbf e_{(a)}^i-\nabla_k\nabla_l\mathbf e_{(a)}^i=R_{mikl}\mathbf e_{(a)}^m$ onto the basis, Riemann curvature tensor can be calculated as \cite{Survey}
\[R_{(a)(b)(c)(d)}=-\gamma_{(a)(b)(c),(d)}+\gamma_{(a)(b)(d),(c)}+2\gamma_{(a)(b)(f)}\gamma^{(f)}_{[(c)(d)]}+\gamma_{(a)(f)(c)}\gamma^{(f)}_{(b)(d)}-\gamma_{(a)(f)(d)}\gamma^{(f)}_{(b)(c)}.\]
From these components it is easy to recover the usual ones, as one we proceed with any tensor, by means of the relation
\begin{equation}
R_{iklm}=\mathbf{e}_i^{(a)}\mathbf{e}_k^{(b)}\mathbf{e}_l^{(c)}\mathbf{e}_m^{(d)}R_{(a)(b)(c)(d)}.
\label{ricciid}
\end{equation}
Then, to compute (covariant) derivatives in Newman-Penrose formalism, it is simpler to evaluate derivatives along the basis vector, defined as 
\begin{equation}
A_{(a),(b)}=\mathbf e_{(b)}^i\frac{\partial}{\partial x^i}A_{(a)}=\mathbf e_{(b)}^i\frac{\partial}{\partial x^i}(\mathbf e_{(a)}^kA_k)=\mathbf e_{(b)}^i[e_{(a)}^kA_{k,i}+A_k\mathbf e_{(a)\ ,i}^k],
\label{howtoderive}
\end{equation}
and relate with usual covariant derivatives by means of the identity
\begin{equation}
A_{(a),(b)}=\mathbf e_{(a)}^i\nabla_kA_i\mathbf e_{(b)}^k+\gamma_{(a)(b)(c)}A^{(c)},
\label{howtoderive2}
\end{equation}
whose proof is very straightforward. Because the connection is assumed to be symmetric in General Relativity, equation (\ref{howtoderive}) remains valid if one replace ordinary derivatives by covariant derivatives. The identity then follows from the definition of spin coefficients.

Bianchi's identity $\nabla_{[i}R_{kl]m}^{\ \ \ \ n}=0$ expressed in terms of intrinsic derivatives is
\begin{equation}
R_{(a)(b)[(c)(d)|(f)]}=0.
\label{bianchi}
\end{equation}

Now we shall restrict our attention to the Kerr metric. It can be expressed in Boyer-Lindquist coordinates by
\begin{equation}
\ud s^2=\rho^2\left(\frac{\ud r^2}{\Delta}+\ud\theta^2\right)+(r^2+a^2)\sin^2\theta\ud\phi^2-\ud t^2+\frac{2Mr}{\rho^2}(a\sin^2\theta\ud\phi-\ud t)^2,
\tag{Kerr metric}
\end{equation}
where
\[\rho^2\equiv r^2+a^2\cos^2\theta\quad \text{and}\quad\Delta\equiv r^2-2Mr+a^2.\]

From this expression, it is clear the existence of both a timelike Killing filed $\left(\frac{\partial}{\partial t}\right)^a$, which we will represent generally by $\xi^a$ and a spacelike Killing filed $\left(\frac{\partial}{\partial\phi}\right)^a$ we shall denote generally by $\psi^a$. Because the coordinate $\phi$ is periodic with period $2\pi$, we conclude the orbits of the Killing filed $\psi^a$ are closed, representing axial symmetry. In fact, the derivation of Kerr metric imposes the existence of these isometries.

A possible Newman-Penrose basis for Kerr metric is given by  $(\mathbf{l,n,m,\bar m})$, where
\begin{equation}
\begin{array}{c}
\displaystyle\mathbf{l}=\frac{1}{\Delta}\left[(r^2+a^2)\frac{\partial}{\partial t}+\Delta\frac{\partial}{\partial r}+a\frac{\partial}{\partial\phi}\right]\\
\displaystyle\mathbf{n}=\frac{1}{2\rho^2}\left[(r^2+a^2)\frac{\partial}{\partial t}-\Delta\frac{\partial}{\partial r}+a\frac{\partial}{\partial\phi}\right]\\
\displaystyle\mathbf{m}=\frac{1}{\sqrt{2}\bar{\rho}}\left(\ui a\sin\theta\frac{\partial}{\partial t}+\frac{\partial}{\partial\theta}+\csc\theta\frac{\partial}{\partial\phi}\right),\\
\displaystyle\text{where} \quad\bar{\rho}\equiv r+\ui a\cos\theta,
\end{array}
\label{KerrNPbasis}
\end{equation}
that can be readily checked to satisfy the requirements (\ref{onNP}) for $\Delta\neq0$, otherwise the first vector is not defined. When we need a Newman-Penrose basis valid on $\Delta=0$, as it is the case for the horizon (on chapter four we shall prove this statement), we need to construct another basis satisfying the same orthogonality relations. Fortunately we can keep the vectors $\mathbf m$ and $\mathbf{\bar m}$ unchanged and adopt the new basis $(\mathbf{l',n',m,\bar m})$ with
\begin{equation*}
\begin{array}{c}
\displaystyle \mathbf l'=\frac{\Delta}{2(r^2+a^2)}\mathbf l\\
\displaystyle \mathbf n'=\frac{2(r^2+a^2)}{\Delta}\mathbf n
\end{array}
\end{equation*}
and express these two vectors not in Boyer-Lindquist coordinates, but by means of Kerr-Schild frame $(u_+,r,\theta,\phi_+)$, defined by $\ud u_+=\ud t+\frac{r^2+a^2}{\Delta}\ud r$ and $\ud\phi_+=\ud\phi+\frac{a}{\Delta}\ud r$, for which Kerr metric is written as
\begin{multline*}
\ud s^2=\rho^2\ud\theta^2-2a\sin^2\theta\ud r\ud\phi_++2\ud r\ud u_++\frac{(r^2+a^2)^2-\Delta a^2\sin^2\theta}{\rho^2}\sin^2\theta\ud\phi_+^2\\
-4\frac{aMr}{\rho^2}\sin^2\theta\ud\phi_+\ud u_+-\left(1-\frac{2Mr}{\rho^2}\right)\ud u_+^2.
\end{multline*}
Then,
\begin{equation}
\begin{array}{c}
\displaystyle \mathbf l'=\frac{\partial}{\partial u_+}+\frac{\Delta}{2(r^2+a^2)}\frac{\partial}{\partial r}+\frac{a}{a^2+r^2}\frac{\partial}{\partial\phi_+}\\
\displaystyle \mathbf n'=-\frac{r^2+a^2}{\rho^2}\frac{\partial}{\partial r}.
\end{array}
\label{KerrNPbasisHor}
\end{equation}

\subsection{Perturbations}

Since $[\mathsterling_\xi+\ui\omega,\mathsterling_\psi-\ui m]=0$, where $\mathsterling$ denotes the Lie derivative along the vector in its subscript, we can find simultaneous eigenvectors $\varphi$ for these two operators:
\begin{equation}
\mathsterling_\xi\varphi=-\ui\omega\varphi\quad\text{and}\quad\mathsterling_\psi\varphi=+\ui m\varphi,
\label{normalmodedef}
\end{equation}
and, thanks to periodicity of orbits of $\psi^a$, we require the azimuthal number $m$ to be an integer.

Our final goal is to determine the reflection coefficient for different spin waves incident on the black hole. Therefore, taking advantage of the possibility of satisfying (\ref{normalmodedef}), we consider perturbations in Kerr metric with time and azimuthal dependence given by the wave form
\begin{equation}
e^{\ui(m\phi-\omega t)}.
\label{wavedep}
\end{equation}
When one applies the Newman-Penrose basis (\ref{KerrNPbasis}) to tangent vector with the dependence above (\ref{wavedep}), one gets

\begin{equation}
\begin{array}{c}
\displaystyle\mathbf{l}=\mathscr{D}_0\\
\displaystyle\mathbf{n}=\frac{-\Delta}{2\rho^2}\mathscr{D}^\dagger_0\\
\displaystyle\mathbf{m}=\frac{1}{\sqrt{2}\bar{\rho}}\mathscr{L}^\dagger_0\\
\displaystyle\mathbf{\bar{m}}=\frac{1}{\sqrt{2}\bar{\rho}^*}\mathscr{L}_0.\\
\end{array}
\label{KNPbasisWave}
\end{equation}
where we have defined the set of operators
\begin{equation}
\begin{array}{c}
\displaystyle\mathscr{D}_n=\frac{\partial}{\partial r}+\ui\frac{K}{\Delta}+2n\frac{r-M}{\Delta}\\
\displaystyle\mathscr{D}^\dagger_n=\frac{\partial}{\partial r}+\ui\frac{K}{\Delta}+2n\frac{r-M}{\Delta}\\
\displaystyle\mathscr{L}_n=\frac{\partial}{\partial\theta}+Q+n\cot\theta\\
\displaystyle\mathscr{L}^\dagger_n=\frac{\partial}{\partial\theta}-Q+n\cot\theta,
\end{array}
\label{operadores}
\end{equation}
where $Q=-a\omega\sin\theta+m\csc\theta$ and $K=-(r^2+a^2)\omega+am$.

Because $R=0$ on Kerr geometry, any coupling between fields and curvature are all equivalent. We shall study some representative fields.

In Newman-Penrose formalism, it is convenient to work with three complex scalars instead of the usual field tensor $F_{ab}$ to write and solve Maxwell's equations. These scalars are $\phi_0\equiv F_{ab}l^am^b$, $\phi_1\equiv\frac{1}{2}F_{ab}(l^an^b+\bar{m}^am^b)$ and $\phi_2\equiv F_{ab}\bar{m}^an^b$. To obtain (homogeneous) Maxwell's equations in terms of these scalars, it is enough to write the expressions
\[\nabla_{[c}F_{ab]}=0\Rightarrow F_{[ij|k]=0},\]
\[\nabla_bF^{ab}=0\Rightarrow\eta^{ij}F_{ki|j}=0,\]
in components to get
\begin{equation}
\phi_{1|1}-\phi_{0|4}=\phi_{2|1}-\phi_{1|4}=\phi_{1|3}-\phi_{0|2}=\phi_{2|3}-\phi_{1|2}=0.
\label{MaxwellNP}
\end{equation}

In order to write (\ref{MaxwellNP}) explicitly, which means writing down the derivatives above accordingly to (\ref{howtoderive2}), e.g. $\phi_1=\frac{1}{2}(F_{12}+F_{43})\Rightarrow\phi_{1|1}=\frac{1}{2}[F_{12,1}-\eta^{ik}(\gamma_{i11}F_{k2}+\gamma_{i21}F_{1k})+F_{43,1}-\eta^{ik}(\gamma_{i41}F_{k3}+\gamma_{i31}F_{4k})]=\phi_{1,1}-(\gamma_{131}F_{42}+\gamma_{241}F_{13})$, we have the somewhat labourious task to compute all spin coefficients necessary to evaluate the intrinsic derivatives involved. It is pointless to show all intermediate calculations, since they are all straightforward. The final result is
\begin{equation}
\begin{array}{c}
\displaystyle\frac{1}{\sqrt{2}\bar{\rho}^*}\left(\mathscr{L}_1-\frac{\ui a\sin\theta}{\bar{\rho}^*}\right)\phi_0=\left(\mathscr{D}_0+\frac{2}{\bar{\rho}^*}\right)\phi_1\\
\displaystyle\frac{1}{\sqrt{2}\bar{\rho}^*}\left(\mathscr{L}_0+\frac{2\ui a\sin\theta}{\bar{\rho}^*}\right)\phi_1=\left(\mathscr{D}_0+\frac{1}{\bar{\rho}^*}\right)\phi_2\\
\displaystyle\frac{1}{\sqrt{2}\bar{\rho}^*}\left(\mathscr{L}_1^\dagger+\frac{\ui a\sin\theta}{\bar{\rho}^*}\right)\phi_2=-\frac{\Delta}{2\rho^2}\left(\mathscr{D}_0^\dagger+\frac{2}{\bar{\rho}^*}\right)\phi_1\\
\displaystyle\frac{1}{\sqrt{2}\bar{\rho}^*}\left(\mathscr{L}_0^\dagger+\frac{2\ui a\sin\theta}{\bar{\rho}^*}\right)\phi_1=-\frac{\Delta}{2\rho^2}\left(\mathscr{D}_1^\dagger-\frac{1}{\bar{\rho}^*}\right)\phi_0.
\end{array}
\label{maxwelltrabalhada}
\end{equation}
It is possible to decouple the equations above. First, note by direct calculation that the operators in the form $\mathscr{D}+\frac{m}{\bar{\rho}^*}$ and $\mathscr{L}+\frac{m\ui a\sin\theta}{\bar{\rho}^*}$ commute. $\mathscr{D}$ and $\mathscr{L}$ representing any of those operators $\mathscr{D}_n$ or $\mathscr{D}_n^\dagger$ and $\mathscr{L}_n$ or $\mathscr{L}_n^\dagger$, respectively. Then by means of the first and fourth of (\ref{maxwelltrabalhada}),
\[\left[\underbrace{
\left(\mathscr{L}_0^\dagger+\frac{\ui a \sin\theta}{\bar{\rho}^*}\right)\left(\mathscr{L}_1-\frac{\ui a \sin\theta}{\bar{\rho}^*}\right)}_{\mathscr{L}_0^\dagger\mathscr{L}_1+\frac{2\ui Qa\sin\theta}{\bar{\rho}^*}
}+\Delta\underbrace{\left(\mathscr{D}_1+\frac{1}{\bar{\rho}^*}\right)\left(\mathscr{D}_1^\dagger-\frac{1}{\bar{\rho}^*}\right)}_{\mathscr{D}_1\mathscr{D}_1^\dagger-\frac{2\ui K}{\bar{\rho}^*}}\right]\phi_0=0.\]
Employing the definitions of $K$ and $Q$,
\begin{equation}
[\mathscr{L}_0^\dagger\mathscr{L}_1+\Delta\mathscr{D}_1\mathscr{D}_1^\dagger+2\ui\omega(r+\ui a\cos\theta)]\phi_0=0.
\label{phi0}
\end{equation}
Similarly, by second and third of (\ref{maxwelltrabalhada}), for $\Phi_2=2(\bar{\rho}^*)^2\phi_2$,
\[\left[\underbrace{\left(\mathscr{L}_0+\frac{\ui a\sin\theta}{\bar{\rho}^*}\right)\left(\mathscr{L}_1^\dagger-\frac{\ui a\sin\theta}{\bar{\rho}^*}\right)}_{\mathscr{L}_0\mathscr{L}_1^\dagger-\frac{2\ui Qa\sin\theta}{\bar{\rho}^*}}+\Delta\underbrace{\left(\mathscr{D}_0^\dagger+\frac{1}{\bar{\rho}^*}\right)\left(\mathscr{D}_0-\frac{1}{\bar{\rho}^*}\right)}_{\mathscr{D}_0^\dagger\mathscr{D}_0+\frac{2\ui K}{\Delta\bar{\rho}^*}}\right]\Phi_2=0,\]
which leads to
\begin{equation}
[\mathscr{L}_0\mathscr{L}_1^\dagger+\Delta\mathscr{D}_0^\dagger\mathscr{D}_0-2\ui\omega(r+\ui a\cos\theta)]\Phi_2=0.
\label{phi2}
\end{equation}
Once equations (\ref{phi0}) and (\ref{phi2}) are solved, one can solve for $\phi_1$ by means of (\ref{maxwelltrabalhada}).

Proceeding similarly as we proceeded with electromagnetism, with aid of (\ref{NPPauli}) and (\ref{spinorderivative}) or (\ref{friedmancollorary}) we may write explicitly Dirac equations $\nabla_{AA'}P^A+\ui\mu\bar Q_{A'}=\nabla_{AA'} Q^A+\ui\mu\bar P_{A'}=0$ as (\ref{DiracSpinor}) in terms of the spin coefficients. The result is
\begin{equation}
\begin{array}{c}
\displaystyle\mathbf{e}_1(P^0)+\left(\frac{1}{2}(\gamma_{211}+\gamma_{341})-\gamma_{314}\right)P^0+\mathbf{e}_4(P^1)+\left(\gamma_{241}-\frac{1}{2}(\gamma_{214}+\gamma_{344})\right)P^1=\ui\mu\bar{Q}^{1'},\\
\displaystyle\mathbf{e}_3(P^0)+\left(\frac{1}{2}(\gamma_{213}+\gamma_{343})-\gamma_{312}\right)P^0+\mathbf{e}_2(P^1)+\left(\gamma_{243}-\frac{1}{2}(\gamma_{212}+\gamma_{342})\right)P^1=-\ui\mu\bar{Q}^{0'},\\
\displaystyle-\mathbf{e}_1(\bar{Q}^{0'})-\left(\frac{1}{2}(\gamma_{211}^*+\gamma_{341}^*)-\gamma_{314}^*\right)\bar{Q}^{0'}-\mathbf{e}_3(\bar{Q}^{1'})-\left(\gamma_{241}^*-\frac{1}{2}(\gamma_{214}^*+\gamma_{344}^*)\right)\bar{Q}^{1'}=\ui\mu\ P^1,\\
\displaystyle-\mathbf{e}_4(\bar{Q}^{0'})-\left(\frac{1}{2}(\gamma_{213}^*+\gamma_{343}^*)-\gamma_{312}^*\right)P^0+\mathbf{e}_2(\bar{Q}^{1'})+\left(\gamma_{243}^*-\frac{1}{2}(\gamma_{212}^*+\gamma_{342}^*)\right)\bar{Q}^{1'}=-\ui\mu P^0.
\end{array}
\label{DSNP}
\end{equation}
Then, we substitute the spin coefficients for azimuthal and temporal dependence given by (\ref{wavedep}) to give
\begin{equation}
\begin{array}{c}
\displaystyle\left(\mathscr{D}_0+\frac{1}{\bar{\rho}^*}\right)P^0+\frac{1}{\sqrt{2}\bar{\rho}^*}\mathscr{L}_{1/2}P^1=\ui\mu\bar{Q}^{1'}\\
\displaystyle\frac{\Delta}{2\rho^2}\mathscr{D}_{1/2}^\dagger P^1-\frac{1}{\sqrt{2}\bar{\rho}}\left(\mathscr{L}_{1/2}^\dagger+\frac{\ui a\sin\theta}{\bar{\rho}^*}\right)P^0=\ui\mu\bar{Q}^{0'}\\
\displaystyle\left(\mathscr{D}_0+\frac{1}{\bar{\rho}}\right)\bar{Q}^{0'}+\frac{1}{\sqrt{2}\bar{\rho}}\mathscr{L}_{1/2}^\dagger\bar{Q}^{1'}=-\ui\mu P^1\\
\displaystyle\frac{\Delta}{2\rho^2}\mathscr{D}_{1/2}^\dagger\bar{Q}^{1'}-\frac{1}{\sqrt{2}\bar{\rho}^*}\left(\mathscr{L}_{1/2}-\frac{\ui a\sin\theta}{\bar{\rho}}\right)\bar{Q}^{0'}=-\ui\mu P^0.
\end{array}
\label{diracpretrabalhada}
\end{equation}
Equations (\ref{diracpretrabalhada}) can be written more succinctly by introducing the abbreviations $f_1=\bar{\rho}^* P^0$, $g_2=-\bar{\rho}\bar{Q}^{0'}$, $f_2=P^1$ and $g_1=\bar{Q}^{1'}$, the last two just for the sake of symmetry.
\begin{equation}
\begin{array}{c}
\displaystyle\mathscr{D}_0 f_1+\frac{1}{\sqrt{2}}\mathscr{L}_{1/2}f_2=(\ui\mu r+a\mu\cos\theta)g_1\\
\displaystyle\Delta\mathscr{D}_{1/2}^\dagger f_2-\sqrt{2}\mathscr{L}_{1/2}^\dagger f_1=-2(\ui\mu r+a\mu\cos\theta)g_2\\
\displaystyle\mathscr{D}_0g_2-\frac{1}{\sqrt{2}}\mathscr{L}_{1/2}^\dagger g_1=(\ui\mu r-a\mu\cos\theta)f_2\\
\displaystyle\Delta\mathscr{D}_{1/2}^\dagger g_1+\sqrt{2}\mathscr{L}_{1/2}g_2=-2(\ui\mu r-a\mu\cos\theta)f_1,
\end{array}
\label{diractrabalhada}
\end{equation}
where we have already substituted the definition of $\bar{\rho}$ and its complex conjugate.

Gravitational perturbations are more complicated, but the same techniques are up to the task. Part of its complications arise because our considerable gauge freedom. In Newman-Penrose formalism, it is more convenient to work with spin connections and curvature tensor components instead of metric components itself. 

The first step towards solving the problem of perturbations is to calculate up to the first order originally algebraically independent vanishing quantities, which can be taken to be $\Phi_0=-C_{abcd}l^am^bl^cm^d$, $\Phi_1=-\sqrt2\bar\rho^*C_{abcd}l^an^bl^cm^d$, $k=\frac{\gamma_{311}}{\sqrt2(\bar\rho^*)^2}$, $s=\frac{\gamma_{313}\bar\rho}{(\bar\rho^*)^2}$, $\Phi_3=-\frac{(\bar\rho^*)^3}{\sqrt2}C_{abcd}l^am^b\bar m^cn^d$ and $\Phi_4=-(\bar\rho^*)^4C_{abcd}n^a\bar m^bn^c\bar m^d$, the set of equations (\ref{gravNP}) is a consequence of components of (\ref{ricciid}) and (\ref{bianchi}). Details of the tedious, but simple, calculations can be found in \cite{chandrasekhar}. Fortunately, we will not need go any step further here.

\begin{equation}
\begin{array}{c}
\displaystyle\left(\mathscr{L}_2-\frac{3\ui a\sin\theta}{\bar{\rho}^*}\right)\Phi_0-\left(\mathscr{D}_0+\frac{3}{\bar{\rho}^*}\right)\Phi_1=-6Mk\\
\displaystyle\Delta\left(\mathscr{D}_2^\dagger-\frac{3}{\bar{\rho}^*}\right)\Phi_0+\left(\mathscr{L}_{-1}^\dagger+\frac{3\ui a\sin\theta}{\bar{\rho}^*}\right)\Phi_1=6Ms\\
\displaystyle\left(\mathscr{D}_0+\frac{3}{\bar{\rho}^*}\right)s-\left(\mathscr{L}_{-1}^\dagger+\frac{3\ui a\sin\theta}{\bar{\rho}^*}\right)k=\frac{\bar{\rho}}{(\bar{\rho}^*)^2}\Phi_0\\
\displaystyle\left(\mathscr{D}_0-\frac{3}{\bar{\rho}^*}\right)\Phi_4-\left(\mathscr{L}_{-1}+\frac{3\ui a\sin\theta}{\bar{\rho}^*}\right)\Phi_3=6Ml\\
\displaystyle\left(\mathscr{L}_2^\dagger-\frac{3\ui a\sin\theta}{\bar{\rho}^*}\right)\Phi_4+\Delta\left(\mathscr{D}_{-1}^\dagger+\frac{3}{\bar{\rho}^*}\right)\Phi_3=6Mn\\
\displaystyle\left(\mathscr{D}_{-1}^\dagger+\frac{3}{\bar{\rho}^*}\right)l-\left(\mathscr{L}_{-1}+\frac{3\ui a\sin\theta}{\bar{\rho}^*}\right)n=\frac{\bar{\rho}}{(\bar{\rho}^*)^2}\Phi_4.\\
\end{array}
\label{gravNP}
\end{equation}

Once more, making use of the commutativity we noticed in the case of electromagnetism and in the case of Dirac field, we can decouple these equations. Making use of the definitions of $K$ and $Q$, we can simplify the resulting equations. It is worth mentioning that this is the third time this procedure is employed. This time the calculations are a little longer, but the algebraic difficulty is not changed. The result is
\begin{equation}
\begin{array}{c}
\displaystyle[\Delta\mathscr{D}_1\mathscr{D}_2^\dagger+\mathscr{L}_{-1}^\dagger\mathscr{L}_2+6\ui\omega(r-\ui a\cos\theta)]\Phi_0=0\\
\displaystyle[\Delta\mathscr{D}_{-1}^\dagger\mathscr{D}_0+\mathscr{L}_{-1}\mathscr{L}_2^\dagger-6\ui\omega(r+\ui a\cos\theta)]\Phi_4=0
\end{array}
\label{gravtrabalhada}
\end{equation}

To study massless scalar fields, there is no need to resort to Newman-Penrose formalism. We can directly make use of the identity \cite{landau2}
\begin{equation}
\square=\frac{1}{\sqrt{-g}}\frac{\partial}{\partial x^a}\left(\sqrt{-g}g^{ab}\frac{\partial}{\partial x^b}\right).
\label{dalembertian}
\end{equation}
And, because of its simplicity, we postpone the discussion of this case to chapter five, since we are interested in not only analysing the problem of reflection and transmission, but also on quantising the field. Nonetheless, we will conclude this field to possess superradiant regime in the last section of this chapter, and later on chapter four, by two other means.

\section{Teukolsky Equations and Scattering Problem for Different Spin Waves}

\subsection{Equations and its Solutions}
It is possible to separate variables of the equations (\ref{phi0}) and (\ref{phi2}). As usual, we suppose $\phi_0=R_{+1}(r)S_{+1}(\theta)$ and $\Phi_2=R_{-1}(r)S_{-1}(\theta)$. Here and henceforward, we choose normalisation of angular function $S$, whatever its subscript such as $\int_0^\pi\ud\theta\sin\theta S^2(\theta)=1$.

After the substitution on the partial differential equations we get
\begin{subequations}
\begin{align}
(\Delta\mathscr{D}_1\mathscr{D}_1^\dagger+2\ui\omega r)R_{+1}=\blambda R_{+1}\\
(\mathscr{L}_0^\dagger\mathscr{L}_1-2a\omega\cos\theta)S_{+1}=-\blambda S_{+1}
\end{align}
\label{teum1}
\end{subequations}
and
\begin{subequations}
\begin{align}
(\Delta\mathscr{D}_0^\dagger\mathscr{D}_0-2\ui\omega r)R_{-1}=\blambda R_{-1}\\
(\mathscr{L}_0\mathscr{L}_1^\dagger+2a\omega\cos\theta)S_{-1}=-\blambda S_{-1}.
\end{align}
\label{teum2}
\end{subequations}
Equations (\ref{teum1}) and (\ref{teum2}) are called \emph{Teukolsky equations} for electromagnetism.

Intentionally we used the same separation constant $\blambda$ for both pair of equations. This is allowed by comparing both angular equations (labeled by (b)). The operators on the left hand side are the same in both equations apart from a reflection through equatorial plane, that is, by interchanging $\theta$ and $\pi-\theta$. It means that if one of these equations, viewed as an eigenvalue equation, determine a root $\blambda_0$, this root will also be a root for the other equation, with the difference that it will be associated to another eigenfunction, related to the first by replacing $\theta$ by $\pi-\theta$. It remains still to fix the relative normalisation between $R_{+1}$ and $R_{-1}$. It is chosen such $\Delta\mathscr D_0\mathscr D_0R_{-1}=\mathscr C\Delta R_{+1}$ and $\Delta \mathscr D_0^\dagger\mathscr D_0^\dagger\Delta R_{+1}=\mathscr C^*R_{-1}$, where $\mathscr C$ is a constant. This choice is possible, as can be seen by applying the operator $\mathscr D_0$ twice on (\ref{teum2}a) and some algebraic rearrangements.

Equations (\ref{diractrabalhada}) can also have its variables separated. $f_1=R_{-1/2}(r)S_{-1/2}(\theta)$, $f_2=R_{1/2}(r)S_{1/2}(\theta)$, $g_1=R_{1/2}(r)S_{-1/2}(\theta)$, $g_2=R_{-1/2}(r)S_{1/2}(\theta)$. The same radial (angular) dependence has been chosen for $f_1$ and $g_2$ ($f_2$ and $g_1$) because this functions appears under the action of identical operators in those variables. After substitution on the partial differential equations,
\begin{subequations}
\begin{align}
\mathscr{D}_0R_{-1/2}-\ui\mu rR_{1/2}=\blambda R_{1/2}\\
\frac{1}{\sqrt{2}}\mathscr{L}_{1/2}S_{1/2}-a\mu\cos\theta S_{-1/2}=-\blambda S_{-1/2}\\
\Delta\mathscr{D}_{1/2}^\dagger R_{1/2}+2\ui\mu rR_{-1/2}=\blambda R_{-1/2}\\
\sqrt{2}\mathscr{L}_{1/2}^\dagger S_{-1/2}-2a\mu\cos\theta S_{1/2}=\blambda S_{1/2}.
\end{align}
\label{teud}
\end{subequations}
We shall refer to (\ref{teud}) as Teukolsky equations for spin-\textonehalf\ fields. It consists again as two pair of equations, the first, radial, composed by equations with label (a) and (c), and the second, angular, composed by (b) and (d). The separation constant holds for both pairs, for the same reason as it does for (\ref{teum1}--\ref{teum2}). These equations can be decoupled, giving a second-order equations, which are useful to decide whether or not superradiance is present. For instance, for -\textonehalf\ components,
\[\left[\Delta\mathscr{D}_{1/2}^\dagger\mathscr{D}_0-\frac{\ui\mu\Delta}{\blambda+\ui\mu r}\mathscr{D}_0-2(\blambda^2+\mu^2r^2)\right]R_{-1/2}=0,\]
\[\left[\mathscr{L}_{1/2}\mathscr{L}_{1/2}^\dagger+\frac{a\mu\sin\theta}{\blambda+a\mu\cos\theta}\mathscr{L}_{1/2}^\dagger+2(\blambda^2-a^2\mu^2\cos^2\theta)\right]S_{-1/2}=0.\]
The +\textonehalf\ components are obtained by the replacement $\theta$ by $\pi-\theta$.

At this point, we summarise  the dependence of the solutions of these equations with the original spinors appearing in (\ref{DiracSpinor}):
\begin{equation}
\begin{array}{c}
\displaystyle P^0=\frac{1}{\sqrt2\bar\rho^*}R_{-1/2}S_{-1/2}\quad\bar Q^{0'}=-\frac{1}{\sqrt2\bar\rho}R_{-1/2}S_{+1/2}\\
\displaystyle P^1=R_{+1/2}S_{+1/2}\quad\bar Q^{1'}=R_{+1/2}S_{-1/2}.
\end{array}
\label{volta}
\end{equation}

Finally, we separate variables for (\ref{gravtrabalhada}) by means of the substitutions $\Phi_0=R_{2}(r)S_2(\theta)$ and $\Phi_4=R_{-2}(r)S_{-2}(\theta)$ to give
\begin{subequations}
\begin{align}
(\Delta\mathscr{D}_1\mathscr{D}_2^\dagger+6\ui\omega r)R_2=\blambda R_2\\
(\mathscr{L}_{-1}^\dagger\mathscr{L}_2-6a\omega\cos\theta)S_2=-\blambda S_2\\
(\Delta\mathscr{D}_{-1}^\dagger\mathscr{D}_0-6\ui\omega r)R_{-2}=\blambda R_{-2}\\
(\mathscr{L}_{-1}\mathscr{L}_2^\dagger+6a\omega\cos\theta)S_{-2}=-\blambda S_{-2}.
\end{align}
\label{teug}
\end{subequations}

We shall refer to (\ref{teug}) as Teukolsky equations for gravitational perturbations. Once more, the same separation constant has been used for (a-b) and (c-d), because of the very same reason as the two previous cases.

The radial Teukolsky equations for different spins can be written in a unified expression below, where the spin $s$ is assumed to be \textonehalf, 1 or 2, accordingly if we are treating spin-\textonehalf\ massless Dirac field, electromagnetic waves or gravitational waves, respectively. By inspection, we see that (\ref{teus}) is the same as (\ref{teum1}--\ref{teum2}), (\ref{teud}) with $\mu=0$ or (\ref{teug}), by appropriate substitution for the value of $s$.

\begin{subequations}
\begin{align}
[\Delta\mathscr{D}_{1-s}\mathscr{D}_0^\dagger+2(2s-1)\ui\omega r]\Delta^sR_{+s}=\blambda\Delta^sR_{+s}\\
[\Delta\mathscr{D}_{1-s}^\dagger\mathscr{D}_0-2(2s-1)\ui\omega r]R_{-s}=\blambda R_{-s}.
\end{align}
\label{teus}
\end{subequations}

Or, equivalently, by written the operators explicitly,

\begin{subequations}
\begin{align}
\left\{\Delta^{-s}\frac{\ud}{\ud r}\Delta^{s+1}\frac{\ud}{\ud r}+\frac{1}{\Delta}[K^2+2\ui sK(r-M)-\Delta(\blambda-2s-4\ui s\omega r)]\right\}R_{+s}=0\\
\left\{\Delta^s\frac{\ud}{\ud r}\Delta^{1-s}\frac{\ud}{\ud r}+\frac{1}{\Delta}[K^2-2\ui sK(r-M)-\Delta(\blambda+4\ui s\omega r)]\right\}R_{-s}=0.
\end{align}
\label{teu}
\end{subequations}

The separation constant $\blambda$ is to be determined as a function of $s, m$ and $\omega$ from the angular equation
\footnote{Once again, the validity of equation (\ref{angularteukolsky}) is verified case by case. For instance, the proof for $s=1$ it is equivalent to (\ref{teum1}b) is as follows:
\[\mathscr L_0^\dagger\mathscr L_1-2a\omega\cos\theta+\blambda=\frac{\ud^2}{\ud\theta^2}+\frac{\ud Q}{\ud\theta}-Q^2-Q\cot\theta+\cot\theta\frac{\ud}{\ud\theta}+\frac{\ud}{\ud\theta}\cot\theta-2a\omega\cos\theta+\blambda.\]
The equality follows after substitutions of the identities $$\frac{\ud}{\ud\theta}\cot\theta=\cot^2\theta-1,\quad \frac{\ud Q}{\ud\theta}=-\frac{m\cos\theta}{\sin^2\theta}-a\omega\cos\theta$$ and $$\frac{\ud^2}{\ud\theta^2}+\cot\theta\frac{\ud}{\ud\theta}=\frac{1}{\sin\theta}\frac{\ud}{\ud\theta}\left(\sin\theta\frac{\ud}{\ud\theta}\right).$$
Similar calculations are performed for $s=$\textonehalf\ with $\mu=0$ and $s=2$.}
\begin{multline}
\left[\frac{1}{\sin\theta}\frac{\ud}{\ud\theta}\left(\sin\theta\frac{\ud}{\ud\theta}\right)+\left(a^2\omega^2\cos^2\theta-\frac{m^2}{\sin^2\theta}-2a\omega s\cos\theta-\right.\right.\\
\left.\left.\frac{2ms\cos\theta}{\sin^2\theta}+s^2\cot^2\theta-s-a^2\omega^2+2am\omega+\blambda\right)\right]S_s=0,
\label{angularteukolsky}
\end{multline}
requiring its solutions to be regular at $\theta=0$ and $\theta=\pi$. The particular result corresponding to Schwarzschild space-time ($a=0$), or equivalently vanishing frequency, has already been established \cite{abramowitz}, leading to $$\blambda=(\ell-s)(\ell+s+1)+2s+a^2\omega^2-2am\omega, \quad\ell\in\mathbb Z.$$
These functions $S_{s\ell m}e^{\ui m\phi}$ in this particular case are usually referred as spin-weighted spherical harmonics. Different solutions will be labeled by the number $\blambda$ for convenience, but they could be labeled by the more familiar number $\ell$.

Again, we are going to suppose these angular functions are normalised accordingly to the criteria above.

We are now able to cast these differential equations in a form we can compute the reflection coefficient. To do so, following \cite{Survey}. Consider the new independent variable $r_*$ so that
\begin{equation}
\frac{\ud}{\ud r_*}=\frac{\Delta}{\varrho^2}\frac{\ud}{\ud r},
\label{difr*}
\end{equation}
where $\displaystyle\varrho^2\equiv r^2+a^2-\frac{am}{\omega}$, not to be confused with $\rho^2$ or even $\bar{\rho}$. Define also the operators $\displaystyle\Lambda_\pm=\frac{\ud}{\ud r_*}\pm\ui\omega$ and $\displaystyle\Lambda^2=\Lambda_\pm\Lambda_\mp=\frac{\ud^2}{\ud r_*^2}+\omega^2$, and the new dependent variables $Y_{+s}\equiv|\varrho^2|^{1/2-s}\Delta^sR_{+s}$ and $Y_{-s}=|\varrho^2|^{1/2-s}R_{-s}$. Finally, introduce the abbreviations
\[P=\frac{2s}{\varrho^4}[2r\Delta-\varrho^2(r-M)]\]
and
\[\mathcal Q=\frac{\Delta}{\varrho^4}\left\{\blambda-(2s-1)\left[\frac{\Delta-2(s-1)r(r-M)}{\varrho^2}+(2s-3)\frac{r^2\Delta}{\varrho^4}\right]\right\}\]
to obtain
\begin{equation}
\Lambda^2Y_{\pm s}+P\Lambda_\mp Y_{\pm s}-\mathcal QY_{\pm s}=0.
\label{eqY}
\end{equation}

Here it is crucial to note directly from (\ref{eqY}), $Y_{-s}$ satisfy the complex conjugate equation for $Y_{+s}$. That means $Y_{+s}=Y_{-s}^*$.

The use of the formal change of dependent variable has led us to a much simpler differential equation, but the relation (\ref{difr*}) remains to be integrated. This is a simple task, after substituting the definition of $\Delta$. Defining the radii of horizons on Kerr metric as usual in the literature, $r_\pm=M\pm\sqrt{M^2-a^2}$, for $r>r_+$, which is the region we are concerned with, since we are treating the differential equation outside the black hole and the boundary conditions are to be imposed on the horizon,

\begin{equation}
r_*=r+\frac{2Mr_+-\frac{am}{\omega}}{r_+-r_-}\log\left(\frac{r}{r_+}-1\right)-\frac{2Mr_--\frac{am}{\omega}}{r_+-r_-}\log\left(\frac{r}{r_-}-1\right).
\label{r*}
\end{equation}

Suppose the coefficient of the first logarithm term on (\ref{r*}) is positive. Then, because $r_+>r_-$, so will be the second. The expression will then have the form $r_*(r)=r+A\log(r/r_+-1)-B\log(r/r_--1)=r+\log\left[\frac{(r/r_+-1)^A}{(r/r_--1)^B}\right]$, for $A$ and $B$ positive. Because the logarithm is a crescent function, $r_*$ will be a crescent function of $r$ as well.

\begin{figure}[h]
\centering
\includegraphics[width=.45\textwidth]{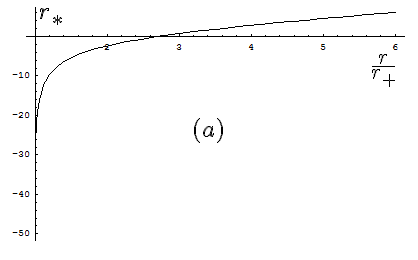}
\includegraphics[width=.45\textwidth]{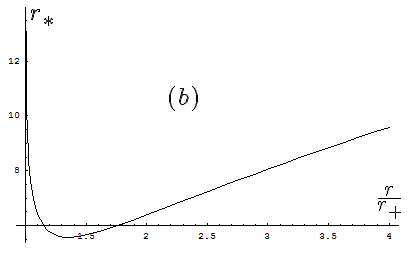}
\caption{Typical $r_*(r)$ relation. (a) for $\omega>\omega_s$ and (b) for $\omega<\omega_s$.}
\label{r*r}
\end{figure}

But it may not be the case, since $m$ is allowed to large. In this case, when $\displaystyle\omega<\omega_s\equiv\frac{am}{2Mr_+}$, that is, precisely when the coefficient of the first logarithm on (\ref{r*}) is negative, the derivative
\[\frac{\ud r_*}{\ud r}=\frac{am+\omega[2Mr+(r-r_+)(r-r_-)]}{\omega(r-r_+)(r-r_-)}\]
shows us $r_*$ it will decrease until it reaches a minimum value at $r=\sqrt{a^2-(am/\omega)}$, and increase indefinitely afterwards. A typical graph of $r_*(r)$ in this case is shown on figure (\ref{r*r}) (b).

In both cases, $r_*\rightarrow\infty$ as $r\rightarrow\infty$. When $\omega<\omega_s$, $r_*\rightarrow\infty$ as $r\rightarrow r_++0$, whereas when $\omega>\omega_s$, $r_*\rightarrow-\infty$.

We now show that $\omega_s$ has a special meaning that will be related to superradiance later. First, consider observers with four-velocity proportional to $\nabla^at$, where $t$ is the Killing parameter associated with $\xi^a$. In Boyer-Lindquist coordinates; for which $g_{tt}=g(\xi,\xi), g_{\phi\phi}=g(\psi,\psi)$ and $g_{t\phi}=g(\xi,\psi)$; we have in particular for these observers $0=\nabla_\phi t=g_{\phi t}u^t+g_{\phi\phi}u^\phi$. They rotate with coordinate angular velocity given by \cite{WaldGR} $\displaystyle\Omega=\frac{\ud\phi}{\ud t}=\frac{\ud\phi/\ud\tau}{\ud t/\ud\tau}=\frac{u^\phi}{u^t}=-\frac{g_{t\phi}}{g_{\phi\phi}}=\frac{a^2}{r_+^2+a^2}=$
\begin{equation}
\Omega=\frac{a}{2Mr_+}, 
\label{omegaa}
\end{equation}
valid on the horizon. So $$\omega_s=m\Omega.$$ We shall see that, when present, superradiant will occur for $\omega<\omega_s$. It will be important to compare with results from chapter four, that we consider the vector $\displaystyle\chi=\frac{\partial}{\partial t}+\tilde{\Omega}\frac{\partial}{\partial\phi}$. Using the expression for Kerr metric, the condition $\chi^a$ be a null vector leads to $\displaystyle\tilde{\Omega}=-\frac{g_{t\phi}}{g_{\phi\phi}}\pm\sqrt{\frac{g_{t\phi}^2-g_{tt}g_{\phi\phi}}{g_{\phi\phi}^2}}$. Restricting over the horizon $\Delta=0$, the numerator of the argument of the square root becomes $\rho^{-4}[a^2\sin^2\theta(r^2+a^2)^2]-\rho^{-4}[a^2\sin^2\theta(r^2+a^2)^2]=0$ and the equation simplifies to $\tilde{\Omega}=\Omega$. This result is crucial and the reader must bear it in mind when accompanying chapter four.

Further, making $\Delta=0$, to locate the horizon\footnote{In chapter four, the notion of horizon will be made more precise as well as its location.}, from (\ref{KerrNPbasisHor}) and (\ref{omegaa}), we see $\mathbf\chi=\mathbf{l'}$, since $\xi^a=\left(\frac{\partial}{\partial u_+}\right)^a$ and $\psi^a=\left(\frac{\partial}{\partial\phi_+}\right)^a$, an immediate consequence of applying chain rule on the definition of these coordinates in terms of Boyer-Lindquist's.

In order to cast (\ref{eqY})  in the form of (\ref{EDO1}), we employ the same trick from \cite{chandrasekhar}. Seek for a function $Z$ satisfying both a (\ref{EDO1})-like equation $\Lambda^2Z=V(r_*)Z$ and
\[Y=\mathcal{F}VZ+(\mathcal{W}-2\ui\omega\mathcal{F})\Lambda_+Z\Rightarrow\Lambda_-Y=\underbrace{-\frac{\Delta^s}{\varrho^{4s}}M\beta Z}_{\equiv\left(\frac{\ud}{\ud r_*}\mathcal{F}V+\mathcal{W}V\right)Z}+\underbrace{\mathcal{F}V+\frac{\ud}{\ud r_*}(\mathcal{W}-2\ui\omega\mathcal{F})}_{\equiv\mathcal{R}}\Lambda_+Z\]
for some functions $\mathcal{F,W}$.
Applying once more the operator $\Lambda_-$ on this last expression, and substituting on (\ref{eqY}), with $Y=Y_{+s}$, substituting the result in this first expression and finally treating $Z$ and $\Lambda_+Z$ as independent fields,
\[\mathcal{R}V-M\frac{\Delta^s}{\varrho^{4s}}\frac{\ud\beta}{\ud r_*}=Q\mathcal{F}V\quad\text{and}\quad\frac{\ud}{\ud r_*}\left(\frac{\varrho^{4s}}{\Delta^s}\mathcal{R}\right)=\frac{\varrho^{4s}}{\Delta^s}[Q(\mathcal{W}+2\ui\omega\mathcal{F})+2\ui\omega\mathcal{R}]+M\beta.\]
By inspection, note that $$\mathcal{K}\equiv\frac{\Delta^s}{\varrho^{4s}}\mathcal{RF}V+M\beta(\mathcal{W}-2\ui\omega\mathcal{F})$$ is a first integral of the equations relating $Y$ and $Z$, so it is enough to find $\mathcal{R}$, $\mathcal{F}$ and $\mathcal{W}$ to express $Z$ in terms of $Y$:
\[Z=\frac{\varrho^{4s}}{\mathcal{K}\Delta^s}[\mathcal{R}Y-(\mathcal{W}-2\ui\omega\mathcal{F})\Lambda_-Y]\quad\text{and}\quad\Lambda_+Z=\frac{\beta M}{\mathcal K}Y+\frac{\varrho^{4s}}{\Delta^s\mathcal K}\mathcal FV\lambda_-Y.\]
As in chapter two, Wronskian relations play a crucial role on investigating presence of superradiance. We can express the Wronskian of two solutions $Y_{1,2}$ in terms of the corresponding $Z_{1,2}$:
\begin{equation}
W(Y_1,Y_2)=\left[\mathcal{RF}V+\frac{\Delta^2}{\varrho^{4s}}M\beta(\mathcal{W}-2\ui\omega\mathcal{F})\right](Z_1\Lambda_+Z_2-Z_2\Lambda_+Z_1)=\frac{\mathcal{K}\Delta^2}{\varrho^{4s}}\tilde{W}(Z_1,Z_2),
\label{wywz}
\end{equation}
where the $\tilde{W}$ represents the Wronskian with respect to the equation for $Z$, $(\Lambda^2-V)Z=0$. Note that we have been using that $\Lambda_\pm$ appearing in expressions like $Z_1\Lambda_\pm Z_2-Z_2\Lambda_\pm Z_1$ act simply like a derivative, since the constant terms cancel one another. Eliminating $Y$ by going back to the radial functions, the left-hand side of this equation is $|\varrho^2|^{1-2s}W(\Delta^2R_{+s,(1)},\Delta^2R_{+s,(2)})$. Still, (\ref{difr*}) allow us to write the Wronskian not with respect to $r_*$ as we have been doing, but in terms of $r$ itself, so this expression becomes $\frac{\Delta}{\varrho^2|\varrho^2|^{2s-1}}W_r(\Delta^2R_{+s,(1)},\Delta^2R_{+s,(2)})$. From (\ref{wywz}),
\begin{equation}
\mathcal{K}\tilde{W}(Z_1,Z_2)=\frac{\varrho^{2(2s-1)}}{|\varrho^2|^{2s-1}}\Delta^{1-s}W_r(\Delta^2R_{+s,(1)},\Delta^2R_{+s,(2)}).
\label{wzwr}
\end{equation}

If the radial functions solutions to Teukolsky equations and its first derivatives are regular outside the black hole, so will be the Wronskian appearing on the right-hand side. But, because $m$ is allowed to assume arbitrary values, on the point $r^2=-a^2+\frac{am}{\omega}$, $\varrho^2$ vanishes. So the right-rand side may still be singular (the possibility of vanishing $\Delta$ is of no concern, since it happens only at $r=r_\pm$, and we are dealing with differential equations only on the outside of the black hole) This singularity divides the region to be consider in the differential equations in two regular parts, one for each sign of $\varrho^2$. In each of these regions separately, the constancy of the Wronskian for $Z$ holds. However, there is no reason why to expect the value of the Wronskian in these two branches should be equal. In fact, the constancy of the Wronskian for $R_{s}$, which is ensured by the regularity of the Teukolsky equations' solutions (which are physically meaningful), together with (\ref{wzwr}) teach us the Wronskian for $Z$ in these to branches are connected to each other by a factor of $(-1)^{2s-1}$. Here the role of spin is absolutely clear. We concluded that for half-integral $s$, that is, for fermions, the Wronskian for $Z$ is constant throughout all the region of interest, while for bosons, integral spins, this Wronskian changes its sign when passing trough the region where $\varrho^2$ vanishes.

For $s=$\textonehalf, corresponding to neutrino\footnote{Here, massless spin \textonehalf field, not exactly realistic according to our current understanding of these particles. It is justified because we are more interested in general properties of fermionic perturbations, not on specific field properties. This also justifies why we considered massless scalar fields from time to time.} waves, evaluating $P$ and $\mathcal Q$, and then $\mathcal{W,R,F}$, we find two different solutions for $\mathcal{R}$,  
which means two potentials $V$, 
that is, there are two possible one-dimensional problems whose correspondent problem of reflection and transmission, expressed in terms of $Y_{\pm s}$ are the same
\footnote{In fact, for this value of spin, we could have obtained $Z$ much more quickly by writing (\ref{teud}) with $\mu=0$ explicitly to obtain $\Lambda_-P_{+1/2}=\blambda\frac{\sqrt\Delta}{\varrho^2}P_{-1/2}$ and $\Lambda_+P_{-1/2}=\blambda\frac{\sqrt\Delta}{\varrho^2}P_{+1/2}$. Combining both equations and letting $Z_\pm=P_{+1/2}\pm P_{-1/2}$, we would have find $Z_\pm$ to satisfy the equation we wished with the potentials below.}. Hence, it is enough to treat only one of them.

\[{}_{s=\frac{1}{2}}V_\pm=\frac{1}{\varrho^4}\left\{\frac{\blambda}{\sqrt{2}\Delta}\mp\sqrt{\frac{\blambda\Delta}{2}}\left[(r-M)-\frac{2r\Delta}{\varrho^2}\right]\right\}.\]

As in chapter two, we seek solutions for $Z$ generally in the form
\begin{equation}
Z\rightarrow
\begin{cases}
e^{\ui\omega r_*}+
R^{\frac{1}{2}}
e^{-\ui\omega r_*} & \text{for}\ r_*\rightarrow+\infty , \text{(corresponding to $r\rightarrow\infty$)}\\
T^{\frac{1}{2}}
e^{\ui\omega r_*} & \text{for}\  r_*\rightarrow-\infty, \text{(corresponding to $r\rightarrow r_++0$, that is, over the horizon).}
\end{cases}
\label{asympbehav}
\end{equation}
The notation $r\rightarrow r_++0$ means that the lateral limit for values of $r\geqslant r_+$, that is, from outside the black hole.

To interpret physically these modes, we first must eliminate the frequency dependence on the radial coordinate. If we had defined
\[\frac{\ud}{\ud r^*}=\frac{\Delta}{r^2+a^2}\frac{\ud}{\ud r},\]
which gives, after integration,
\[r^*=r+\frac{2Mr_+}{r_+-r_-}\log\left(\frac{r}{r_+}-1\right)-\frac{2Mr_+}{r_+-r_-}\log\left(\frac{r}{r_-}-1\right).\]
A comparison with the $r_*(r)$ relation shows it shares the same limits as $r_*$ for $\omega>\omega_s$, that means, $r\rightarrow-\infty$ for $r\rightarrow r_++0$ and $r^*\rightarrow\infty$ as $r\rightarrow\infty$. Moreover, near horizon, $r^*\rightarrow\left(1+\frac{am}{2Mr_+\omega}\right)^{-1}r_*=\left(1-\frac{\omega_s}{\omega}\right)^{-1}r_*$.

Then, near horizon  our mode dependence with new radial coordinate is $\exp[\ui(kr^*+\omega t)]$, where $k=\omega-\omega_s$. We are going to make use of this relation on chapter five.

Then, the group velocity is given by $\displaystyle-\frac{\ud\omega}{\ud k}=-1$, hence this solution form impose the boundary condition prohibiting outgoing modes from the horizon. Notwithstanding, the phase velocity $\displaystyle-\frac{\omega}{k}=-\left(1-\frac{\omega_s}{\omega}\right)^{-1}$ can change sign accordingly with the regime allows (bosonic) superradiance or not. It in perfect accordance with our expectation on physical grounds. An observer near horizon will never see outgoing modes, the group velocity near horizon shows he will see ingoing radiation with unit speed. We have chosen modes to respect this boundary condition. For an observer at infinity, the group velocity will coincide with the phase velocity, since $r^*\to r_*$ when $r\to\infty$\footnote{It can be seen comparing the integrated formulas for the two radial coordinates in terms of the Boyer-Lindquist's.}, showing he may see outgoing radiation in some circumstances.

Because the potentials are real, we can choose $Z$, as like $Y$ to be a real quantity. Then, we may also allows one to interpret $|R|=|R^\frac{1}{2}|^2\geqslant0$ and $|T|=|T^\frac{1}{2}|^2\geqslant0$ as the reflection and transmission coefficients respectively.

Besides, because the Wronskian retains its value in all the interval, as we have seen, we may set its value for $r_*\rightarrow-\infty$ with its value for $r_*\rightarrow+\infty$ for solutions respecting (\ref{asympbehav}). So, $-2\ui\omega(|R|-1)=2\ui\omega |T|\Leftrightarrow |R|+|T|=1$. This conservation law shows that \emph{no superradiance occurs in any frequency}. Absence of superradiance for massive fermions will be deduced by other means later on section 3.4, but using the very same techniques, only letting calculations be a little longer, \cite{chandrasekhar} could deduce the reflection and transmission coefficients associated with spin-\textonehalf field with mass $m_e$ satisfies $1-|R_+|=\frac{\omega}{\sqrt{\omega^2-m_e^2}}|T_+|$, showing mass does not change our conclusions.

For $s=1$ (correspondingly to electromagnetic perturbations), also two possible potentials are found.

\[{}_{s=1}V_\pm=\frac{\Delta}{\varrho^4}\left[\blambda-\left(a^2-\frac{am}{\omega}\right)\frac{\Delta}{\varrho^4}\mp\ui\sqrt{a^2-\frac{am}{\omega}}\varrho^2\frac{\ud}{\ud r}\frac{\Delta}{\varrho^4}\right],\]
not necessarily real.

Our observation $Y_{-s}^*=Y_{+s}$ now plays an important role as follows.
We must interpret the reflection and transmission coefficients in terms of these functions, not in terms of $Z$. It appears we have to repeat the construction of the potential which describes the differential equation satisfied by $Z\equiv Z_+$ corresponding to $Y_{+s}$ to construct the potential which describes the equation for $Z_-$ related to $Y_{-s}$. Fortunately this labour is dispensable, thanks to our very observation. For this $s=1$, noting that complex conjugation can be done by reversing the sign of $\omega$ (but keeping $\varrho^2$ unchanged), we conclude that $Z_-$ satisfies precisely the same equation as $Z_+$. Consequently the asymptotic behaviour we seek is

\[Z_\pm\rightarrow
\begin{cases}
e^{\pm\ui\omega r_*}+\mathscr R_\pm e^{\mp\ui\omega r_*} & \text{for}\ r_*\rightarrow+\infty\\
\mathscr T_\pm e^{\pm\ui\omega r_*} & \text{for}\ r_*\rightarrow-\infty,
\end{cases}\]
again only outgoing modes are permitted over the horizon. The difference is simply how we must find the reflection and transmission coefficients. They are found by $R=\mathscr R_+\mathscr R_-$ and $T=\mathscr T_+\mathscr T_-$ respectively. Because they can be written as a square modulus of a complex number (the coefficients appearing in the corresponding form of $Y$s) these quantities are real and non-negative.

Now, we must consider these potentials in separate cases.
\begin{enumerate}
\item For $0<\omega<\omega_s<\frac{m}{a}$, the argument of the square root is negative, so the potentials are real. Again, we employ (\ref{asympbehav}), but, this time, remember the Wronskian changes its sign for finite $r_*$, so we equate the Wronskian between the \emph{solutions} $Z_+$ and $Z_-$ for $r_*\rightarrow-\infty$ with the opposite of the one for $r_*\rightarrow+\infty$, so we obtain the conservation law $R-T=1$, which shows that \emph{superradiance occurs} for this interval.
\item The case $\displaystyle\frac{m}{a}\leqslant\omega_s=\frac{am}{2M(M+\sqrt{M^2-a^2})}$ is of no concern, since we are supposing $M>a$, so $\displaystyle a^2<2M^2<2M(M+\sqrt{M^2-a^2})\Rightarrow\frac{a}{2M(M+\sqrt{M^2-a^2})}<\frac{1}{a}$.
\item If $\omega_s\leqslant\omega\leqslant\frac{m}{a}$, the potential is also real, but in this case the potential is non-singular because $\varrho^2$ does not vanish, so Wronskian equality leads to $R+T=1$ and \emph{no superradiance occurs}.
\item If $\omega>\frac{m}{a}$, the potential has an imaginary part. It is not a problem for us now, since we are dealing with two solutions $Z_+$ and $Z_-$, their Wronskian will be kept constant, and its sign will not be reversed in this interval of frequencies. The only difference is that these two solutions are not related by complex conjugate operation. Anyway, this Wronskian relation also leads to $R+T=1$, so no superradiance is to be observed.
\end{enumerate}

For $s=2$, four different potentials $V$ are found:
\begin{multline*}
{}_{s=2}V=\frac{\Delta}{\varrho^8}\left[q-\frac{\varrho^2}{(q\pm3(a^2-\frac{am}{\omega}))^2}\left\{[q\pm3\left(a^2-\frac{am}{\omega}\right)\left[\Delta\varrho^2q''-2\varrho^2q-2r(q'\Delta-\Delta'q)\right.\right.\right.+\\
+\left.\left.\left.\left.\varrho^2\left(\kappa\varrho^2-q'\pm3\left(a^2-\frac{am}{\omega}\right)\Delta'(q'\Delta-\Delta'q)\right)\right]\right]\right\}\right],
\end{multline*}
where prime denotes differentiation with respect to $r$, $q=\blambda\varrho^4+3\varrho^2(r^2-a^2)-3r^2\Delta$ and \\ $\kappa=\pm\sqrt{36M^2-2\blambda\left[\left(a^2-\frac{am}{\omega}\right)(5\blambda+6)-12a^2\right]\pm6\left(a^2-\frac{am}{\omega}\right)\blambda(\blambda+2)}$, where the sign in front of the square root is to be chosen independently of the choice of the others. Although it has a cumbersome form, it shares the property we verified for $s=1$ that $Z_+$ and $Z_-$ are both solutions of the same differential equation. Consequently, the analysis concerning the presence or absence of superradiance is analogous, and the conclusion is the same: superradiance is present (when and only when) $\omega<\omega_s$.

\subsection{Physical Interpretation and Reasonableness of Definitions of Reflection and Transmission Coefficients}
Now, after having found the asymptotic behaviour of perturbations, we are apt to extract further physical significance from the problem of transmission and reflection following \cite{chandrasekhar}. For the electromagnetic case, we may write the energy-momentum tensor $\propto F^{ac}F^b_c-\frac{1}{4}g^{ab}F_{cd}F^{cd}$ in terms of the three scalars we wrote Teukolsky equations with:
\[T_{ab}\propto\{|\phi_0|^2n_an_b+|\phi_2|^2l_al_b+2|\phi_1|^2[l_{(a}n_{b)}+m_{(a}\bar m_{b)}]-4\phi_0^*\phi_1n_{(a}m_{b)}-4\phi_1^*\phi_2l_{(a}m_{b)}+2\phi_2\phi_0^*m_am_b\}+\mathrm{c.c.},\]
where $\mathrm{c.c.}$ denotes the complex conjugate of the former term.

Our aim is to prove the reflection (transmission) coefficient we defied above by $R=\mathscr R_-\mathscr R_+$ ($T=\mathscr T_-\mathscr T_+$) represent the ratio between reflected (transmitted) and incident energies per unit of time and solid angle across certain closed surface. In case of reflected wave, this surface can be taken  to be orthogonal to $\left(\frac{\partial}{\partial r}\right)^a$ for convenience; whilst for transmitted wave, we can take the horizon itself, whose normal is $\chi^a$, defined above.

For reflection coefficient, we contract the energy momentum tensor with $\xi^a$ and $\left(\frac{\partial}{\partial r}\right)^a$ before integration over the surface orthogonal to these vectors. To calculate it explicitly from the solutions we found above, recall $\phi_0=R_{+1}S_{+1}$ and $\phi_2=\frac{1}{(\bar\rho^*)^2}R_{-1}S_{-1}$ with $R_{+1}=\frac{|\varrho^2|^{\frac{1}{2}}}{\Delta}Y$ and $R_{-1}=|\varrho^2|^{\frac{1}{2}}Y^*$. From the asymptotic behaviours of functions $Z$, we may write the corresponding asymptotic behaviours of functions $Y$ and subsequently for the radial functions required for evaluation. The calculations are somewhat lengthly. The results are
\begin{equation*}
R_{+1}\to
\begin{cases}
-4\omega^2\frac{e^{\ui\omega r_*}}{r}-\left(\blambda+2\omega\sqrt{a^2+\frac{am}{\omega}}\right)\mathscr R_+\frac{e^{-\ui\omega r_*}}{r^3} & r\to\infty\\
-4\omega^2\sqrt{\varrho^2(r_+)}\mathscr T_+\frac{e^{\ui\omega r_*}}{\Delta} & r\to r_++0
\end{cases}
\end{equation*}
and
\begin{equation*}
R_{-1}\to
\begin{cases}
-\left(\blambda+2\omega\sqrt{a^2+\frac{am}{\omega}}\right)\frac{e^{\ui\omega r_*}}{r}-4\omega^2\mathscr R_-re^{-\ui\omega r_*} & r\to\infty\\
-\frac{(\blambda+2\omega\sqrt{a^2+\frac{am}{\omega}})\sqrt{\varrho^2(r_+)}}{\varrho^2(r_+)\left[\varrho^2(r_+)-\frac{\ui(r_+-M)}{\omega}\right]}\mathscr T_-\Delta e^{\ui\omega r_*} & r\to r_++0.
\end{cases}
\end{equation*}

As before, we are allowed to distinguish incident, reflected and transmitted parts of radiation and calculate the energy flux corresponding to reflected and incident parts separately. The ratio between these quantities is found to be exactly $R$, which justifies our interpretation of reflection coefficient. Also, it is found $\lim_{r\to\infty}r^2\xi^aT_{ab}\left(\frac{\partial}{\partial r}\right)^b<\infty$ as desired.

To calculate the transmission coefficient, these integrations are more properly evaluated at the horizon rather than infinity. Therefore $\xi^aT_{ab}$ should be contracted with the surface element given by $\chi_b2Mr_+\sin\theta\ud\theta\ud\phi_+\ud u_+$ in Kerr-Schild coordinates, which accordingly to what we saw $\chi_b$ is identified with the covector $l'_b$ in Newman-Penrose basis. Hence, $$2Mr_+T_{ab}\chi^a\chi^b=\left.\frac{\ud^2E}{\ud t \ud \mathfrak o}\right|_{r_+}+\Omega\left.\frac{\ud^2 L_z}{\ud t\ud \mathfrak o}\right|_{r_+},$$ where the first term in the right hand side denotes the energy per unit of time and solid angle $\ud \mathfrak o$ and similarly $L_z$ denotes the axial component of angular momentum. Imposing the ratio between the axial component of angular momentum and energy is $m/\omega$, valid for fields with dependence $\exp[\ui(m\phi-\omega t)]$, we obtain $$\left.\frac{\ud^2E}{\ud t\ud \mathfrak o}\right|_{r_+}=\frac{2Mr_+}{1+\Omega/\omega} T_{ab}\chi^a\chi^b.$$ The contraction above is simply $\frac{1}{2\pi}|\tilde\phi_0|^2$, with $\tilde\phi_0$ is defined as $\phi_0$, bow with respect to the new Newman-Penrose basis (\ref{KerrNPbasisHor}), i.e., $$\tilde\phi_0=F_{ab}l^{\prime a}m^b=\frac{\Delta}{2(r^2+a^2)}\phi_0.$$

The remaining calculations are straightforward. The ratio between transmitted and incident portions of energy flux is precisely $T$, as we wished to show.

Likewise, we can follow this procedure to other spin fields, namely the neutrino and gravitational we solved in the previous section. The evaluation of fermionic energy-momentum tensor can be followed accordingly to the prescription we shall utilise on chapter four below and for gravitational perturbation, one may in principle resort to one of the well-known mechanisms to compute energy carried by gravitational waves, such as an energy-momentum pseudo-tensor of gravitational field or an evaluation of second-order correction of Einstein tensor. Both strategies of gravitational case lead to enormous labour and will not be explored further here. Yet, we believe the physical meaning we have firstly attributed to reflection and transmission coefficients are sufficient to predict the presence or absence of superradiance, although it is not sufficient to prove it rigorously.

\section{Another Approach}
At this point, a natural question arises. In the first section of this chapter, we noticed, that, in Kerr geometry, every spinorial component of a Dirac spinor satisfies Klein-Gordon equation. At first glance, it may appear that as a consequence of the analysis of spin-0 field, fermions should also present superradiance, in discordance to what we have deduced using Newman-Penrose formalism. On that formalism, equations seem too complicated to address this question, since Teukolsky equations are of second order already when we separate its variables, and these second-order equations associated with different spins do not coincide. Therefore, we use another approach, based on \cite{WaldGR} to convince oneself about the existence of superradiance for bosonic modes and not for fermionic ones.

First, consider a scalar filed $\varphi$, for which there is a conserved current ($\nabla^aj_a=0$ as a direct consequence of Klein-Gordon equation) given by $j^a=-\ui(\varphi^*\nabla^a\varphi-\varphi\nabla^a\varphi^*)$. Then, by Gauss theorem, the integral of $j^an_a$ along any closed surface vanishes, in particular along a 3-surface consisting on the union of
\begin{description}
\item[(i)] A timelike surface placed at large distances from the black hole over which the flux of $j^a$ can be interpreted as the net `number particle' across it during a time interval $\Delta t$, say the parameter of timelike isometry has changes from $t$ to $t+\Delta t$ along this piece of surface;
\item[(ii)] a pair of spacelike surfaces, one in the future of the other, the second being obtained from the first by carrying it by the timelike isometry along parameter change $\Delta t$; and
\item[(iii)] a portion of the horizon necessary to close the entire surface.
\end{description}
Supposing, as before, a wave form $\varphi=\varphi_0e^{\ui(m\phi-\omega t)}$ for the scalar field, the integral over \emph{(ii)} vanishes, since, by symmetry, the integral over each of the spacelike surfaces has to be the same, but, since the surface is closed, we must reverse the orientation in one of these surfaces in order to preserve the convention of outward orientation along the entire surface. Therefore, the integral over \emph{(i)}, which is the integral we seek is simply the integral over \emph{(iii)} with its sign reversed. Fortunately, we know explicitly the vector field $n^a$ along \emph{(iii)} to be $-\chi^a$, defined above. Hence,
\[-\chi_aj^a=\ui[\varphi^*(-\ui\omega+\ui m\Omega)\varphi-\varphi(\ui\omega-\ui m\Omega)\varphi^*]=2|\varphi_0|^2(\omega-m\Omega),\]
which shows there is a net flux of particles leaving the black hole whenever $\omega<m\Omega$, characterising superradiance.

The same calculation could be repeated with the energy current $T_{ab}\xi^b$ --- which is also divergent-free since $T_{ab}=T_{ba}$, $\xi^a$ obeys the Killing equation and $T_{ab}$ itself is divergent-free --- instead of the number current as we did. The only difference is that a time average must be taken in order to obtain the equivalent result.

The solution of the question that opens this section is answered by the following reasoning. We saw that each component of a Dirac spinor satisfies an independent Klein-Gordon equation. However, this does not mean a general solution of the corresponding Klein-Gordon system is also a solution of Dirac equation, instead, the solutions of the later forms a subspace of the solutions of the former. This can be easily checked by noticing the dimension of the space of solutions of Klein-Gordon system is eight (it has eight constants to be determined from boundary conditions, since each equation is of second-order), whilst the dimension of solutions of Dirac equation is four.

Conservation of (\ref{DiracCurrentSpinor}) imposes a condition on the solutions that may not be satisfied by the corresponding Klein-Gordon problem. And, as we have seen, this current is null or timelike and future-directed vector. Therefore $-\chi^aj_a$ is non-negative everywhere on horizon\footnote{This can be proved by the same techniques we pointed on the previous footnote explaining why $j^a$ is timelike for massive fermions.} and, by the same reasoning, we conclude no superradiance occurs on the subspace of solutions of Klein-Gordon system for which Dirac equations (\ref{DiracSpinor}) holds. It is worth mentioning that this approach has the significant advantage that it proves superradiance to be absent for \emph{any} half-integral spin radiation, not only for massless neutrinos as we did solving Teukolsky equations. See remark on section 3.1 about the existence of well posed initial value problem for higher spins.

Curiously, on chapter five, it will become apparent that Dirac number current conservation also is enough to prevent superradiance in a completely different system: charged Dirac particles in Minkowski space-time with a sharp electromagnetic field. Even though in this case the `second order Dirac equation' in presence of an external field does \emph{not} coincide with the corresponding Klein-Gordon system of equations, fact which could be foreseen by physical grounds, since spin carries dipole moment, which interacts with the external field.

\chapter{Relations with Black Hole Thermodynamics}

\section{Superradiance and Ordinary Thermodynamics}

It is possible to construct heuristic arguments to justify that superradiance is thermodynamically favourable. We follow mostly \cite{beke}. This reference considers changes in its energy $E$, momentum $\vec{P}$ as measured by an observer on the laboratory after the emission (or absorption) of a photon with energy $\hslash\omega$ and momentum $\hslash\vec{k}$\footnote{Here we kept the constant $\hslash$ to simplify the identification of arguments of quantum origin. Note it does not appear in final expressions of stimulated emission, but it does on spontaneous version.}. Primed quantities refer after the emission, whilst unprimed refer to quantities before emission. Up to the first order on $\omega$, $\vec{k}$, the proper energy $\tilde{E}$, connected by a Lorentz transformation with $E$: $\ud\tilde{E}=\frac{1}{\sqrt{1-v^2}}(\ud E-\vec{v}\cdot\vec{P})$ changes accordingly to
\begin{equation}
\tilde{E}'-\tilde{E}=\mp\ \frac{\hslash(\omega-\vec{v}\cdot\vec{k})}{\sqrt{1-v^2}},
\label{bek0}
\end{equation}
where the upper sign refers to emission and the lower to absorption, the velocity $\vec{v}$ is given by $\frac{\partial E}{\partial\vec{P}}$. The point is when $\omega<\vec{v}\cdot\vec{k}$ (a body propagating through some media with velocity greater than light in that transparent media, for example), emission is accompanied by excitation and absorption by de-excitation.

Bekenstein call this process just described as spontaneous superradiance, opposed to superradiant amplification as follows.

If you consider incident radiation on a body with intensity $I(\theta,\phi,\omega)$, where $\theta, \phi$ are the polar and azimuthal angles and $\omega$ is the frequency the energy per unit time absorbed by the body within a solid angle $\ud \mathfrak o$ will often be possible to be written as $\frac{\ud E}{\ud t}=aI\ud \mathfrak o-W$, where the coefficient $a$ may depend the geometry of the body, on the frequency and direction of incident wave, but not on its intensity. $W$ takes into account any kind of radiation spontaneously emitted. If $\hat{n}$ denotes the unit vector pointing toward the direction of propagation, the absorption of radiation will cause the momentum to be changed by a rate of $n(\omega)aI\ud \mathfrak o\ \hat{n}$, where $n(\omega)>0$ is the index of refraction. Denoting by $\vec{U}$ the rate  of spontaneous emission of momentum, $\frac{\ud\vec{P}}{\ud t}=n(\omega)aI\ud \mathfrak o\ \hat{n}-\vec{U}$. The change on entropy $S$ of a reversible process can be calculated by means of $T\ud S=\ud\tilde{E}$, that can be seen for us as a definition of an `effective temperature' $T$. When the concept of entropy is well posed, the effective temperature \emph{is} the temperature in the usual sense and therefore satisfies $T>0$ and is the same for different objects when they are in thermal equilibrium and are allowed to exchange energy. The proper energy and the energy $E$ are connected by a Lorentz transformation $\ud\tilde{E}=\frac{1}{\sqrt{1-v^2}}\left(\ud E-\vec{v}\cdot\ud\vec{P}\right)$. So
\begin{equation}
\frac{\ud S}{\ud t}=\frac{1}{T\sqrt{1-v^2}}\left[aI(1-n(\omega)\vec{v}\cdot\hat{n})\ud \mathfrak o-W+\vec{U}\cdot\vec{v}\right].\label{bek1}
\end{equation}

For sufficiently high intensities, only the first term contributes significantly ($W$ and $\vec{U}$ do not depend on the incident radiation, by hypothesis), and, because the entropy of the radiation itself is proportional to the logarithm of the intensity\footnote{Remember radiation particles are treated as identical.} plus the entropy contribution of spontaneous emission, which is supposed to be independent of incident intensity, we may neglect the change in entropy of radiation comparing to the change in entropy of the body, which increases linearly with intensity. Therefore, whenever $\omega-\vec{v}\cdot{k}<0$, where $\vec{k}=n(\omega)\omega\hat{n}$, the requirement that the total entropy ought not to decrease, we are enforced to conclude $a<0$, that is, superradiance occurs. Recall that we supposed $a$ to be independent on the intensity, so the conclusion must be the same even for low intensities.

Bekenstein \cite{beke} gave examples of application of his method. In one of which, he managed to describe Cherenkov radiation in terms of these two phenomena described. Imposing no change in proper energy, in (\ref{bek0}), whose the right hand side cannot vanish for subluminal motion, and vanishes on superluminal motion precisely when the root for $\vartheta$ of $\vec v\cdot\vec k=vk\cos\vartheta$ is the one to create the usual Mach's cone. And arguments from (\ref{bek1}) predict all modes for which $\cos\vartheta>n^{-1}(\omega)$ are amplified, yet to be detected by experiments.

Similar arguments holds when the object is rotating as a whole with angular velocity $\Omega$ even if its is surrounded by vacuum, as the Zel'dovich cylinder we studied on the second chapter. To avoid precession, we assume axial symmetry. The changes on energy and angular momentum $L$ are calculated in an exactly analogous way, only it may be worthwhile bearing in mind that the coefficient $a$ may clearly depend on the azimuthal number $m$ for modes proportional to $e^{\ui(m\phi-\omega t)}$. Using the fact that angular momentum and energy are on the proportion $m:\omega$, and the relation between $\tilde{E}$ and $E$ from classical mechanics, $\ud\tilde{E}=\ud E-\vec{\Omega}\cdot\ud\vec{L}$,
\begin{equation}
\frac{\ud S}{\ud t}=\frac{1}{T}\left(\frac{\omega-m\Omega}{\omega}\ aI\ \ud \mathfrak o-W+\Omega U\right),
\label{bek2}
\end{equation}
from which, we conclude superradiance to be present ($a<0$) whenever $\omega<m\Omega$, in perfectly accordance with the results from chapter two.

Formulas (\ref{bek1}) and (\ref{bek2}) show that superradiance is not only allowed, but also thermodynamically favourable. 

At first glance, it may seem Bekenstein's semi-classical argument apply equally well for both bosons and fermions. But, although never remarked by Bekenstein himself, it is not true. When we neglected effects of both spontaneous emission $W$ and $U$ and the entropy carried by radiation itself, we were supposing it were possible to construct beans of arbitrary high intensities for a given set of quantum numbers, which is false for (the intrinsically quantum mechanical) fermionic radiation thanks to Pauli exclusion principle. Therefore, we must bear in mind this arguments is only applicable to bosons. Indeed, as we have seen from other means, superradiance is absent for fermions, and if this argument were applicable, we would conclude that we could violate the second law of thermodynamics from absence of fermionic superradiance. In order to ensure second law of thermodynamics is preserved, however, we need to calculate the details of spontaneous and stimulated emission of the rotating device.

\newpage
\section{Superradiance and Black Hole Thermodynamics}

Before using the so-called laws of black hole thermodynamics as a framework for investigating superradiance, we shall first explain their meanings in details.

\subsection{Area Theorem}

In order to prove the area theorem, which will soon be stated, one crucial ingredient is the Raychaudhuri's equation for null geodesics, which we shall derive, following \cite{HE}. Consider a family of null geodesics paremetrised by an affine parameter $\lambda$ and let $V^a$ denote its tangent vector\footnote{It is worth mentioning that, unlike timelike geodesics, this parameter and its tangent vector are not unique.}. Our object of study is the subspace orthogonal to $V^a$. Since this three-dimensional subspace includes the vector $V^a$ itself ($g(V,V)=0$), we are interested only in the set of equivalence classes of the equivalent relation given by $K^a\sim k^a$ if $K^a-k^a=\mu V^a,\ \mu\in\mathbb{R}$. To endow this set of equivalence classes with the corresponding two-dimensional vector space structure along the whole geodesics congruence, we choose a basis $(V^a,L^a,T^a,S^a)$ of $T_pM$ for a point $p$ in the congruence, where $T^a$ and $S^a$ are spacelike vectors normalised to unit, orthogonal to each other and to $V^a$ with respect to $g_{ab}$, and $L^a$ is a (well-chosen) null vector orthogonal to $S^a$ and $T^a$ and such that $g(V,L)=-1$. We construct a basis of $T_qM$ for other points $q$ of the null geodesics by simply parallelly transporting the basis of $T_pM$ along this geodesics\footnote{Here we make use of the well-known fact that the Fermi derivative coincides with covariant derivative, since any curve of the family is a geodesic. See \cite{HE} for details.}. The tensor $h_{ab}\equiv g_{ab}+2V_{(a}L_{b)}$ projects a vector onto this two-dimensional subspace orthogonal both to $V^a$ and $L^a$. Written in term of the basis $(V^a,L^a,T^a,S^a)$, we have in components $h_{\mu\nu}=\mathrm{diag}(0,0,1,1)$.

We define the \emph{expansion tensor} by $\theta_{ab}\equiv h^c_ah^d_b\nabla_{(d}V_{c)}$, its trace $\theta\equiv\nabla^aV_a$ is called the \emph{expansion}; its (clearly symmetric) trace-free part, $\sigma_{ab}\equiv\theta_{ab}-\frac{1}{2}h_{ab}\theta$, is called \emph{shear tensor} of the congruence, also, we define the (clearly antisymmetric) \emph{vorticity} or the \emph{twist} by $\omega_{ab}\equiv h^c_ah^d_b\nabla_{[d}V_{c]}$. It is easily checked that
\begin{equation}
B_{ab}\equiv h^a_ch^d_b\nabla^cV^d=\frac{1}{2}\theta h_{ab}+\sigma_{ab}+\omega_{ab}.
\label{exptensor}
\end{equation}
Now we can state the Raychaudhuri's equation for null geodesics as follows
\begin{equation}
\frac{\ud\theta}{\ud\lambda}=-R_{ab}V^aV^b+2\omega_{cd}\omega^{cd}-2\sigma_{cd}\sigma^{cd}-\frac{1}{2}\theta^2
\label{raychaudhuri}
\end{equation}
\begin{proof}
To prove it, let $\mathfrak B_{ab}\equiv\nabla_b V_a$ and consider its (covariant) derivative along the tangent vector
\begin{multline*}
V^c\nabla_c(\mathfrak B_{ab})=V^c\nabla_c\nabla_bV_a=V^c(\nabla_b\nabla_cV_a-R_{adbc}V^d)=\nabla_b(V^c\nabla_cV_a)-\nabla_cV_a\nabla_bV^c-R_{adbc}V^dV^c=\\
=-\mathfrak B_{ac}\mathfrak B^c_b-R_{adbc}V^cV^d.
\end{multline*}
Now we project the indexes $a$ and $b$ on the subspace spanned by $(T^a,S^a)$ using the positive definite metric $h_{ab}$. This leads to $V^c\nabla_c B_{ab}=-B_{ac}B^c_b-h^e_ah^f_bR_{efbc}V^cV^d$, from which much information can be extracted. For our purpose, we it is sufficient to take its trace, noticing that the trace of $B_{ab}$ is the same as the trace of $\theta_{ab}$, to obtain
\[\frac{\ud\theta}{\ud\lambda}=-\theta_{ab}\theta^{ba}-R_{ab}V^aV^b.\]
The proof is complete by contracting (\ref{exptensor}) to give $B^{ab}B_{ba}=\frac{1}{2}\theta^2+\sigma^{ab}\sigma_{ab}-\omega^{ab}\omega_{ab}$.
\end{proof}

Now, by Frobenius' theorem (see, for example, \cite{WaldGR} or \cite{epoisson}. We shall not reproduce a proof of this well known theorem for sake of brevity) that in the case we are dealing with null geodesics that generates the event horizon, which is a null surface, the vorticity vanishes. If we suppose the null energy condition holds, then the right hand side of (\ref{raychaudhuri}) is non-positive, since the square of shear is non-negative\footnote{It is easily verified taking into account its symmetry property, so we can diagonalise the matrix $\Sigma$ formed by its components with real eigenvalues. Since trace of an operator is basis-independent, $\sigma_{cd}\sigma^{cd}=$Tr$(\Sigma^2)$ is a sum of squares of real numbers.}. The null energy condition states that for all null vectors $k^a$, $T^{ab}k_ak_b\geqslant0$. Contracting Einstein's equations with $k^ak^b$, we conclude that $R_{ab}k^ak^b\geqslant0$.

The null energy condition is a particular case of the weak energy condition, that is, $T_{ab}\xi^a\xi^b\geqslant0\ \forall\xi\in \mathscr X(M), \xi$ timelike. This energy condition can be interpreted by saying that the energy density as measured by any observer is non-negative. If weak energy condition, which may seem reasonable for some readers, holds, then, by continuity, null energy condition must hold. It is important to note, however, that weak energy condition does \textbf{not} imply the analogous relation $R_{ab}\xi^a\xi^b\geqslant0$, since $g(\xi,\xi)\neq0$. In order to that relation to hold, we should require the strong energy condition, that is $\left(T_{ab}-\frac{1}{2}T^c_cg_{ab}\right)\xi^a\xi^b\geqslant0$. But we emphasise that the null energy condition is the most general one, since it does not imply any other energy condition.

Back to (\ref{raychaudhuri}), we can state that
\begin{equation}
\frac{\ud\theta}{\ud\lambda}\leqslant-\frac{1}{2}\theta^2\Rightarrow\theta^{-1}(\lambda)\geqslant\theta_0^{-1}+\frac{\lambda}{2},
\label{theta-1}
\end{equation}
where $\theta_0$ denotes an initial value of the expansion.

To see a geometrical meaning for the expansion, consider an infinitesimal change in the tangent vector, based on \cite{epoisson}, $V^a(\lambda+\ud \lambda)+(\delta^a_b+B^a_b\ud \lambda)V^b(\lambda)$. The Jacobian of this transformation is $\det(\delta^a_b+B^a_b\ud\lambda)=1+\theta\ud\lambda$.\footnote{Here, we used the identity $\det(\mathbf{1}+At)=1+t\ \text{Tr}A+\mathscr{O}(t^2)$, for $A\in\text{Mat}(\mathbb{C},n)$, $t\in\mathbb{C}$ and $n\in\mathbb{N}$ \cite{barata}}.
Thus $\theta$ gives the relative change in the cross-sectional area with the affine parameter,
\begin{equation}
\theta=\frac{1}{A}\frac{\ud A}{\ud\lambda}.
\label{geomtheta}
\end{equation}

Another way to proceed, useful to us later is to consider a Jacobi field $J^a$ (see \cite{barata} for a definition and an excellent discussion) with initial condition $J^a(0)=0$. From construction the Jacobi field representing separation between neighbouring geodesics are Lie dragged along each geodesics, hence $\mathsterling_VJ=0$. We are interested only on evolution of the orthogonal part $J^a$ with respect with the tangent vector. This is because $V^b\nabla_b(V^aJ_a)=V^aV^b\nabla_bJ_a+V^bJ_a\nabla_bV^a=0$, since the second from vanishes from geodesics equation, the first term can be rewritten as $V_b(\mathsterling_VJ)^b+J^aV_b\nabla_aV^b=0$, provided $g(V,V)$ is constant. That means that the projection $V^aJ_a$ is constant along the geodesics.

Jacobi equation reads $\frac{\ud^2J^a}{\ud\lambda^2}=-R^a_{\ bcd}V^bJ^cV^d$, where $\lambda$ is an affine parameter and $V^a$ is the tangent vector, as before. So, the evolution of this field with $\lambda$ can be obtained by
\begin{equation}
J^\alpha(\lambda)=E^\alpha_\beta\left.\frac{\ud J^\beta}{\ud\lambda}\right|_{\lambda=0},
\label{defE}
\end{equation}
supposing $J^a(0)=0$.
This equation is to be understood only on the subspace spanned by $(T^a,S^a)$, this is the reason why we have written (only here) its indexes in Greek alphabet, to avoid confusion.
Let the set of $e_\alpha$ denote a basis for this subspace at some point on the congruence and paralleling transporting it along the geodesics that contains that point, as before. Then
\begin{equation}
\frac{\mathrm D J^\alpha}{\ud\lambda}=V^a\nabla_a[J^b(e_\alpha)_b]=(e_\alpha)_bV^a\nabla_aJ^b=(e_\alpha)_b\mathfrak B^b_{\ a}J^a=B^\alpha_{\ \beta} J^\beta,
\label{JwB}
\end{equation}
where on the second step we used that the basis has been parallelly transported and on the third the result $\mathsterling_VJ=0\Rightarrow V^b\nabla_bJ^a=J^b\nabla_bV^a$. Comparing (\ref{JwB}) with the derivative of (\ref{defE}), we conclude, in matrix notation
\[\mathbf{B}=\frac{\ud\mathbf{E}}{\ud\lambda}\ \mathbf{E}^{-1},\]
whose trace gives
\[\theta=\frac{1}{\det \mathbf{E}}\frac{\ud}{\ud\lambda}\det\mathbf{E},\]
from which not only (\ref{geomtheta}) follows again, but also we see that if there is a point $p$ where the expansion becomes infinite, on this point $\mathbf{E}$ is singular, that means that neighbouring geodesics of that particular congruence collapses into a point. When this happens, we say that $p$ is a \emph{conjugate point} with respect to the point where the initial condition for the congruence is met. An equivalent definition is that two points are said to be conjugate to each other if there is a non-identically zero Jacobi field that vanishes on these points. This definitions are equivalent. The Jacobi field of the second definition is simply the one chosen to describe the congruence for which the matrix $\mathbf{E}$ becomes singular and vice-versa.

Adopting \cite{HE} terminology for the causal boundary of space-time, the (future\footnote{In this chapter, all horizons are to be understood as future horizons. The analogous definitions replacing past by future, is referred in the literature as past horizon, and the region enclosed as white hole.}) \emph{event horizon} $\mathcal{H}$ is defined as the boundary\footnote{Here it is meant the boundary in topological sense, not to be confused with the boundary of a manifold. The former will be denoted by a dot over the set, the latter by the symbol $\partial$.} of the black hole region with respect to $M\cup\partial M$, that is, $(M\setminus J^-(\mathscr{I}^+))\dot{}=\dot{J}^-(\mathscr{I}^+)$. The black hole region itself $\mathcal{B}\equiv M\setminus J^-(\mathscr{I}^+)$ is clearly a past set, that is, $I^-(\mathcal{B})\subset\mathcal{B}$, which means, accordingly to what we are going to show (following \cite{WaldGR,HE}), that the event horizon is achronal, that is, $I^+(\mathcal{H})\cap\mathcal{H}=\emptyset$.

\begin{proof}
Indeed, suppose $q\in\mathcal{H}$, then $I^-(q)\subset M\setminus\mathcal{B}$, even though it is quite intuitive, it can be rigorously proved: let $p\in I^-(q)$, then there exists a neighbourhood $\mathcal{V}$ of $q$ in $I^+(p)$, since it is open. But $q\in\dot{\mathcal{B}}$, so $\mathcal{V}$ intercepts $M\setminus\mathcal{B}$, consequently, $p\in I^-(\mathcal{V}\cap(M\setminus\mathcal{B}))\subset M\setminus\mathcal{B}$. See figure (\ref{83m}). Analogously, $I^+(q)\subset\mathcal{B}$.

\begin{figure}[h]
\centering
\includegraphics[width=.35\textwidth]{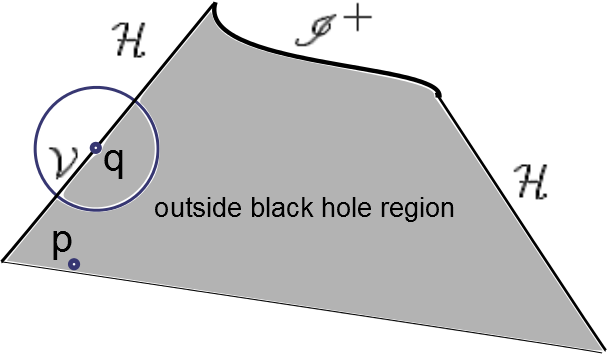}
\caption{Diagram showing the black hole region and auxiliary structures to prove achronality of event horizon.}
\label{83m}
\end{figure}

Suppose, by absurd, that exists $r\in\mathcal{H}$ such that $r\in I^-(q)$. Again, because $I^-(q)$ is open, there is an open neighbourhood $\mathcal{V}'$  such that $r\in\mathcal{V}'\subset I^-(q)\subset I^-(\mathcal{H})$, which is impossible, since $I^-(\mathcal{H})$ is open and therefore it does not intercept its frontier. The contradiction proves the achronality of the event horizon.
\end{proof}

We can go further and endow $\mathcal{H}$ with manifold structure. It is an imbedded, closed submanifold.
\begin{proof}
To prove it, following once more \cite{HE, WaldGR}, we choose Riemann normal coordinates $(x^0,x^1,x^2,x^3)$ at $q\in\mathcal{H}$ for the manifold $M$. From the continuity of the metric on $M$, there is a neighbourhood $\mathcal{U}$ of $q$ such that $\frac{\partial}{\partial x^0}$ is timelike and whose integral curves (or surfaces with $x^i$, $i\in\{1,2,3\}$ constant) intersects both $I^+(q)$ and $I^-(q)$. From the very inclusions we just saw above, each of these integral curves intercepts $\mathcal{H}$, and only once, since it is achronal. The value of $x^0$ at which the interception occurs defines a (Lipschitz continuous, with Lipschitz constant $1$: $|x^0(q_1)-x^0(q_2)|^2\leqslant\sum_{i=1}^3[x^i(q_1)-x^i(q_2)]^2$, $(q_1,q_2)\in\mathcal{H}^2$ since there is not a pair of points on $\mathcal{H}$ separated by a timelike interval) function of the coordinates $x^i$. Therefore, expressing the fact that it is enough to specify the integral curve (characterised by the value of $(x^1,x^2,x^3)$) to determine the point $q$, we consider the map from $\mathcal{U}\cap\mathcal{H}$ onto $\mathbb{R}^3$  responsible for this identification. Because of the fact that the interception occurs in only one point, the existence of the inverse function is guaranteed, as well as its continuity, by the inverse function theorem. Consequently, the above map is a homeomorphism in the induced topology on the horizon. Repeating this procedure for other points on the horizon, we construct an atlas for $\mathcal{H}$.
\end{proof}

A sequence $\lambda_n$ of curves is said to converge to a point $p$, if $\exists N\in\mathbb{N}$ such that all neighbourhoods of $p$ intercepts $\lambda_n$ $\forall n>N$. A point is said to be a limit point of a sequence of causal curves $\lambda_n$ if all its neighbourhoods intercepts an infinite number of curves in the sequence, and a causal curve is said to be a limit curve $\lambda$ if exists a subsequence $\lambda'_n$ of $\lambda$ such that $\forall p\in\lambda$, $\lambda'_n$ converges to $p$.

Next, we state the following intuitive lemma adapted directly from \cite{HE}: Let $\lambda_n$ be an infinite sequence of future inextendible causal curves. Through any limit point $p$ of $\lambda_n$, there is a future inextendible causal curve $\lambda$ limit curve of the sequence.
\begin{proof}
Denote the open ball centred on $q$ with normal coordinate radius $a$ by $\mathscr{B}(q,a)$ and a convex\footnote{The existence of this neighbourhood is guaranteed because, as any differential manifold, $M$ is a metric space, where all balls are convex, by an immediate consequence of the triangular inequality.} neighbourhood about $p$ by $\mathcal{U}_1$ and adopt Riemann normal coordinates. Because $p$ is a limit point of $\lambda_n$, there is a ball $\mathscr{B}(p,b)$ and a subsequence of $\lambda_n\cap\mathcal{U}_1$, $\lambda(1,0)_n$ which converges to $p$. Since the closure of $\mathscr{B}(p,b)$ is compact, it will contain at least a limit point $y$ of $\lambda(1,0)_n$, which must lie both in $\mathcal{U}_1$ and in or on the light cone from $p$ , since it is guaranteed the existence of causal curves between any neighbourhoods of the limit points $p$ and $y$ of subsequences of a sequence of causal curves. Because $\lambda_n$ is \emph{future} inextendible, Let $x_{11}=y$ and suppose, without lost of generality, $x_{11}\in J^+(p)\cap\mathcal{U}_1\cap\overline{\mathscr{B}(p,b)}$, since we are sure about the existence of such a point on the \emph{future} of $p$. Next, let $\lambda(1,1)_n$ denote a subsequence of $\lambda(1,0)_n$ which converges to $x_{11}$. Then repeat this procedure to find other points within the ball $\mathscr{B}(p,b)$, $\displaystyle x_{ij}\in J^+(p)\cap\mathcal{U}_1\cap\overline{\mathscr{B}\left(p,\frac{j}{i}b\right)}$ for $j\leqslant i$ will be a limit point of the subsequence $\lambda(i,j)_n$, or $\lambda(i-1,i-1)_n$ when $j=0$. The inclusion on $\mathcal{U}_1$ is guaranteed because it is convex. The closure of the union of all $x_{ij}$ will be a causal curve from $x_{10}=p$ to $x_{11}$, henceforth called $\lambda$. See figure (\ref{186m}) to help visualisation of this construction.
\begin{figure}[h]
\centering
\includegraphics[width=.35\textwidth]{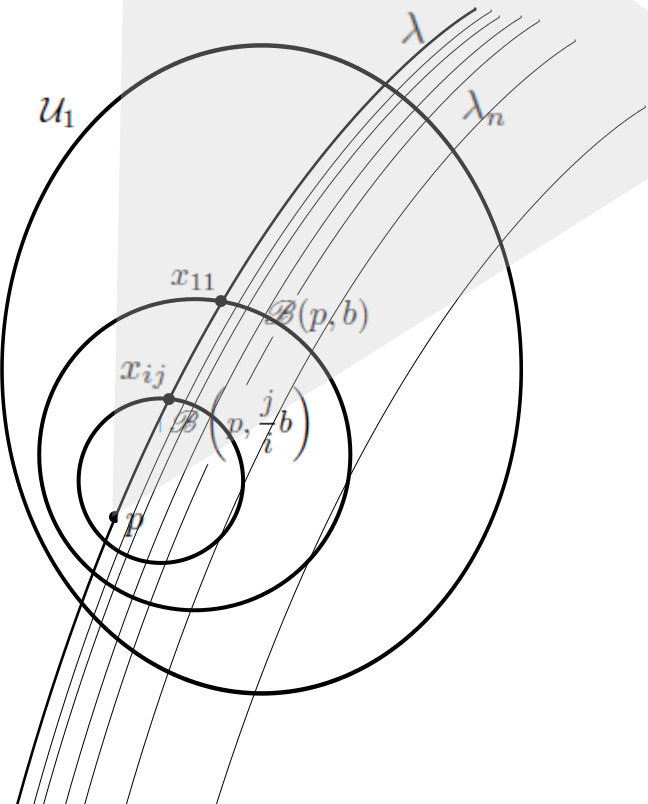}
\caption{Diagram showing the sequence of curves $\lambda_n$ and auxiliary structures to construct the limit casual curve $\lambda$. The shaded region represents the casual future of $p$, $J^+(p)$.}
\label{186m}
\end{figure}

The claim that $\lambda$ is a limit curve is proved by picking a point $x_{mj}\in\lambda$ and noting that the subsequence $\lambda(m,m)_n$ intercepts all balls $\mathscr{B}(x_{mj},\frac{1}{m}b)$ $\forall j\in\mathbb{N},\ j\leqslant m$, So $x_{mj}$ is a limit point. The arbitrariness of $x_{mj}$ shows $\lambda$ is indeed a limit curve. If we had chosen a point of $\lambda$ not on its interior, so that could not be expressed in the form $x_{mj}$, the argument above can be employed remembering that \emph{any} neighbourhood of this point contains points of the form $x_{mj}$, and therefore will include a neighbourhood of some of these points, which will intercept infinitely many $\lambda_n$. To extend $\lambda$ indefinitely, just repeat this procedure to a convex normal neighbourhood $\mathcal{U}_2$ of $x_{11}$ and so on.
\end{proof}

The next result, that teaches us that every point on the horizon lies on a \emph{future inextendible} null geodesics entirely contained on the horizon itself, without future endpoints (on $M$\footnote{They can, of course, have endpoints on $\mathscr{I}^+\subset\partial M$, but, roughly speaking, this just a consequence of the definition of the null infinity. For an observer in space-time or for a light ray, there is no future endpoints, only the meaningless ones `at infinity'.}), follows as a corollary from the above lemma.
\begin{proof}
Choose a sequence of points $p_n$ outside the black hole (on $I^-(\mathscr{I}^+)$) which converges to $p\in\mathcal{H}=(M\setminus J^-(\mathscr{I}^+))\dot{}$. From the definition of null infinity $\mathscr{I}^+$, there is, for each point on this sequence a future directed null curve connecting it to $\mathscr{I}^+$. From the lemma we proved above, it is guaranteed the existence of the limit curve $\lambda\ni p$. If there were a point of $\lambda$ in $I^-(\mathscr{I}^+)$, then this same geodesic $\lambda$ would connect it with $p$, this fact is in contradiction with the hypothesis that $p$ lies on the frontier of this region. This shows that $\lambda\subset\mathcal{H}$.
\end{proof}

If a null geodesics $\gamma(\lambda)$ with affine parameter $\lambda$ orthogonal to a (partial) Cauchy surface $\Sigma$ from $\Sigma\cap\mathcal{H}$ to $p\in\mathcal{H}$. Then there is no conjugate point on this interval.
\begin{proof}
Without loss of generality, suppose the value of the affine parameter at $q$ is zero. It is always possible since the result of adding an arbitrary constant to an affine parameter is again an affine parameter. Suppose by absurd, there is such conjugate point $r$ to some point $q\in\Sigma\cap\mathcal{H}$. So let $J^a$ be a non-trivial Jacobi field that vanishes on $q$ and $r$. The Jacobi equation for $J^a$ is
\[\frac{\ud^2}{\ud\lambda^2}J^a=-R^a_{\ bcd}V^bJ^cV^d.\]
This field is viewed as a field on the space spanned by the vectors $(T^a,S^a)$ as above\footnote{The reader must bear in mind this space is spacelike and orthogonal to the null geodesics, then $g(\hat J,\ud/\ud\lambda)=0$ below. We are making use of these facts on what follows.}, then we may write $J^a=f(x^b)\hat{J}^a(x^b)$ with $g(\hat{J},\hat{J})=1\Rightarrow g(\hat{J},\frac{\ud\hat{J}}{\ud \lambda})=0$, substituting on the Jacobi equation, contracting the result with $\hat{J}_a$ and using the relation just found, we obtain
\[\frac{\ud^2f}{\ud\lambda^2}+\underbrace{\left[g\left(\hat{J},\frac{\ud^2\hat{J}}{\ud\lambda^2}\right)+R^a_{\ bcd}V^b\hat J^cV^d\hat J_a\right]}_{\equiv h}f=0.\]
Because $r$ and $q$ are conjugate points, and $J^a$ at least for some $\xi\in[r,p]$, $J^a$ is not zero. Since $J^a$ is continuous, it will be different from zero on the interval $(r,p)$. Define $b=[-f(e^{a\lambda}-1)^{-1}]|_\xi$ and the field
\[Z^a=[b(e^{a\lambda}-1)+f]\hat{J}^a,\]
with $a>0$ chosen so that $a^2+\min_{y\in[r,\xi]}h(y)>0$.
With this construction for $b$, clearly $Z^a(\xi)=0$, and also $Z^a(q)=0$, since $J^a(q)=0\Rightarrow f=0$ and the term in parenthesis vanish thanks to our choice of the affine parametrisation, $\lambda=0$ on $q$. Also, making use of Jacobi equation for $f\hat J^a$,
\[Z_a\frac{\ud^2Z^a}{\ud\lambda^2}+Z_ aR^a_{\ bcd}V^bZ^cV^d=be^{a\lambda}\left(a^2\hat J^aZ_a+Z_a\frac{\ud^2 \hat J^a}{\ud\lambda^2}+R^a_{\ bcd}V^b\hat J^cV^dZ_a\right)\]
therefore non-negative everywhere on the interval $(q,\xi)$, thanks to our choice of $a$. The aim of this field is to serve as auxiliary one to construct a casual but non-null curve from $q$ to $p$. This curve will be in the family of curves $\gamma_u$, one for each $u\in(-\epsilon,\epsilon)$ given by $\exp_{\gamma(\lambda)}(uS|_{\gamma(\lambda)})$, where $\exp$ denotes the geodesic exponential map (see \cite{barata} for definition). The field $S=\frac{\ud}{\ud u}$ is required to satisfy $h^a_bS^b(u_0)=Z^a$, where $h_{ab}$ is the projector onto the spacelike subspace of interest, as defined in the beginning of this section (so $\xi$ and $q$ will be unaltered since $Z^a$ vanishes on these points) and
\begin{equation}
\left.g\left(S^b\nabla_bS,\frac{\ud}{\ud\lambda}\right)\right|_{u=u_0}+\left.g\left(S,\frac{\mathrm D}{\ud\lambda}S\right)\right|_{u=u_0}=
\begin{cases}
-\epsilon\lambda & 0\leqslant\lambda\leqslant\frac{1}{4}\lambda_\xi\\
\epsilon\left(\lambda-\frac{1}{2}\lambda_\xi\right) & \frac{1}{4}\lambda_\xi\leqslant\lambda\leqslant\frac{3}{4}\lambda_\xi\\
\epsilon(\lambda_\xi-\lambda) & \frac{3}{4}\lambda_\xi\leqslant\lambda\leqslant\lambda_\xi,
\end{cases}
\label{intervals}
\end{equation}
where $\lambda_\xi$ denotes the value of $\lambda$ at the point $\xi$. This is considered only on the region where $\displaystyle0<\epsilon<\min_{\frac{1}{4}\lambda_\xi\leqslant\lambda\leqslant\frac{3}{4}\lambda_\xi}\left(Z_a\frac{\ud^2Z^a}{\ud\lambda^2}+Z_ aR^a_{\ bcd}V^bZ^cV^d\right)$. The reason for all these requirements will become apparent when we calculate the first and second variations in $g\left(\frac{\ud}{\ud\lambda},\frac{\ud}{\ud\lambda}\right)$, which will play the role of first and second variations of arc length of timelike curves.
\begin{multline*}
\left.\frac{\ud}{\ud u}\right|_{u=0}\int_\xi^q\ud\lambda\ g\left(\frac{\ud}{\ud\lambda},\frac{\ud}{\ud\lambda}\right)=\int_\xi^q\ud\lambda\ S^a\nabla_ag\left(\frac{\ud}{\ud\lambda},\frac{\ud}{\ud\lambda}\right)=2\int_\xi^q\ud\lambda\ (\ud\lambda)_bS^a\nabla_a\left(\frac{\ud}{\ud\lambda}\right)^b=\\
2\int_\xi^q\ud\lambda\ (\ud\lambda)_b\left(\frac{\ud}{\ud\lambda}\right)^a\nabla_aS^b=2\cancel{\int_\xi^q\ud\lambda\ \left(\frac{\ud}{\ud\lambda}\right)^a\nabla_a\left[(\ud\lambda)_bS^b\right]}-
2\int_\xi^q\ud\lambda\ S^b\frac{\mathrm D}{\ud\lambda}(\ud\lambda)_b,
\end{multline*}
where on the third step, we interchanged derivatives using that $\left[S^a,\left(\frac{\ud}{\ud\lambda}\right)^a\right]=0$, by hypothesis. This hypothesis is usually expressed by saying we can choose basis with these two vectors. On the cancellation, we used the field $S^a$ vanishes at the endpoints $\xi$ and $q$. This variation clearly vanishes for geodesics, this is a widely known result. We derived it here only because it is an intermediate step to compute the second variation. In what follows, we already suppose the curve is geodesic.
\begin{multline}
\frac{1}{2}\frac{\partial^2}{\partial u^2}g\left(\frac{\ud}{\ud\lambda},\frac{\ud}{\ud\lambda}\right)=\frac{\partial}{\partial u}g\left(\frac{\mathrm D}{\partial u}\frac{\ud}{\ud\lambda},\frac{\ud}{\ud\lambda}\right)=\frac{\partial}{\partial u}g\left(\frac{\mathrm D}{\partial\lambda}\frac{\partial}{\partial u},\frac{\ud}{\ud\lambda}\right)=\frac{\partial}{\partial u}\left[\frac{\partial}{\partial\lambda}g\left(\frac{\partial}{\partial u},\frac{\ud}{\ud\lambda}\right)-g\left(\frac{\partial}{\partial u},\frac{\mathrm D}{\partial\lambda}\frac{\ud}{\ud\lambda}\right)\right]\\
=\frac{\partial^2}{\partial u\partial\lambda}g\left(\frac{\partial}{\partial u},\frac{\ud}{\ud\lambda}\right)-\frac{\partial}{\partial u}g\left(\frac{\partial}{\partial u},\frac{\mathrm D}{\partial\lambda}\frac{\ud}{\ud\lambda}\right)=\frac{\partial^2}{\partial u\partial\lambda}g\left(\frac{\partial}{\partial u},\frac{\ud}{\ud\lambda}\right)-\cancel{g\left(\frac{\mathrm D}{\partial u}\frac{\partial}{\partial u},\frac{\mathrm D}{\partial\lambda}\frac{\ud}{\ud\lambda}\right)}-g\left(\frac{\partial}{\partial u},\frac{\mathrm D^2}{\partial u\partial\lambda}\right)=\\
\frac{\partial}{\partial\lambda}\left[g\left(\frac{\mathrm D}{\partial u}\frac{\partial}{\partial u},\frac{\ud}{\ud\lambda}\right)+g\left(\frac{\partial}{\partial u},\frac{\mathrm D}{\partial\lambda}\frac{\partial}{\partial u}\right)\right]-g\left(\frac{\partial}{\partial u},\frac{\mathrm D^2}{\partial\lambda^2}\frac{\partial}{\partial u}-\mathbf R\left(\frac{\ud}{\ud\lambda},\frac{\partial}{\partial u}\right)\frac{\ud}{\ud\lambda}\right),
\label{secondvariation}
\end{multline}
which can be evaluated at $u=0$ and integrated in the variable $\ud\lambda$ immediately to obtain the second variation of null geodesics. On the last step, the first term has been obtained by making use of Leibniz' rule and the variation vector and once again, tangent vectors commute; and the second by interchanging the order of the derivatives, including the curvature tensor In our case, $V^a=(\ud/\ud\lambda)^a$.

Now, with all the choices made, including the range of $\epsilon$, we ensure the integrand of (\ref{secondvariation}) is larger (within the region of integration, of course) than the derivative of the boundary term with respect to $\lambda$, since the later is bounded by $\pm\epsilon$, and is positive only on the middle part of the interval of integration. Consequently, comparing (\ref{intervals}) with (\ref{secondvariation}), we conclude that $\frac{\ud^2}{\ud u^2}g\left(\frac{\ud}{\ud\lambda},\frac{\ud}{\ud\lambda}\right)<0\Rightarrow g\left(\frac{\ud}{\ud\lambda},\frac{\ud}{\ud\lambda}\right)<0$ for sufficiently small $\epsilon$, (that means sufficiently small deformations) since the first derivative vanishes. The final curve will be this curve in the interval $[q,\xi]$ and $\gamma$ in $[\xi,p]$.

The curve we just constructed is causal and non-null, and therefore (at least part of it, a non-inextendible curve, more precisely) lies on the interior of $J^-(\mathscr{I}^+)$, and also still on $\mathcal{H}$, which is clearly a contradiction since $\mathcal{H}$ is a frontier, $\mathcal{H}=\dot{J}^-(\mathscr{I}^+)$.
\end{proof}

The hypothesis that there exists a partial Cauchy surface, whose future domain of dependence includes $\mathscr{I}^+$ is called \emph{future predictability} of space-time. How could it not be so? If one had a singularity from which one can draw a null curve that extends until $\mathscr{I}^+$, that is, a naked singularity. So, if one believes the cosmic censorship conjecture, this hypothesis is not of concern. We will assume this hypothesis to hold.

As we have already discussed, the absence of conjugate points on the congruence of null geodesics generators of the horizon, means this congruence have \emph{finite expansion everywhere}. Because we have proved this geodesics are inextendible, (\ref{theta-1}) obliges it to be simply \emph{non-negative} everywhere. Consider a family of (partial) Cauchy surfaces $\Sigma_t$ parametrised by $t\in\mathbb{R}$. (\ref{geomtheta}) teach us that locally the cross-sectional area of the congruence of future-directed generators of the horizon locally increase with an increase on its affine parameter. We learned also that existing generators can never ceases to exist, since they cannot have future endpoints on $M$, but new generators can appear, all generators that reached $\Sigma_{t_1}$ will also reach $\Sigma_{t_2}$ for $t_2>t_1$, that is $\Sigma_{t_2}\subset I^+(\Sigma_{t_1})$. Consequently,

\begin{center}
\framebox[1.1\width]{\large{the area of $\mathcal{H}\cap\Sigma_{t_1}$ is no greater than the area of $\mathcal{H}\cap\Sigma_{t_2}$.}}
\end{center}

This is the statement of the area increase theorem. We may interpret the intersection of a Cauchy surface indexed by $t$ with the event horizon (here understood as a global structure) as the `event horizon' at `some instant $t$', which justifies the usual informal statement `the area of the event horizon can never decrease'.

\begin{figure}[h]
\centering
\includegraphics[width=.45\textwidth]{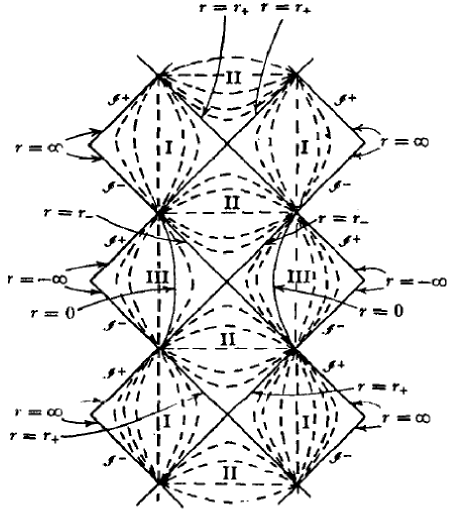}
\caption{Reproduced from \cite{HE}. Carter-Penrose diagram for maximally extended Kerr solution for $M^2>a^2$. Dotted lines represent surfaces of constant coordinate $r$. From this diagram, we can identify the event horizon at $r=r_+$ as well as interpret the surface $r=r_-$ as a Cauchy horizon with respect to a (partial) Cauchy surface for the asymptotically flat region I, the one physically meaningful.}
\label{CarterPenroseDiagram}
\end{figure}

Before finishing this section, let us identify precisely the event horizon on Kerr metric, not only to suit as a example of the structure we have been working with, but also this result will be extremely useful, as, in fact, we have already implicitly used it on chapter three.  To accomplish that, we must extend Kerr solution for $M^2>a^2$ across the coordinate singularities at $\Delta=0$. Defining, as in chapter three, the coordinate $r^*$ by $\ud r^*=\frac{r^2+a^2}{\Delta}\ud r$ and introducing the new coordinates
\[u\equiv
\begin{cases}
t-r^* & r<r_-\ \text{or}\ r>r_+;\\
t+r_* & r_-<r<r_+
\end{cases}\]
and
\[v\equiv
\begin{cases}
t+r^* & r<r_-\ \text{or}\ r>r_+;\\
-t+r^* & r_-<r<r_+,
\end{cases}\]
Kerr metric along the symmetry axis is $\displaystyle\ud s^2=\frac{|\Delta|}{r^2+a^2}\ud u\ud v$. This patch is analytic on the surfaces where $\Delta$ vanishes.  In order to represent the structure of space-time in a Carter-Penrose diagram, we introduce the conformal transformation to $(U,V)$ and the conformal metric: $\tan U=e^{\alpha u},\ \tan V=e^{\alpha v}$ for $r_-<r<r_+$ and  $\tan U=-e^{-\alpha u},\ \tan V=-e^{-\alpha v}$ otherwise, where $\alpha\equiv\frac{r_+-r_-}{2(r_+^2+a^2)}$. Hence, in both regions, $\displaystyle\ud s^2=-4\frac{|\Delta|}{\alpha^2}\csc(2U)\csc(2V)\ud U\ud V$. The diagram depicted on figure (\ref{CarterPenroseDiagram}) is plotted in the plane $(T,X)$, such as $V=T+X$ and $U=T-X$. Repeating this procedure, the maximally extended solution includes infinitely many asymptotically flat regions similar to the one we began with.

The future null infinity $\mathscr I^+$ can be identified with the region where $v\to\infty$ for finite $u$, corresponding to $V\to+\frac{\pi}{2}$, a segment (since, as defined above $v$ dependence on $r$ varies accordingly to whether $r$ is smaller or larger than $r_+$, for instance) of bisector of the even quadrants on the diagram. Similarly, we identify the past null infinity $\mathscr I^-$ as a segment of bisector of odd quadrants. They are represented on the diagram explicitly, from which it is clear the black hole region with respect to the asymptotically flat region I, the complement of $J^-(\mathscr I^+)$ is represented by region II.

Therefore, now we can safely identify the surface determined by $r=r_+$ as the event horizon, as we made use in chapter three.

As it may be noted, the definition of the black hole, and consequently the event horizon, as global structures has led to derive powerful results, the area theorem as one of its most outstanding ones. Nevertheless, in order to identify them on specific examples, we need to know all the global structure of space-time.

\subsection{`Zeroth Law'}

Two other important results are the relation between variations in area, angular momentum and energy of a black hole, we shall derive following \cite{bch} and a especial property of one of the coefficients appearing in that relation, referred as the `zeroth law', we shall derive following \cite{WaldGR}.
We define the Killing vector $\chi^a$ to be normal to the horizon of the black hole. It must be a combination of the timelike Killing vector $\xi^a$, which is required to be normalised to $g(\xi,\xi)=-1$ at infinity and tangent to the horizon, since it is mapped into itself under timelike isometry, and the axial Killing vector $\psi^a$, required to be normalised such that its closed orbits have period $2\pi$: $\chi^a=\xi^a+\Omega\psi^a$. In order to $\chi^a$ satisfy Killing equation, it is necessary that $\psi_{(b}\nabla_{a)}\Omega=0$.
Then, by definition, on the horizon $g(\chi,\chi)=0\Rightarrow\nabla_ag(\chi,\chi)=0$. So, we may define the \emph{surface gravity} of the black hole by $\nabla^a(\chi_b\chi^b)=-2\kappa\chi^a$. Taking the Lie derivative along $\chi^a$ on both sides of this equation,
$0=-2[\chi^a\mathsterling_\chi\kappa+\kappa\mathsterling_\chi\chi^a],$
so $\mathsterling_\chi\kappa=0$. This means of course $\kappa$ is constant along the orbits of $\chi^a$.

Also follows the identity
\begin{equation}
\chi^b\nabla_a\chi_b=-\chi^b\nabla_b\chi_a=-\kappa\chi_a.
\label{step1}
\end{equation}
This is the geodesic equation with a non-affine parameter. Replacing the original parameter, we call $v$, that is, $\frac{\ud}{\ud v}=\chi^a\nabla_a$ by another, we call $\lambda$ and the corresponding tangent vector $k^a$, $\frac{\ud}{\ud\lambda}=k^a\nabla_a$ satisfying
\begin{equation}
k^a=e^{-\kappa v}\chi^a,
\label{kchi}
\end{equation}
we get $k^b\nabla_bk^a=e^{-2\kappa v}[\chi^b\nabla_b\chi^a-\chi^a\chi^b\nabla_b(\kappa v)]=0$, which means $\lambda$ is indeed a corresponding affine parameter. Using $\chi_{[a}\nabla_b\chi_{c]}=0$, an immediate result from Frobenius theorem and \ref{kchi}, we find $k_{[a}\nabla_bk_{c]}=-e^{2\kappa v}\left[\frac{1}{2}\nabla_a\chi_b+\chi_{[a}\nabla_{b]}(\kappa v)\right]$. The contraction of this equation with two vectors tangent to the horizon, say any vector $n$ with $g(k,n)=0$ vanishes. This means $B_{ab}=0$, consequently $\theta=\omega_{ab}=\sigma_{ab}=0$. From (\ref{raychaudhuri}),
\begin{equation}
R_{ab}k^ak^b=0.
\label{stephalf}
\end{equation}

Applying $\chi_{[d}\nabla_{c]}$ to both sides of equation (\ref{step1}),
\begin{equation}
\chi_a\chi_{[d}\nabla_{c]}\kappa+\kappa\chi_{[d}\nabla_{c]}\chi_a=\chi_{[d}\nabla_{c]}(\chi^b\nabla_b\chi_a)=(\chi_{[d}\nabla_{c]}\chi^b)(\nabla_b\chi_a)+\chi^b\chi_{[d}\nabla_{c]}\nabla_b\chi_a.
\label{step2}
\end{equation}
By using the definition of the Riemann curvature tensor, the Killing equation, and the cyclic property of curvature tensor, then, for a generic  Killing filed $\chi^a$ as in our case, we find
\begin{equation}
\nabla_c\nabla_d\chi_e=-R_{dec}^{\ \ \ f}\chi_f.
\label{step5halves}
\end{equation}
Bearing this in mind the second term on the right-hand side of (\ref{step2}) becomes $-\chi^bR_{ba[c}^{\ \ \ \ e}\chi_{d]}\chi_e$. To evaluate the first term, apply the Frobenius's theorem to the hyper-surface orthogonal to the horizon tangent to $\chi^a$ to find
\begin{equation}
\chi_{[a}\nabla_b\chi_{c]}=0\Rightarrow\chi_c\nabla_a\chi_b=-2\chi_{[a}\nabla_{b]}\chi_c.
\label{step3}
\end{equation}
With these substitutions,
\[(\chi_{[d}\nabla_{c]}\chi^b)(\nabla_b\chi_a)=-\frac{1}{2}(\chi^b\nabla_d\chi_c)\nabla_b\chi_a=-\frac{\kappa}{2}\chi_a\nabla_d\chi_c=\kappa\chi_{[d}\nabla_{c]}\chi_a,\]
where we have used (\ref{step1}). So,
\begin{equation}
\chi_a\chi_{[d}\nabla_{c]}\kappa=\chi^bR_{ab[c}^{\ \ \ \ e}\chi_{d]}\chi_e.
\label{step4}
\end{equation}
Applying $\chi_{[d}\nabla_{e]}$ to (\ref{step3}), using (\ref{step3}) and (\ref{step5halves}), and after a little algebra,
\[-\chi_{[a}R_{b]}^{\ \ f}\chi_f\chi_d=\chi_{[a}R_{b]cd}^{\ \ \ \ f}\chi^c\chi_f,\]
contracting with $g^{ec}$ and comparing with (\ref{step4}),
\begin{equation}
\chi_{[d}\nabla_{c]}\kappa=-\chi_{[d}R_{c]}^{\ \ f}\chi_f.
\label{step8}
\end{equation}

For the area theorem, we hypothesised the null energy condition. Here, for the `zeroth law of black hole thermodynamics', we make use of a much stronger version of this hypothesis, the \emph{dominant energy condition}, which states that not only does the weak energy condition hold, but also for every timelike or null vector $W^a$, $T^{ab}W_a$ is a non-spacelike vector. Physically this means that the energy flow is causal. This also means, by appropriate contractions, that the energy density is greater than pressure, for instance. But, from Einstein equations, together with (\ref{stephalf}), $T^{ab}\chi_a\chi_b=0$ on the horizon. So $T^{ab}\chi_b$ is zero or parallel to $\chi^a$, so $\chi_{[c}T_{a]b}\chi^b=0$. Once again using Einstein equations, from (\ref{step8}),
\[\chi_{[a}\nabla_{b]}\kappa=0,\]
this equation means that the $\kappa$ has vanishing derivative, i.e., is constant, throughout the hyper-surface orthogonal to $\chi^a$, that is, that $\kappa$ is constant throughout all the horizon, not only on each separate orbit of $\chi^a$, as we had conclude from $\mathsterling_\chi\kappa=0$. This statement is called `zeroth law of black hole thermodynamics' because the (formal, at this point) analogy with the zeroth law of ordinary thermodynamics $T$ is constant between bodies in equilibrium, and throughout the body itself as a consequence.

\subsection{`First Law'}
Accordingly to Komar (\cite{WaldGR}), for a asymptotically flat space-time, its mass is defined by
\begin{equation}
M=-\frac{1}{4\pi}\oint_{\partial S_\infty}\nabla^b\xi^a\ud\sigma_{ab}
\label{komassa}
\end{equation}
Integrating the identity $\nabla_b\nabla^b\xi^a=-R^a_b\xi^b$, obtained from the definition of the curvature tensor and the Killing equation, therefore valid for any Killing field, we obtain
\begin{equation}
\oint_{\partial S}\nabla^b\xi^a\ud\sigma_{ab}=-\int_SR^a_b\xi^b\ud\sigma_a.
\label{kidin}
\end{equation}
Choosing $S$ to be spacelike, tangent to $\psi^a$, asymptotically flat, $\partial S\supset\partial S_\infty$, and can be used to evaluate the mass as in (\ref{komassa}). For convenience, the remaining part of $\partial S$ consists of $\partial B\equiv\mathcal{H}\cap S$. So that $\partial S=\partial S_\infty\cup\partial B$.
From (\ref{kidin}), (\ref{komassa}), and Einstein field equations,

\begin{equation}
M=\int_S\underbrace{(2T_a^b-T_c^c\delta_a^b)}_{=\frac{1}{4\pi}R^b_a}\xi^a\ \ud\sigma_b+\frac{1}{4\pi}\oint_{\partial B}\nabla^b\xi^a\ud\sigma_{ab}.
\label{komass}
\end{equation}

Also from (\ref{kidin}), replacing $\xi^a$ by $\psi^a$ and the Komar definition of angular momentum, $J_T=+\frac{1}{8\pi}\int_{\partial S_\infty}\nabla^b\psi^a\ud\sigma_{ab}$, we obtain similarly, for the same choice of surface $S$,
\begin{equation}
J_T=-\int_ST^a_b\psi^b\ud\sigma_a\underbrace{-\frac{1}{8\pi}\oint_{\partial B}\nabla^b\psi^a\ud\sigma_{ab}}_{\equiv J}.
\label{koj}
\end{equation}

The physical interpretation of (\ref{komass}) and (\ref{koj}) are evident. The first term on the expressions represent the contribution of matter and the second can be interpreted as the black hole itself contribution for mass and angular momentum respectively measured at infinity. Making use of the definition of the Killing field $\chi^a$ and the result for the black hole angular momentum $J$,
\begin{equation}
M=\int_S(2T^b_a-T^c_c\delta^b_a)\xi^a\ud\sigma_b+2\Omega J+\frac{1}{4\pi}\oint_{\partial B}\nabla^b\chi^a\ud\sigma_{ab}.
\label{bch9}
\end{equation}

Choose $n^a$ to be another null vector orthogonal to $\partial B$ with $g(\chi,n)=-1$. Then $\ud\sigma_{ab}=\chi_{[a}n_{b]}\ud\sigma$ (see \cite{lee} or \cite{Nakahara} for discussion of integration on manifolds). Contracting (\ref{step1}) with $n^a$, we find
\begin{equation}
\kappa=-\chi^an^b\nabla_b\chi_a.
\label{kappan}
\end{equation}
So the last term on (\ref{bch9}) becomes $\frac{1}{4\pi}\int_{\partial B}\kappa\ud\sigma$. Because $\kappa$ is constant over the horizon, we rewrite equation (\ref{bch9}) as
\begin{equation}
M=\int_S(2T^b_a-T^c_c\delta_a^b)\xi^a\ud\sigma_b+2\Omega J+\frac{\kappa}{4\pi}A\stackrel{\text{\tiny{Einstein Eq.}}}{=}\int_S\left(2T^b_a-\frac{R}{8\pi}\delta_a^b\right)\xi^a\ud\sigma_b+2\Omega J+\frac{\kappa}{4\pi}A,
\label{smarr}
\end{equation}

As mentioned before, we are interested in comparing two close solutions and relate their parameters changes. Instead of comparing any two solutions, it is meaningful comparing only solutions that share the event horizon, the timelike and axial Killing fields, $\delta\psi^a=\delta\xi^a=0\Rightarrow\delta\chi^a=\psi^a\delta\Omega$ and, also, by gauge choice (that is the freedom we have when mapping the two manifolds corresponding to the two solutions), the surface $S$. Furthermore,
\begin{equation}
\delta\chi_{[a}\chi_{b]}=\delta n^{[a}n^{b]}=0 \footnote{This condition means that variations of the vectors $\chi^a$ and $n^a$ are parallel to the vectors $\chi^a$ and $n^a$ themselves. We require also the vector $\delta\chi^a$ to be Lie dragged along $\chi^a$, that is, $\mathsterling_\chi\delta\chi^a=0$.},
\label{paralelismo}
\end{equation}

Therefore, for these solutions, variation of (\ref{kappan}), gives
\begin{multline}
\delta\kappa=\frac{1}{2}n^c\nabla_c(\chi^a\delta\chi_a+\chi_a\delta\chi^a)+\frac{1}{2}\delta\chi^c\nabla_c(\chi_a\chi^a)=\nabla_b\delta\chi_a\chi^{(a}n^{b)}+\cancel{\delta\chi_an^b\nabla_b\chi^a}+\\
\chi_an^b\nabla_b\psi^a\delta\Omega+\cancel{\chi_a\delta n^b\nabla_b\chi^a}.
\label{varkappa}
\end{multline}
Repeating a technique already employed here for slightly different purpose, complete a basis for the tangent space to some point on the horizon by adding two other complex null vectors $m$ and $\bar{m}$ with $g(m,\bar{m})=-1$. Then $g_{ab}=-2n_{(a}l_{b)}+2m_{(a}\bar{m}_{b)}$ and, because on the horizon $\delta\chi_a\propto\chi_a$, the first term on the right-hand side of (\ref{varkappa}) becomes $-\frac{1}{2}\nabla^a(\delta\chi_a)$. Finally, by using $\delta\chi_a=(\ud g_{ab})\chi^b+g_{ab}\psi^b\delta\Omega$, this term can be written as $-\frac{1}{2}\chi^b\nabla^a(\delta g_{ab})$.
The terms were canceled in (\ref{varkappa}) because in they together add $\nabla_b\chi^a(\delta\chi_an^b+\chi_a\delta n^b)$, and, as already pointed out, $\chi^b\nabla_b(\delta\chi_a)+\delta\chi_b\nabla_a\chi^b=\mathsterling_\chi\delta\chi=0$, we can substitute on the first term in parenthesis, and thanks to the properties (\ref{paralelismo}), they become $\chi^a\nabla_b\chi_a\delta\eta^b$, where $\delta\eta^b$ is a vector parallel to the horizon and orthogonal to $\chi^a$ as can be seen from the decomposition of $g_{ab}$ above in terms of the vectors $(\chi,n,m,\bar{m})$, reason why we canceled those terms.
Now we integrate (\ref{varkappa}). For this integration we follow closely Carter on \cite{Survey}.
\[A\delta\kappa=\oint_{\partial S}\frac{1}{2}\chi^a\nabla^b(\delta g)_{ab}\ \ud\sigma-\oint_{\partial S}\chi_an_b\nabla^a\psi^b\delta\Omega\ \ud\sigma,\]
because $\nabla^a\psi^b$ is antisymmetric thanks to Killing equation, we may retain only the antisymmetric part of $\chi_an_b$ in the second term, and, since $\chi_{[a}n_{b]}\ud\sigma=\ud\sigma_{ab}$, this integral can be identified with angular momentum.
\[A\delta\kappa=\oint_{\partial S}\chi^b\nabla^{[c}(\delta g)^{a]}_{\ \ c}\ud\sigma_{ab}-8\pi J\delta\Omega,\]
where the integrand can be expanded together with $\ud\sigma_{ab}=\chi_{[a}n_{b]}$, and, by using $g(\chi,\chi)=0$ and $g(n,\chi)=-1$ and also the symmetry $(\delta g)_{ab}=(\delta g)_{ba}$, is found to coincide with $\frac{1}{2}\chi^a\nabla^b(\delta g)_{ab}\ud\sigma$. Then make use of the freedom of choosing $S$ to impose $\psi^a\ud\sigma_{ab}=0$, without changing $\partial S$, we have already fixed. This condition means $S$ is invariant under the orbits of $\psi^a$. So the $\psi^a$ term does not contribute, and we may substitute $\chi\leftrightarrow\xi$ on the integral term, whence it becomes
\begin{equation}
\oint_{\partial S}\xi^b\nabla^{[c}(\delta g)^{a]}_c\ \ud\sigma_{ab}=\int_S\xi^a\nabla_b\nabla^{[b}(\delta g)^{c]}_c\ \ud\sigma_a-4\pi\delta M,
\label{variationalprinciple}
\end{equation}
Where $\frac{1}{4\pi}\oint_{\partial S_\infty}\xi^a\nabla^{[c}(\delta g)^{b]}_c\ \ud\sigma_{ab}$ is identified with a particular case of variation of mass (energy) accordingly to Arnowitt, Deser and Misner (ADM) when $\partial S_\infty\rightarrow i^0$. This definition is inspired on constructing Hamiltonian formalism (and energy is defined as common on Hamiltonian mechanics) to General Relativity from foliations of parametrised Cauchy surfaces. In ADM formalism $\xi^a$ is just a timelike field orthogonal to the foliation surfaces, but in case of the existence of such timelike isometry, and $\xi^a$ coincides on infinity with the generator of that, as is in our case above, ADM and Komar notions of mass coincide. A simple proof of this coincidence is found in \cite{beig}. Its approach consists on writing in components a sufficiently generic metric tensor and comparing the two definitions of mass. See Fischer and Marsden article on \cite{Survey} for details of ADM approach and \cite{brewin} for details of definition of mass itself.

It only remains the first term in the right-hand side of (\ref{variationalprinciple}) to be analysed. For variation of $g_{ab}$, and also on the element of integration \cite{landau2} provide us a plenty of useful identities (particularly when deriving Einstein's equations from Einstein-Hilbert action, §95). We refer the reader to this reference for proof of these identities, that were adapted for our purpose. Using the notation of this reference also,
\begin{equation*}
\begin{array}{c}
\displaystyle\nabla_b\nabla^{[b}(\delta g)^{c]}_c=-\frac{\delta(R\sqrt{|g|})}{2\sqrt{|g|}}-\frac{1}{2}(\delta g)_{ab}\left(R^{ab}-\frac{1}{2}g^{ab}R\right)\\
\displaystyle\delta(\ud\sigma_a)=\frac{\delta\sqrt{|g|}}{\sqrt{|g|}}\ud\sigma_a=\frac{1}{2}(\delta g)^c_c\ud\sigma_a.
\end{array}
\end{equation*}
Therefore
\[4\pi\delta M+8\pi J\delta\Omega+A\delta\kappa=-4\pi\int_S T^b_b\xi^a\ \ud\sigma_a-\frac{1}{2}\delta\int_SR\xi^a\ \ud\sigma_a.\]
From (\ref{smarr}) we can relate the variation of the total mass with the variation of the black hole mass. If we intend to consider vacuum, $T_{ab}=0$, then,
\begin{equation}
\frac{\kappa}{2\pi}\ \delta \left(\frac{A}{4}\right)=\delta M-\Omega\delta J
\label{firstlaw}
\end{equation}

This formula, which is possible to be obtained more generally by other means, see \cite{iyerwald}, are given the name of `first law of black hole thermodynamics'. The reason for this is its resemblance with the first law of ordinary thermodynamics, $T\delta S=\delta U+P\delta V$.

The effect of considering matter is the presence of the variations of the term including the energy-momentum tensor on (\ref{smarr}). It is a little simpler if all the matter is electromagnetic field, since it has vanishing trace of energy-momentum tensor, and, therefore $R=0$. Details of the possibility of charged black holes can be explored further on B. Carter chapter on \cite{Survey}. This reference shows how formula (\ref{smarr}) is altered in this case. The final result would be changed in such a way the analogy with thermodynamics is totally preserved: by the addition of $-\Phi\ \ud Q$ on the right-hand side of (\ref{firstlaw}), where $\Phi$ represents the electric potential as measured by the observer along the timelike Killing field $\xi^a$ and $Q$ is the charge of the black hole.

Of course relation (\ref{firstlaw}) could be derived more rapidly and directly from the knowledge of the precise form of Kerr metric (or Kerr-Newman metric
\footnote{We will refer to this exact solution to Einstein-Maxwell equations again. The importance of this solution relies on uniqueness and no-hair theorems, proved mostly by Carter. In Boyer-Lindquist coordinates, Ker-Newman metric is \cite{WaldGR}
\[\ud s^2=-\left(\frac{\Delta-a^2\sin^2\theta}{\rho^2}\right)\ud t^2-\frac{2a\sin^2\theta(r^2+a^2-\Delta)}{\rho^2}\ud t\ud\phi+\left[\frac{(r^2+a^2)^2-\Delta a^2\sin^2\theta}{\rho^2}\right]\sin^2\theta\ud\phi^2+\frac{\rho^2}{\Delta}\ud r^2+\rho^2\ud\theta^2,\]
and
\[A_a=-\frac{er}{\rho^2}[(\ud t)_a-a\sin^2\theta(\ud\phi)_a],\]
where $\rho^2\equiv r^2+a^2\cos^2\theta$ as before and $\Delta\equiv r^2+a^2+e^2-2Mr$. $e$ represents the electric charge and $A_a$ the four-potential.}
, if electric charge is allowed). The advantage of our presentation is, not only because addition of matter is well on track (see \cite{bch} for inclusion of matter in the form of perfect fluid), but also because this derivation make the role of each hypothesis much clearer.

\subsection{Superradiance}

Here, we are going to make use only of the above results, which are analogous to the second, zeroth and first law of thermodynamics. The analogy between the black hole physics and thermodynamics goes far beyond, namely, there is an analogue for the third law (at least to some of its possibles versions), and, after the discovery of Hawking's radiation, the analogy has been promoted to something much deeper. The quantities appearing in (\ref{firstlaw}) are indeed associated with temperature ($T=\frac{\kappa}{2\pi}$), entropy ($S=\frac{Ac^3}{4G\hslash}$ in conventional units), and obviously energy ($E=M$) of black hole, which resembles the first law of Thermodynamics, as already pointed out. The area theorem, after extremely interesting discussions (which include, for example, whether or not black holes can be used to violate the second law of thermodynamics), may be promoted, as suggested Bekenstein, to what is called the generalised second law of thermodynamics, which states that black hole entropy should be added to the total entropy as if it were the entropy of any other physical system, and this total entropy can never decrease in an isolated system. We are going back to this question and its relation with superradiance latter.

Now, we apply the results above to superradiance. This argument is originally due to Bekenstein (\cite{beke73}). If there is an incident radiation on a black hole, changes on angular momentum and energy will be in the proportion of energy and angular momentum of the radiation carried \emph{radially} away from the black hole. We dealt with radiation in the form $e^{\ui(m\phi-\omega t)}$. For this wave form, this proportion will be $m:\omega$. This can be argued from the particle level picture for which the energy and angular momentum projection on azimuthal direction are respectively $\hslash\omega$ and $\hslash m$, or can be verified explicitly for each filed from its energy momentum tensor, by computing the ratio $\displaystyle\frac{T_{ab}\psi^a(\partial/\partial r)^b}{-T_{ab}\xi^a(\partial/\partial r)^b}$. For example, for scalar field $\varphi$, $T_{ab}=\nabla_a\varphi\nabla_b\varphi-\frac{1}{2}g_{ab}\nabla_c\varphi\nabla^c\varphi$, the ratio reduces to $-\displaystyle\frac{\psi(\varphi)}{\xi(\varphi)}=\frac{m}{\omega}$, using the wave-form dependence on $\phi$ and $t$.

Then, $\frac{m}{\omega}=\frac{\ud J}{\ud M}$. By requiring that the left-hand side of (\ref{firstlaw}) is non-negative, we have
\begin{equation}
\ud M\left(1-\Omega\frac{m}{\omega}\right)\geqslant0.
\label{beke3}
\end{equation}

From this equation, we see that if the incident wave is on superradiant regime $\omega<m\Omega$, we are driven to the conclusion that $\ud M<0$, which means that energy is extracted from the black hole. By energy conservation arguments, it follows that superradiant scattering is present.

It is remarkable the coincidence of (\ref{beke3}) with (\ref{bek2}) with vanishing $W$ and $\vec U$, corresponding to the case with no spontaneous emission, as we would expect for a purely classical black hole. Both arguments have thermodynamical nature in its interpretations, but they have completely different deductions.

If we had included the possibility of the field and the black hole to be charged and we had modified the `first law', we could deduce superradiance to be present whenever $\omega<m\Omega+e\Phi$, where $e$ is the electric charge of the incident field and $\Phi$ is the electrical potential at the horizon as measured by an observer following the timelike isometry. It follows that superradiance is predicted even for a non-rotating black hole, as long as it is charged.

A natural question may arise: the above argument would seem to apply to fermions as well as to bosons, and, as we saw above, superradiance is absent for fermions. The reason that the argument fails is that the energy-momentum tensor for fermions does not obey the null energy condition, so the area theorem cannot apply.
\footnote{With purpose to give an example, we prove that a real scalar filed $\phi$, whose energy-momentum tensor is $T_{ab}=\nabla_a\phi\nabla_b\phi-\frac{1}{2}g_{ab}[\nabla_c\phi\nabla^c\phi+V(\phi)]$ satisfy both the null and the weak energy condition. Contracting this tensor twice with a timelike vector $\xi$ we obtain $\left[\frac{1}{2}(D_\xi\phi)^2+(D_\perp\phi)^2+V(\phi)\right]|g(\xi,\xi)|$, where $D_\xi$ and $D_\perp$ denotes the (covariant) derivative along $\xi$ and the projection of $\nabla_a\phi$ onto the spacelike subspace orthogonal to $\xi$. So, if $V$ is non-negative and, since the metric induced on that spacelike subspace is positive definite which ensures the non-negativity of the second term, the weak energy condition holds, and, therefore, the null energy condition as well.}
Let me show explicitly this failure for the example of a massive, although if the field were massless our conclusions would be the same, spin-\textonehalf \ field, whose energy-momentum written in Newman-Penrose formalism is given by (following \cite{chandrasekhar})
\begin{multline}
T_{AA'BB'}=\frac{\ui}{2}[P_A\nabla_{BB'}\bar{P}_{A'}-\bar{P}_{A'}\nabla_{BB'}P_A+P_B\nabla_{AA'}\bar{P}_{B'}-\bar{P}_{B'}\nabla_{AA'}P_B-\\
Q_A\nabla_{BB'}\bar{Q}_{A'}+\bar{Q}_{A'}\nabla_{BB'}Q_A+Q_B\nabla_{AA'}\bar{Q}_{B'}-\bar{Q}_{B'}\nabla_{AA'}Q_B]
\label{emtspinor}
\end{multline}
Contracting with $n^an^b$ identified with $\iota^A\bar{\iota}^{A'}\iota^B\bar{\iota}^{B'}$, as we know from (\ref{spinorNP}), we get after substituting the results of (\ref{volta}) from chapter three, near the horizon, on the notation of that chapter,
\[\omega\frac{\varrho^2(r_+)}{\rho^4}|R_{-1/2}|^2(S_{+1/2}^2+S_{-1/2}^2),\]
which is negative on the region $0<\omega<m\Omega=\omega_s$, because $\varrho^2(r_+)=r_+^2+a^2-\frac{am}{\omega}=2Mr_+\left(1-\frac{\omega_s}{\omega}\right)<0$ for $\omega<\omega_s$. This region where the area increase law ceases to hold is precisely the superradiant interval! If we wish to find an explicit violation of weak energy condition it is enough to contract energy momentum tensor twice with the timelike vector $k^a=r^{-2}[(\Delta/2) l^a+\rho^2n^a]$, for which $g(k,k)=-\frac{\Delta\rho^2}{r^4}<0$ as long as $r>r_+$.

This example shows clearly that energies conditions are not to be taken for granted. This statement is reinforced by quantum mechanics. Even in Minkowski space-time, there are states for which the expectation value of the normal ordered energy momentum tensor violates the null energy condition even if the classical version is successfully applied! \cite{ANEC}. For this reason, one usually weaken the energy condition, by imposing the non-negativity to hold only after an average is taken along a complete null geodesic. For this averaged energy condition, reference \cite{ANEC} has shown that it is applied for any globally hyperbolic two-dimensional space-time for a massless scalar field minimally coupled with curvature. Several questions remain open on this topic.

Also from the area theorem, we can derive an upper limit from the amount of energy that can be extracted from a black hole either by superradiant scattering or by any other means, such as the Penrose process, for instance (see chapter one). We can compute explicitly the area of $\mathcal{H}\cap\Sigma_t$ for a Kerr black hole. Using Boyer-Lindquist coordinates again,
\[A=\int_{r=r_+}\ud\theta\ud\phi\sqrt{g_{\theta\theta}g_{\phi\phi}}=\int_0^\pi\ud\theta\int_0^{2\pi}\!\ud\phi\ (r_+^2+a^2)\sin\theta=4\pi(r_+^2+a^2)=16\pi M_0^2,\]
where
\begin{equation}
M_0^2\equiv\frac{1}{2}[M^2+\sqrt{M^4-J^2}]=\frac{a}{4\Omega}=\frac{J}{4M\Omega},
\label{irredmass}
\end{equation}
where the last two equalities -- obtained under the substitution of (\ref{omegaa}) -- hold supposing, of course, $\Omega\neq0$; is called \emph{irreducible mass} in the literature. The second equality was obtained by expressing $J$ in terms of $a$, $J=Ma$ and the value of $\Omega$.

The reason for this name is obvious, the qualifier `irreducible' is clear by virtue of the area theorem it cannot reduce. Directly from the definition of $M_0^2$, $M^2=M_0^2+\frac{J^2}{4M_0^2}$, which reveals the reason for the noun `mass', this is the least value for the mass of a black hole, that happens when $J=0$, when the superradiant interval shrinks to zero length.

\subsection{Superradiance, Weak Cosmic Censorship Conjecture, and the `Third Law'}

Before I finish this chapter, the existing relation between the `third law of black hole thermodynamics', weak cosmic censorship conjecture and superradiance deserves to be briefly commented. First consider, as we mentioned before, the possibility of changing parameters of Kerr (or Kerr-Newman) black hole by incident radiation on the usual form $e^{\ui(m\phi-\omega t)}$. The change in the parameters will be, as already mentioned, in the proportion $\frac{m}{\omega}=\frac{\ud J}{\ud M}$. So, by the incidence of sufficiently low energy radiation carrying sufficiently high angular momentum, it would seem possible to push the black hole toward extreme case and creating a naked singularity. At least for bosonic radiation, this possibility is ruled out because of superradiance, since this modes are not absorbed by that black hole, on the contrary, there is a net current outgoing the black hole, so these modes pull the black hole \emph{away} from the extreme case.

For a Kerr-Newman black hole, the surface gravity can be computed explicitly because we know from the precise form of the Killing fields $\xi^a$ and $\psi^a$ to be $\left(\frac{\partial}{\partial t}\right)^a$ and $\left(\frac{\partial}{\partial\phi}\right)^a$, respectively.
\[\kappa=\frac{\sqrt{M^2-a^2-e^2}}{2M(M+\sqrt{M^2-a^2-e^2})-e^2},\]
from which we see that for a extreme black hole $\kappa=0$. If, guided by the weak cosmic censorship conjecture one is driven to the belief that extreme black holes cannot be formed from a customary black hole by any process, one will state the impossibility of achieving $\kappa=0$ by any physical process. This is in a obvious analogy with the third law of thermodynamics, reason why it is referred this way. It is important to reinforce, that, differently from analogues of the zeroth, first and second laws of thermodynamics, we are currently in lack of a proof for this statement.

The question of whether or not fermionic waves can be used to contradict this hypothesis is more complicated. \cite{hodccc} claims to have proven that a quantum-mechanical phenomenon prevents this process to succeed. His argument relies on the limiting case of Hawking expression for the expected value of the number of particle created by black holes \cite{Hawking}. On the limit case $T\rightarrow0$, which represents a black hole near the extreme case, for which $\kappa=0$, there is still present radiation proportional to $\theta(m\Omega-\omega)$, which means that Hawking process only `creates particles' that prevent the limit case to be reached\footnote{In fact, it is not accurate to attribute the cause of this emission in this particular case to Hawking radiation. In the next chapter we are going to deduce this result originally discovered by Unruh, before Hawking's discovery. Hawking \cite{Hawking} showed his result reduces to Unruh's (see chapter five) in the limit $\kappa\to0$.}. His conclusion, although seems to rely on imprecise arguments invoking Pauli principle. Later on this work (chapter five), we shall recast this question in a more elucidating way. Either way, this discussion provides us with an insight that the origin of the absence of superradiance for fermions might be quantum-mechanical if the weak cosmic censorship conjecture is true. It is time for quantum mechanics!

\chapter{Quantum Mechanical Considerations}

\section{Klein Paradox}

We begin this chapter by giving a first example of a \emph{quantum} system that exhibits superradiance. Let us consider bosonic and fermionic fields minimally coupled with an electromagnetic field in Minkowski space-time. First choose coordinates such that $g_{\mu\nu}=\eta_{\mu\nu}=\text{diag}(-1,1,1,1)$ and the potential four-vector is $A^\mu=(\varphi,0,0,0)$, that automatically satisfies Lorenz gauge if we suppose constancy of $\varphi$. 

\begin{figure}[h]
\centering
\includegraphics[width=.7\textwidth]{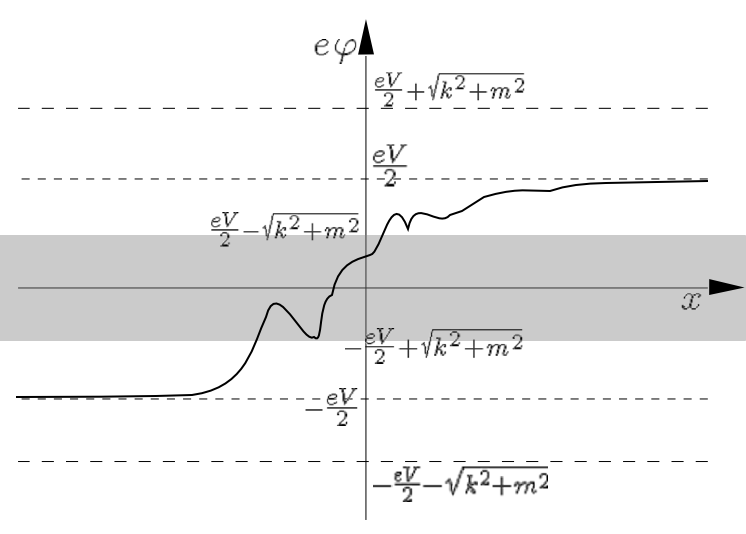}
\caption{Electromagnetic potential barrier. The shaded region is referred on the text as `superradiant region'.}
\label{barreiramanogue}
\end{figure}

We assume $\varphi$ to depend only on $x$ in this coordinate system and to be uniform when $x\rightarrow\pm\infty$, with value $\pm\frac{V}{2}$. This assumption allows us to refer to $\varphi(x)$ as a potential barrier. It will become clear from the assumptions above that the ODEs obtained will respect the hypothesis for the function $V(\xi)$ of chapter two. A picture showing this barrier is represented on figure (\ref{barreiramanogue}).

Besides its name, the phenomenon is not paradoxical by any means. Its name was assigned apparently because of a historical mistake \cite{Manogue}.

\subsection{Bosons}
A massive scalar field $\phi$ minimally coupled satisfies
\[g^{ab}(\partial_a-\ui eA_a)(\partial_b-\ui eA_b)\phi-m^2\phi=\square\phi+2\ui e\varphi\frac{\partial\phi}{\partial t}+e^2\varphi^2-m^2\phi^2=0.\]
Following \cite{Manogue}, we are interested in potentials $\varphi$ which goes to $\pm\frac{V}{2}$ as $x\rightarrow\pm\infty$. We use the in ansatz $w_\text{in}=Ne^{\ui(\vec{k}\cdot\vec{y}-\omega t)}f(x)$, where $\vec{k}\cdot\vec{e_x}=0$, so that the mode is transverse ($k_a\partial^af=0$). So $f$ obeys
\[\frac{\ud^2f}{\ud x^2}+[(\omega-e\varphi)^2-(k^2+m^2)]f=0,\]
which is obviously on the form (\ref{EDO1}). To apply the method developed in the second chapter, we must apply some boundary condition on the asymptotic behaviour. Because we are interested in quantising this field, we need to seek a \emph{complete} set of modes. Again, following \cite{Manogue}, we consider modes given asymptotically by
\begin{equation*}
w_{\text{in},+}=N_+e^{\ui(\vec{k}\cdot\vec{y}-\omega t)}\times
\begin{cases}
e^{\ui qx}+R_+^{\frac{1}{2}}e^{-\ui qx} & x\rightarrow-\infty\\
T_+^{\frac{1}{2}}e^{\ui rx} & x\rightarrow\infty
\end{cases}
\end{equation*}
and
\begin{equation*}
w_{\text{in},-}=N_-e^{\ui(\vec{k}\cdot\vec{y}-\omega t)}\times
\begin{cases}
T_-^{\frac{1}{2}}e^{-\ui qx} & x\rightarrow-\infty\\
e^{-\ui rx}+R_-^{\frac{1}{2}}e^{\ui rx} & x\rightarrow\infty,
\end{cases}
\end{equation*}
which are solutions as long as
\begin{subequations}
\begin{align}
q^2=\left(\omega+\frac{V}{2}\right)^2-(k^2+m^2)\quad\text{and}\\
r^2=\left(\omega-\frac{V}{2}\right)^2-(k^2+m^2).
\end{align}
\label{disprel}
\end{subequations}

It may happen $q$ and $r$ to be imaginary, representing damping.

Although the two asymptotic region are characterised by different wave numbers, the square modulus $R_\pm=|R_\pm^\frac{1}{2}|^2$ \emph{is} the reflection coefficient, since reflected and incident waves share the same wave number. Transmission coefficients are not $T_\pm$, though. One must use equations (\ref{numbert}, \ref{energyt}) to define this coefficient in terms of number or energy, respectively.

In order to interpret physically correctly these modes, it is necessary not only (\ref{disprel}), but also to determine the sign of $q$ and $r$. Therefore, we require the group velocity of modes (+), for which we intend to be propagating from left to right $\frac{\ud\omega}{\ud q}=\frac{q}{\omega+eV/2}$ to be non-negative, which leads to $q>0$ for $\omega>-eV/2$ and $q<0$ for $\omega<-eV/2$. Similarly, we require for (--) modes, $\frac{\ud\omega}{\ud r}=\frac{r}{\omega-eV/2}\geqslant0$, whence $r>0$ for $\omega>eV/2$ and $r<0$ for $\omega<eV/2$.

The normalisation constants are chosen such that
\[(w_{\text{in},\pm},w'_{\text{in},\pm})_{KG}=\mathrm{sgn}\left(\omega\pm \frac{eV}{2}\right)\delta(\omega-\omega')\delta(\vec{k}-\vec{k'})\quad \text{and}\quad (w_{\text{in},\pm},w_{\text{in},\mp})_{KG}=0,\]
where the Klein-Gordon product is defined as usual $$(f,g)_{KG}=-\ui\int_\Sigma\ud\Sigma_a(f^*\nabla^ag-g\nabla^af^*),$$ independent of the choice of Cauchy surface $\Sigma$. From this convention follows $N_+=[2(2\pi)^3|q|]^{-\frac{1}{2}}$ and $N_-=[2(2\pi)^3|r|]^{-\frac{1}{2}}$.

Another complete set of solutions is
\begin{equation*}
w_{\text{out},-}=M_-e^{\ui(\vec{k}\cdot{y}-\omega t)}\times
\begin{cases}
e^{-\ui qx}+Q_-e^{\ui qx} & x\rightarrow-\infty\\
S_-e^{-\ui rx} & x\rightarrow\infty
\end{cases}
\end{equation*}
and
\begin{equation*}
w_{\text{out},+}=M_+e^{\ui(\vec{k}\cdot\vec{y}-\omega t)}\times
\begin{cases}
S_+e^{\ui qx} & x\rightarrow-\infty\\
e^{\ui rx}+Q_+e^{-\ui rx} & x\rightarrow\infty.
\end{cases}
\end{equation*}
If normalised accordingly with the same criteria give $M_-=[2(2\pi)^3|q|]^{-1/2}$ and $M_+=[2(2\pi)^3|r|]^{-1/2}$. These modes, of course, obey the same orthogonality relations as the in modes.

The out modes can be readily interpreted as a kind of time reversal in modes. More precisely, in modes travel from left to the right or from right to the left and split up in two, one part is reflected and the other transmitted. Out modes begin coming both from right and from left, they merge together at $x=0$ and proceed propagating either to the right or to the left. This is the reason why we label the modes this manner. In and out modes are to be interpreted as ingoing or outgoing modes to the electric potential barrier respectively.

Now, using the same technique explored on the first two chapters, the relation between the reflection and transmission coefficients follows by equating the Wronskian of each mode and its complex conjugate in the two asymptotic regions, since the potential is real. Because we have several cases to consider one by one, this task is somewhat laborious.
\begin{enumerate}
\item If $-\frac{eV}{2}+\sqrt{k^2+m^2}\leqslant\omega\leqslant+\frac{eV}{2}-\sqrt{k^2+m^2}$ or $\omega\geqslant\frac{eV}{2}+\sqrt{k^2+m^2}$, or else $\omega\leqslant-\frac{eV}{2}-\sqrt{k^2+m^2}$, $R_+=1-\frac{r}{q}T_+$ and $R_-=1-\frac{q}{r}T_-$.
\item If $\frac{eV}{2}-\sqrt{k^2+m^2}\leqslant\omega\leqslant+\frac{eV}{2}+\sqrt{k^2+m^2}$, $R_+=R_-=1$.
\item If $-\frac{eV}{2}-\sqrt{k^2+m^2}\leqslant\omega\leqslant-\frac{eV}{2}+\sqrt{k^2+m^2}$, $R_-=R_+=1$.
\end{enumerate}
From these cases, the first is more interesting because for $-\frac{eV}{2}+\sqrt{k^2+m^2}\leqslant\omega\leqslant+\frac{eV}{2}-\sqrt{k^2+m^2}$, we see $q>0$ and $r<0$, so \emph{superradiance is present}. Here the nature of superradiance is slightly different from the cases studied on introduction, since it happens even respecting $R+T=1$, but here $T$ may be negative, see (\ref{numbert}, \ref{energyt}) (not to be confused with $T_\pm$, which are non-negative, since they are square modulus).

Denoting by $u$ and $v$ with appropriate indexes positive and negative normed modes of $w$, we write
\begin{equation}
\phi=\sum_\pm\int_{-\infty}^{+\infty}\ud\omega\int\ud^2k\ (a_\text{in}u_\text{in}+b_\text{in}^\dagger v_\text{in}),
\label{expansionfield}
\end{equation}
where the coefficients $a$ and $b$ have been promoted to operators satisfying the same canonical commutation relations $[a_{\text{in},\omega,\vec{k},i},a^\dagger_{\text{in},\omega',\vec{k}',i'}]=\ui\delta(\omega-\omega')\delta^2(\vec{k}-\vec{k'})\delta_{ii'}$, where $i$ denotes the possible values of $+$ and $-$ modes. The vacuum state $|0_\text{in}\rangle$ is determined by $a_{\text{in},\omega,\vec{k},i}|0_\text{in}\rangle=b_{\text{in},\omega,\vec{k},i}|0_\text{in}\rangle=0$.

As usual, the current vector field $j^a=-\ui e[\phi^*(\partial^a-\ui eA^a)\phi-\phi(\partial^a-\ui eA^a)\phi^*]$ is promoted to an operator by inserting (\ref{expansionfield}) on this equation for $\phi$ and $\phi^*$, for the latter, we replace operators by its hermitian conjugate on (\ref{expansionfield}). Then, when computing $\langle0_\text{in}|j^x|0_\text{in}\rangle$ we make use of point-splitting regularisation\cite{Birrel} and only retain terms for which neither annihilation operators appear on the right nor creation on the left, since $\langle0_\text{in}|a^\dagger=0$ is the hermitian conjugate equation for the definition of vacuum state. We left with $bb^\dagger v^*\partial^xv-aa^\dagger u\partial^xu^*$. Employing canonical commutation relations to reverse the order of the operators, we obtain simply $v^*\partial^xv-u\partial^xu^*$, which can be promptly evaluated by conservation of the Wronskian on superradiant interval to give
\begin{equation}
\langle0_\text{in}|j^x|0_\text{in}\rangle=\frac{e}{(2\pi)^3}\int\ud^2k\int\ud\omega\ \frac{r}{q}T_+,
\label{numbert}
\end{equation}
for $x\rightarrow\pm\infty$.

Also, one can compute the flux of energy across a surface $x$ constant $\langle0_\text{in}|T^{tx}|0_\text{in}\rangle$ using the usual expression for the energy momentum tensor for complex scalar field $T^{ab}=2(\partial-\ui eA)^{(a}\phi^*(\partial-\ui eA)^{b)}\phi-\eta^{ab}[(\partial_c-\ui eA_c)\phi^*(\partial^c-\ui eA^c)\phi+m^2\phi^*\phi]$ and again using point-split regularisation. The result on superradiant interval is
\begin{equation}
\langle0_\text{in}|T^{tx}|0_\text{in}\rangle=
\frac{1}{(2\pi)^3}\int\ud^2k\int\ud\omega\ \left(\omega\pm\frac{eV}{2}\right)\frac{r}{q}T_+
\label{energyt}
\end{equation}
for $x\rightarrow\mp\infty$.

Another natural question arises. We may ask how many outgoing particles there are in the vacuum $|0_\text{in}\rangle$. Given the physical interpretation for out modes, \cite{Manogue} refer to this number as the number of created particles, that is, how many outgoing particles are there if there is no incident particle either from left or from right. To answer this question, we must find the corresponding number $N_\text{out}$ operators for these modes. If we find operators $(a,b,a^\dagger,b^\dagger)_\text{out}$, then as we know, we can interpret $a^\dagger a$ and $b^\dagger b$ as the number of particles and antiparticles of that kind\footnote{The choice of normalisation is understood clearly here. If we had chosen a different normalisation criteria, the number operator would carry an inconvenient factor other than unity.}.

Out modes and in modes must be linked by a Bogoliubov transformation
\begin{equation}
\begin{array}{c}
\displaystyle u_{j,\text{out}}=\sum_k\alpha_{jk}u_{k,\text{in}}+\sum_l\beta_{jl}v_{l,\text{in}}\\
\displaystyle v_{m,\text{out}}=\sum_k\gamma_{mk}u_{k,\text{in}}+\sum_l\epsilon_{ml}v_{l,\text{in}},
\end{array}
\label{BTmo}
\end{equation}
where the indexes run over $(i,\vec{k},\omega)$, and the sum is understood to be the sum over $i$ and the integral over $\ud^2k$ and $\ud\omega$, just to simplify notation. Substituting on (\ref{expansionfield}), using the orthogonality relations for both sets of modes we identify the creation and annihilation operators for out modes as
\begin{equation}
\begin{array}{c}
\displaystyle a_{j,\text{out}}=\sum_k\alpha^*_{jk}a_{k,\text{in}}-\sum_l\beta_{jl}^*b_{l,\text{in}}^\dagger\\
\displaystyle b_{m,\text{out}}=-\sum_k\gamma_{mk}a_{k,\text{in}}^\dagger+\sum_l\epsilon_{ml}b_{l,\text{in}}.
\end{array}
\label{BTop}
\end{equation}
Then, from definition of $|0_\text{in}\rangle$ and from (\ref{BTop}),
\begin{equation}
\begin{array}{c}
\displaystyle\langle N_a\rangle\equiv\langle0_\text{in}|a^\dagger_{j,\text{out}}a_{j,\text{out}}|0_\text{in}\rangle=\sum_k\beta_{jk}\beta^*_{kj}\\
\displaystyle\langle N_b\rangle\equiv\langle0_\text{in}|b^\dagger_{j,\text{out}}b_{j,\text{out}}|0_\text{in}\rangle=\sum_k\gamma_{jk}\gamma^*_{kj}.
\end{array}
\label{numberin}
\end{equation}
The Bogoliubov coefficients on superradiant regime may be calculated directly from the definition of the modes themselves and comparing to (\ref{BTmo}) to give, going back to explicit notation $\displaystyle\beta_{\omega\omega',\vec{k}\vec{k'},ii'}=\sqrt{\left|\frac{r}{q}\right|}(T^{1/2}_+)^*\delta(\omega-\omega')\delta^2(\vec{k}-\vec{k'})\delta_{i-}\delta_{i'-}$ and $\displaystyle\gamma_{\omega\omega',\vec{k}\vec{k'},ii'}=-\sqrt{\left|\frac{r}{q}\right|}(T^{1/2}_+)^*\delta(\omega-\omega')\delta^2(\vec{k}-\vec{k'})\delta_{i+}\delta_{i'+}$. Substituting on (\ref{numberin}) we obtain
\begin{equation*}
\displaystyle\langle N_a\rangle=\langle N_b\rangle=\frac{1}{(2\pi)^3}\int\ud\omega\int\ud^2k\left|\frac{r}{q}\right||T_+|\delta(0)\delta^2(0).
\end{equation*}
Of course this number is infinite, but this is just because the potential extends indefinitely on $z$, $y$ and $t$, each of this extensions are responsible for one $\delta(0)$ factor, and whose origin can be interpreted by saying there are particles being created everywhere on $y$ and $z$ and forever. So this is not to be subjected to any kind of concern.

Taking all calculations into consideration, we concluded that on superradiant regime, there are charged particles created by the potential barrier, there is a positive energy flux (and therefore momentum) across a surface orthogonal to $x$ on the right of the barrier and negative flux across a surface orthogonal to $x$ on the left of the barrier and there is a net current of particles traveling from right to the left.

\subsection{Fermions}
Consider now a spin-\textonehalf \ field minimally coupled to the same electromagnetic field. Dirac equation is therefore
\begin{equation}
\gamma^\mu(\partial_\mu-\ui eA_\mu)\psi+m\psi=0.
\label{dchar}
\end{equation}
In order to accurately write modes whose interpretations are clear, we introduce spinors $\chi_{j+}$, $j\in\{1,2,3,4\}$ instead of the usual $\chi_j$ normalised so $\chi^\dagger_j\chi_k=\delta_{jk}$ and simultaneous eigenspinors of chiral $\frac{\mathbf{1}\pm\gamma^5}{2}$ and energy $\frac{\ui\gamma^0+\mathbf{1}}{2}$ operators. The reason for this, is first our interest in helicity rather than chirality and energies corrected by taking into account the interaction with the external field. The operators
\[P_+=-\frac{1}{2m}\left\{\ui\left[q\gamma^1+\vec{k}\cdot\vec{\gamma}-\gamma^0\left(\omega+\frac{eV}{2}\right)\right]-m\right\}\]
and
\[P_-=-\frac{1}{2m}\left\{\ui\left[-r\gamma^1+\vec{k}\cdot\vec{\gamma}-\gamma^0\left(\omega-\frac{eV}{2}\right)\right]-m\right\}\]
play the role of Dirac operators (apart from a $\frac{1}{2m}$ factor) $\slashed{p}-m$ with appropriate frequencies and wave numbers for incident modes to our problem. It would be obviously possible to adapt the operator to reflected (R) or transmitted (T) modes as well\footnote{The modes are written explicitly below.}, this can be accomplished by reversing the sign of $q$ (to be interpreted as adapted to +R or --T modes) on the first expression or reversing the sign of $r$ (+T or --R modes) on the second. For helicity, consider
\[\sigma_+=\frac{\ui\gamma^5}{\sqrt{k^2+m^2}}\left[q\gamma^0-\left(\omega+\frac{eV}{2}\right)\gamma^1\right]\]
and
\[\sigma_-=\frac{\ui\gamma^5}{\sqrt{k^2+m^2}}\left[-r\gamma^0-\left(\omega-\frac{eV}{2}\right)\gamma^1\right].\]
So that $\frac{\mathbf1+\sigma_\pm}{2}$ and $\frac{\mathbf1-\sigma_\pm}{2}$ are projection operators on the subspace of eigenspinors of helicity operator. These operators are found in \cite{sakurai}; the prescription consists on choosing for helicity $\sigma=\ui\gamma^5\gamma^aw_a$, where $w^a$ is spacelike orthogonal to $p^a$, normalised such as $g(w,w)=1$. The justification for the name helicity for the operators constructed accordingly to this prescription is that, when applied on an eigenspinor of the projection of spin operator on the direction of motion, these two operators give the same result. See \cite{sakurai} for details.

The operators above follow this prescription with appropriate $p^a$ for each operator's index.
 
Taking advantage of the fact that each operator $P$ commutes with the corresponding (i.e., carrying the same indeces) operator $\sigma$\footnote{It is easy to prove this assertion using the commutation relation for gamma matrices, the orthogonality $g(p,w)=0$, and the well known identity $\slashed A\slashed B+\slashed B\slashed A=2g(A,B)$ for any vectors $A^a$ and $B^a$: $[\gamma^ap_a,\gamma^5\gamma^aw_a]=-\gamma^5(\slashed p\slashed w+\slashed w\slashed p)$=0.}, we define the constant spinors $\chi_{1+}=\mathcal N_{1+}P_+\frac{\mathbf1+\sigma_+}{2}\chi_1$, $\chi_{2+}=\mathcal N_{2+}P_+\frac{\mathbf1-\sigma_+}{2}\chi_2$, $\chi_{3+}=\mathcal N_{3+}(\mathbf1-P_+)\frac{\mathbf1-\sigma_+}{2}\chi_3$ and $\chi_{4+}=\mathcal N_{4+}(\mathbf1-P_+)\frac{\mathbf1+\sigma_+}{2}\chi_4$. And similar definitions for spinors carrying the label `R' or `T'. The notation is meant to be self-explanatory.

The normalisation choice
\[\mathcal{N}_{i\pm}=\mathcal{N}_{iR\pm}=\mathcal{N}_{iT\mp}=\left[\frac{1}{2m}\left(\frac{m}{\sqrt{k^2+m^2}}+1\right)\cdot\left|\omega\pm\frac{eV}{2}+\sqrt{k^2+m^2}\right|\right]^{-\frac{1}{2}}\]
leads to
\begin{equation}
\displaystyle\chi_{j\pm}^\dagger\gamma^0\chi_{k\pm}=\chi_{jR\pm}^\dagger\gamma^0\chi_{kR\pm}=\chi_{jT-}^\dagger\gamma^0\chi_{kT\mp}=\ui\delta_{jk} \mathrm{sgn}\left(\omega\pm\frac{eV}{2}\right)\mathrm{sgn}\left(\frac{5}{2}-j\right)
\label{+-spinorsR}
\end{equation}
for $q^2>0$ (upper sign) or $r^2>0$ (lower sign), otherwise
\begin{equation}
\displaystyle \chi_{jR\pm}^\dagger\gamma^0\chi_{k\pm}=\ui\delta_{jk}\mathrm{sgn}\left(\omega\pm\frac{eV}{2}\right)\mathrm{sgn}\left(\frac{5}{2}-j\right),
\label{+-spinorsI}
\end{equation}
with upper sign referring to the case of imaginary $r$, and the lower to imaginary $q$.

Similarly to what we did for bosons, we present a complete set of solutions defined in terms of the constant spinors defined above.
\begin{equation*}
\psi_{\text{in},\lambda,+}=N_{\lambda,+}e^{\ui(\vec{k}\cdot\vec{y}-\omega t)}\times
\begin{cases}
e^{\ui qx}\chi_{\lambda,+}+e^{-\ui qx}(\mathcal R_{\lambda,+}\chi_{1R+}+\mathcal R'_{\lambda+}\chi_{2R+}) & x\rightarrow-\infty\\
e^{\ui rx}(\mathcal T_{\lambda+}\chi_{1T+}+\mathcal T'_{\lambda+}\chi_{2T+}) & x\rightarrow\infty
\end{cases}
\end{equation*}
and
\begin{equation*}
\psi_{\text{in},\lambda,-}=N_{\lambda,-}e^{\ui(\vec{k}\cdot\vec{y}-\omega t)}\times
\begin{cases}
e^{-\ui qx}(\mathcal T_{\lambda-}\chi_{1T-}+\mathcal T'_{\lambda-}\chi_{2T-}) & x\rightarrow-\infty\\
e^{-\ui rx}\chi_{\lambda-}+e^{\ui rx}(\mathcal R_{\lambda-}\chi_{1R-}+\mathcal R'_{\lambda+}\chi_{2R-}) & x\rightarrow\infty
\end{cases}
\end{equation*}
where the constant spinors $\chi_\lambda$ are chosen to obey $\chi_\lambda^\dagger\chi_{\lambda'}=\delta_{\lambda\lambda'}$.

The normalisation criteria is
\[(\psi_\lambda,\phi'_{\lambda'})_D=\delta(\omega-\omega')\delta(\vec{k}-\vec{k'})\delta_{\lambda\lambda'}\delta_{+-},\]
the Dirac product is also defined as usual $$(\psi,\psi')_D=-\ui\int_\Sigma\ud\Sigma_\mu(\bar{\psi}\gamma^\mu\psi').$$ From this convention follows $N_{\lambda,+}=(2\pi)^{-\frac{3}{2}}\left(\frac{m}{|q|}\right)^{\frac{1}{2}}$ and $N_{\lambda,-}=(2\pi)^{-\frac{3}{2}}\left(\frac{m}{|r|}\right)^{\frac{1}{2}}$.

Out modes can be defined as we did for bosons, by time reversal of in modes, and its physical interpretation will be the same.

To investigate dispersion relations, we wish to obtain second order differential equation instead. To obtain this differential equation, we proceed in an analogue way one would proceed to prove that solutions to free Dirac equation are also solutions to the Klein-Gordon equation \cite{landau4}, applying the operator $\gamma^\nu(p_\nu-eA_\nu)+m$ to both sides of (\ref{dchar})
\[\left[\underbrace{\gamma^\mu\gamma^\nu}_{=\gamma^{(\mu}\gamma^{\nu)}+\gamma^{[\mu}\gamma^{\nu]}=\eta^{\mu\nu}+\gamma^{[\mu}\gamma^{\nu]}}(p_\mu-eA_\mu)(p_\nu-eA_\nu)-m^2\right]\psi=0,\]
noticing that the antisymmetric part of $(p_\mu-eA_\mu)(p_\nu-eA_\nu)$ is $-\frac{1}{2}\ui eF_{\mu\nu}$, we arrive at
\begin{equation}
\left[(p^\rho-eA^\rho)(p_\rho-eA_\rho)-m^2-\frac{1}{2}\ui eF_{\mu\nu}\gamma^{[\mu}\gamma^{\nu]}\right]\psi=0.
\label{ecampo}
\end{equation}
In the particular coordinate system and gauge we are dealing with, the last term in brackets is reduces to $\displaystyle -\ui e\gamma^0\gamma^1\frac{\ud\varphi}{\ud x}$.

This system of equations has only one (crucial) difference when compared with four decoupled copies of the bosonic one, by the presence of a term containing the electromagnetic field tensor itself, that means the gradient of $\varphi$, but because we assumed $\varphi$ to be uniform at $x\rightarrow\pm\infty$, this term \emph{does not alter the dispersion relations} (\ref{disprel}). No surprises arise from the presence of this term, its origin can be understood by recognising a charged spin-\textonehalf\ particle possesses a magnetic dipole moment, whose interaction with external field changes the Hamiltonian\footnote{The non-relativistic limit make clear the identification with the dipole moment.}.

To investigate the presence or absence of superradiance, it is easier to apply conservation of the Dirac current instead of (\ref{GAI}), although the latter is also perfectly applicable. Wronskian relations are so obtained by comparing the value of $\bar\psi_i\gamma^1\psi_j$ in the two asymptotic regimes. These relations are calculated with aid of the identities $P_+^\dagger P_+=\frac{\ui}{m}\gamma^0\left(\omega+\frac{eV}{2}\right)P_+,\  P_+^\dagger\gamma^0\gamma^1P_+=-\frac{\ui q}{m}P_+,\   P_+^\dagger\gamma^0\gamma^1P_{+R}=0$ and analogous for labels $\pm T$ and $-R$
\footnote{They are proven from well known properties of Dirac gamma matrices. For instance, the first one is proved by writing the definition of the operator $P_+$, whose adjoint is obtained from $\gamma^{0\dagger}=\gamma^0$ and $\gamma^{i\dagger}=-\gamma^i$ for $ i\in\{1,2,3\}$. The only term which changes sign after conjugation is therefore the term appearing $\gamma^0$. Anticommutation relations for Dirac matrices eliminate all terms containing two different gamma matrices in the product $P_+^\dagger P_+$. Finally, using the results for the square of gamma matrices, the first equality is proven.}.
 \[1-\mathcal R_{\lambda+}^{\prime*}\mathcal R'_{\lambda'+}-\mathcal R_{\lambda'+}^*\mathcal R_{\lambda'+}=\frac{\mathrm{sgn}(\omega-eV/2)r}{\mathrm{sgn}(\omega+eV/2)q}(\mathcal T_{\lambda+}^*\mathcal T_{\lambda'+}+\mathcal T_{\lambda+}^{\prime*}\mathcal T'_{\lambda'+})\]
for + modes and the same for -- modes, except by substitution $r\leftrightarrow q$; whilst from the second we obtain relations combining coefficients of both modes. Noting that the numerator of the fraction on the Wronskian relation for + modes above is $|r|$ and the denominator $|q|$ and finally $q^2\geqslant r^2$ from (\ref{disprel}), we conclude \emph{superradiance is absent for fermions}, once again.

It may seem counterintuitive the possibility of unattenuated propagation of fermionic modes for high potentials and not for lower potentials, i.e., for $\frac{eV}{2}-\sqrt{\vec k^2+m^2}\leqslant\omega\leqslant\frac{eV}{2}+\sqrt{\vec k^2+m^2}$, $r$ is imaginary while for higher potentials, it is real. This phenomenon can be understood in terms of negative energy solutions of Dirac equations \cite{sakurai}. For high enough potentials, it is possible to originally negative modes solutions to be shifted until they reach positive energy levels. This threshold is precisely the limit for bosonic superradiance. In this picture, one can think the presence of superradiance for bosons as a consequence of the positivity of energy in solutions of Klein-Gordon equations.

To quantise the field, we expand it in normal modes and promote the coefficients to operators:
\begin{equation}
\Psi=\sum_{i,\lambda}\int\ud\omega\int\ud^2k\ (c_\text{in}u_\text{in}+d^\dagger_\text{in}v_\text{in}).
\label{expansionffield}
\end{equation}

As usual for Dirac fields, the operators $c$ and $d$ now obey the canonical anticommutation relations $\{c_{\text{in},\omega,\vec{k},\lambda,i}, c^\dagger_{\text{in},\omega',\vec{k'},\lambda',i'}\}=\delta(\omega-\omega')\delta^2(\vec{k}-\vec{k'})\delta_{ii'}\delta_{\lambda\lambda'}$ and the same for $d$s. In an analogue way as we did for bosons, in order to give a physical description of the reflection and transmission problem, compute the vacuum expectation value for the current $j^a=-e\bar\Psi\gamma^a\Psi$ to give, after charge symmetrising, assuming $x\rightarrow\pm\infty$ and comprising the integration only to the region of interest, i.e., where bosonic modes are superradiant, (otherwise, fraction on the last integrand would change):
\begin{multline*}
\langle0_\text{in}|j^x|0_\text{in}\rangle=-\frac{e}{(2\pi)^3}\int\ud\omega\int\ud^2k\ (v^\dagger_\text{in}\gamma^0\gamma^1v_\text{in}-u^\dagger_\text{in}\gamma^0\gamma^1u_\text{in})=\\
-\frac{e}{(2\pi)^3}\int\ud^2k\int\ud\omega\frac{\mathrm{sgn}\left(\omega-\frac{eV}{2}\right)}{\mathrm{sgn}\left(\omega+\frac{eV}{2}\right)}\sum_{j=1,2}\left(|\mathcal T_{j+}|^2+|\mathcal T'_{j+}|^2\right),
\end{multline*}
where Wronskian relations restricted to the region of integration has been used in the last passage. The integrand is always positive. That means there is a net current traveling from right to left.

Again, in a similar fashion we did for bosons, compute the vacuum expectation value of the flux of energy across a plane of constant $x$, that is, $t-x$ component of
\[\langle0_\text{in}|T^{ab}|0_\text{in}\rangle=\frac{\ui}{2}\left\langle0_\text{in}\left|\bar\Psi\gamma^{(a}\partial^{b)}\Psi-\partial^{(a}\bar\Psi\gamma^{b)}\Psi\right|0_\text{in}\right\rangle\]
We write this expression in two different points, which eventually will be taken the limit to coincide, substitute in this expression (\ref{expansionffield}) and the definition of in vacuum, and finally the Wronskian relations. The result is
\[\langle0_\text{in}|T^{tx}|0_\text{in}\rangle=-\frac{1}{(2\pi)^3}\int\ud\omega\int\ud^2k\ \frac{\mathrm{sgn}(\omega-eV/2)r}{\mathrm{sgn}(\omega+eV/2)q}\left(\omega\mp\frac{eV}{2}\right)\sum_{i=\pm}\sum_\lambda|T_{\lambda,i}|,\]
for $x\rightarrow\pm\infty$. From the sign of this expression, we conclude that energy and momentum are transported to the right on the right of the barrier and to the left on the left side of the barrier.

To describe particle production, we use again a Bogoliubov transformation (\ref{BTop}) and compute the coefficients from the relation between modes (\ref{BTmo}). After a labourious calculations, we obtain for the non-vanishing $\beta$s and $\gamma$s
\[\gamma_{\omega\omega'\vec k\vec k'++}=\beta_{\omega\omega'\vec k\vec k'- -}=\sqrt{\left|\frac{r}{q}\right|}\sum_{j=1,2}(|\mathcal T_{j+}|^2+|\mathcal T'_{j+}|^2)\delta(\omega-\omega')\delta(\vec k-\vec k')\]
which leads to
\[\langle N_a\rangle=\langle N_b\rangle=\frac{1}{(2\pi)^3}\int\ud\omega\int\ud^2k\left|\frac{r}{q}\right|\sum_{j=1,2}(|\mathcal T_{j+}|^2+|\mathcal T'_{j+}|^2)\ \delta(0)\delta^2(0).\]

We conclude that the behaviour of created particles for fermions is similar to bosons. The origin of the infinite factors are again of no concern.

\emph{The barrier creates charged particles of both types} and the energy flux together with the current indicates simply that negative-charged particles flows to the left while positive-charged to the right, creating a negative net current on both sides of the barrier, independently of the particle's statistics. This description is similar to the quantum description of a black hole.

\section{Quantum Mechanical Black Hole Superradiance}

Our treatment will follow basically \cite{unruh} with aid of \cite{casals} for fermions. Again we are considering the problem of scattering a field on a Kerr black hole, but this time we would like to quantise our fields. We shall deal fields as quantum-mechanical and the Kerr background classically.

\subsection{Bosons}
First, we quantise a spin-0 filed $\Phi$, described by a complex massless scalar field
whose action is given by
\begin{equation}
S=\int\sqrt{-g}\ \ud^4x
\nabla_a\Phi^*\nabla^a\Phi.
\label{cscalaraction}
\end{equation}

\emph{Normal modes}, $\varphi$, are defined by requiring (\ref{normalmodedef}).

The action leads (\ref{cscalaraction}) to the field equation $\square\Phi=0$ and, it is not necessary to invoke Newman-Penrose formalism to write this equation explicitly. Instead, from (\ref{dalembertian}), and introducing the usual ansatz to separate variables as we did on chapter three, $\varphi=\mathcal R(r)S(\theta)e^{\ui(m\phi-\omega t)}$,
\begin{equation*}
\begin{array}{c}
\displaystyle\left[\frac{1}{\Delta}\frac{\ud}{\ud r}\Delta\frac{\ud}{\ud r}+\left(\frac{\omega(r^2+a^2)-am}{\Delta}\right)^2+\frac{k^2}{\Delta}\right]\mathcal R=0\\
\displaystyle\left[\frac{1}{\sin\theta}\frac{\ud}{\ud\theta}\sin\theta\frac{\ud}{\ud\theta}+\left(\omega a\sin\theta-\frac{m}{\sin\theta}\right)^2-k^2\right]S=0,
\end{array}•
\end{equation*}
where $k^2$ is a separation constant. From the equation for $S$ above, we may identify the angular dependence of normal modes $S(\theta)e^{\ui m\phi}$ as the spherical harmonics in Schwarzschild case $a=0$, and in general a particular case of (\ref{angularteukolsky}) with $s=0$. This constant represent the dependence with the total angular momentum labeled by $\ell$.

Let us introduce once more the change of radial variables by $\displaystyle\frac{\ud r^*}{\ud r}=\frac{r^2+a^2}{\Delta}$, the tortoise coordinate which can be integrated explicitly giving $$ r^*=r+\frac{2M^2\mathrm{atan}\left(\frac{r-M}{\sqrt{M^2-a^2}}\right)}{\sqrt{M^2-a^2}}+M\log\Delta.$$ This means, as we saw, $r\rightarrow r_++0\Rightarrow r^*\rightarrow-\infty$ and $r\rightarrow\infty\Rightarrow r^*\rightarrow\infty$.

The radial equation becomes
\begin{equation}
\left[\frac{1}{r^2+a^2}\frac{\ud}{\ud r^*}(r^2+a^2)\frac{\ud}{\ud r^*}+\left(\omega-\frac{am}{r^2+a^2}\right)^2-\frac{k^2\Delta}{(r^2+a^2)^2}\right]\mathcal R(r^*)=0.
\label{nsfradialc}
\end{equation}
If our purpose were only to examine the presence of superradiance, we would only concern with modes satisfying the boundary condition
\begin{equation}
\mathcal R_+(r^*)=\frac{1}{\sqrt{2\pi\omega(r^2+a^2)}}\times
\begin{cases}
e^{-\ui\omega r^*}+R^{\frac{1}{2}}_+e^{\ui\omega r^*} & r^*\rightarrow+\infty\\
T_+^\frac{1}{2}e^{-\ui(\omega-m\Omega)r^*} & r^*\rightarrow-\infty,
\end{cases}
\label{inqbhs+}
\end{equation}
here we also consider
\begin{equation}
\mathcal R_-(r^*)=\frac{1}{\sqrt{2\pi(\omega-m\Omega)(r^2+a^2)}}\times
\begin{cases}
T_-^\frac{1}{2}e^{\ui\omega r^*} & r^*\rightarrow+\infty\\
e^{\ui(\omega-m\Omega)r^*}+R^\frac{1}{2}_-e^{-\ui(\omega-m\Omega)r^*} & r^*\rightarrow-\infty.
\end{cases}
\label{inqbhs-}
\end{equation}

At this point, the reader, guided by our previous discussion on chapter three, can readily interpret physically both modes, since they are similar to several other discussions we have already studied, except because we have already changed the wave number near horizon from $\omega$ to $\omega-m\Omega$. This is by no means surprising since we have already concluded that would be the case on chapter three, after relating the radial coordinates $r_*$ and $r^*$. Even though, wee could foresee the appearance of this term, because we have already concluded that null geodesics are tangent not to $\xi^a$, but to $\chi^a$ near horizon. We did not face this question before on chapter three thanks to our former change of variables.

Equation (\ref{nsfradialc}) are ready to be studied with methods from chapter two. Note the potential is real and it is not difficult to evaluate the derivatives needed to compare Wronskian in the two asymptotic regions, since we can employ the much simpler relation $\frac{\ud\hat r}{\ud r}$ instead of the cumbersome integrated version. We obtain from conservation of $W(\mathcal R_+,\mathcal R_+^*)$, $W(\mathcal R_-,\mathcal R_-^*)$, $W(\mathcal R_+,\mathcal R_-)$ and $W(\mathcal R_+,\mathcal R_-^*)$, respectively
\begin{equation}
1-|R_+|=\frac{\omega-m\Omega}{\omega}|T_+|,\qquad1-|R_-|=\frac{\omega}{\omega-m\Omega}|T_-|
\label{wronskianU1}
\end{equation}
\begin{equation}
(\omega-m\Omega)T^\frac{1}{2}_+=\omega T^\frac{1}{2}_-\quad\text{and}\quad(\omega-m\Omega)T^\frac{1}{2}_+R^{\frac{1}{2}*}_-=-\omega T_-^{\frac{1}{2}*}R^\frac{1}{2}_+.
\label{wronskianU2}
\end{equation}
From the first of these equations, we conclude superradiance to be present when $\omega<m\Omega$. For a scalar field, we have already proved it on section 3.4 and also on chapter four by two other means.

We choose mode normalisation such
\[(\varphi(\omega,m,k,\lambda),\varphi(\omega',m',k',\lambda'))_{KG}=\kappa(\omega,m,\lambda)\delta(\omega-\omega')\delta_{kk'}\delta_{mm'}\delta_{\lambda\lambda'},\]
where $\lambda=\pm$ and
\[\kappa(\omega,m,\lambda)=
\begin{cases}
+1 & (\lambda=+,\ \omega>0)\ \text{or}\ (\lambda=-,\ \omega-m\Omega>0)\\
-1 & \text{otherwise}.
\end{cases}\]

From the action (\ref{cscalaraction}) we know the momenta to be $\pi=\sqrt{-g}\ \xi^a\nabla_a\Phi^*$ and $\pi^*=\sqrt{-g}\ \xi^a\nabla_a\Phi$. We now proceed with canonical quantisation by imposing the canonical commutation relations over the normal modes and the corresponding momenta. The annihilation and creation operators defined by $a(\omega,m,\lambda,k)=(\Phi,\varphi(\omega,m,\lambda,k))_{KG}$ for $\kappa>0$ and $b^\dagger(\omega,m,\lambda,k)=(\Phi,\varphi(-\omega,-m,\lambda,k))_{KG}$ for $\kappa<0$ and their hermitian conjugate respect $[a(\omega,m,\lambda,k),a^\dagger(\omega',m',\lambda',k')]=\delta(\omega-\omega')\delta_{mm'}\delta_{\lambda\lambda'}\delta_{kk'}$ and $[b(\omega,m,\lambda,k),b^\dagger(\omega',m',\lambda',k')]=\delta(\omega-\omega')\delta_{mm'}\delta_{\lambda\lambda'}\delta_{kk'}$, provided the field operators and their momenta respect the canonical commutation relations. With possess of those, the vacuum state is $a(\omega,m,\lambda,k)|0\rangle=b(\omega,m,\lambda,k)|0\rangle=0, \forall(\omega,m,\lambda,k)$. In a manner similar to the one we have already performed on this chapter, we can expand the field in terms of the normal modes and the creation and annihilation operators as follows
\[\Phi=\sum_m\int_{\kappa(\omega,m,\lambda)>0}\ud\omega\sum_{k,\lambda}[a(\omega,m,k)\varphi(\omega,m,k)+b^\dagger(\omega,m,k)\varphi(-\omega,-m,k)],\]
where the integration if restricted to the region $\kappa>0$ as a consequence of definition of operators $a$ and $b^\dagger$.

To write the quantum version of energy-momentum tensor that follows from (\ref{cscalaraction})
\[T_{ab}=\frac{1}{6}[4\nabla_{(a}\Phi^*\nabla_{b)}\Phi-g_{ab}\nabla^c\Phi\nabla_c\Phi^*-\Phi^*\nabla_b\nabla_a\Phi-\Phi\nabla_b\nabla_a\Phi^*],\]
 we require the expression obtained by substitution of the classical $\Phi$ by its quantum version to be symmetric under the interchange of $\Phi$ and $\Phi^\dagger$. This requirement is necessary in order to the quantity to be finite: the operators are evaluated on the same event, hence any term containing the commutator would give a contribution of $\delta(0)$, coming from the canonical commutation relations. We obtain, after contracting with $\xi^a$, to give, as usual the energy current vector\footnote{The classical version of this operator indeed deserves to be called current, since $\nabla_aJ^a=0$, since $T^{ab}$ is divergent-free and $\xi^a$ is a Killing vector.}
\[J^a=\frac{1}{6}g^{ab}\xi^c\left(\{\nabla_b\Phi^\dagger,\nabla_c\Phi\}+\{\nabla_c\Phi^\dagger,\nabla_b\Phi\}-\frac{1}{2}g_{bc}\{\nabla_d\Phi^\dagger,\nabla^d\Phi\}-\frac{1}{2}\{\Phi^\dagger,\nabla_c\nabla_b\Phi\}-\frac{1}{2}\{\nabla_c\nabla_b\Phi^\dagger,\Phi\}\right),\]
whose vacuum expectation value can be used to evaluate the vacuum expectation value of the energy stolen from the black hole per unit of coordinate time, which is the proper time for an inertial observer at infinity. Since its integral gives the energy flux across a spacelike surface, we may integrate $-\frac{\ud E}{\ud t}=\int_S\sqrt{-g}\langle0|J^a|0\rangle \ud S_a$ along a spacelike surface orthogonal both to $\xi^a$ and (for convenience) to $\frac{\partial}{\partial r}$. This quantity can be interpreted as the energy gained by the black hole.
\[-\frac{\ud E}{\ud t}=\sum_m\int\ud\omega\sum_k\omega^2\left(\frac{|R_+|-1}{2\pi|\omega|}-\frac{|T_-|}{2\pi|\omega-m\Omega|}\right)=\frac{1}{\pi}\sum_m\int_{\omega-m\Omega<0}\ud\omega|\omega|\sum_k(1-|R_+|),\]
where on the last step, we made use of the Wronskian relations (\ref{wronskianU1}) and (\ref{wronskianU2}). And, thanks to these same Wronskian relations, we know that the integrand is negative on the entire region of integration, so is the integral itself. This indicates a constant outflow of energy from the black hole.

Similar procedure shows there is a decrease in angular momentum of the black hole as well. We contract the energy-momentum tensor operator with $\psi^a$, take the vacuum expectation value by substituting the field by its expansion in normal modes and integrate over the same spacelike surface, orthogonal to $\frac{\partial}{\partial r}$ for convenience. The result is
\[\frac{\ud L}{\ud t}=\frac{1}{\pi}\sum_m\int_{\omega-m\Omega<0}\ud\omega\ \frac{m|\omega|}{\omega}(1-|R_+|).\]
The negative result shows a rotating black hole is unstable under quantum-mechanical considerations if we wait long enough. This mechanism is in accordance with the area theorem, as can be seen immediately from (\ref{irredmass}). Unfortunately it is currently impossible to observe this effect in nature, since the timescales $(\sim MM_0^2)$ involved are much larger than the age of the Universe unless the black hole is tiny enough.

To understand the quantum-mechanical version superradiance properly, we must consider not vacuum state, but another state which takes into account that superradiance is \emph{stimulated} emission, not spontaneous as we have just considered. A possible state is a `one-particle state', defined by
\[|1\rangle=\int_0^\infty\ud\omega\ \alpha(\omega)a^\dagger(\omega,m,k)|0\rangle,\]
where $\alpha(\omega)$ is the distribution of frequencies contained in our `wave packet'. Normalisation requires $\int\ud\omega|\alpha(\omega)|^2=1$. This state avoids normalisation problems plane waves have.

The notation $|1\rangle$ is intended to be intuitive, because this state is an eigenstate of the `number operator' $\int\ud\omega\ a^\dagger(\omega) a(\omega)$ with unity eigenvalue, in this sense our nomenclature `one-particle state' ought to be understood.

If the state contains frequencies restricted within the superradiant interval, this fact is translated by saying $\mathrm{supp}\ \alpha\subset(0,m\Omega)$. The most simple such $\alpha$ would be $\alpha(\omega)=\delta(\omega-\omega_0)$, for some $0<\omega_0<m\Omega$, representing a monochromatic packet. Naturally, we could have included the possibility of more than one possible value of $k$ or $m$, say $|1\rangle=\sum_m c_m\int\ud\omega\ \alpha(\omega) a(m,\omega)^\dagger|0\rangle$ with $\sum_m |c_m|^2=1$, but the inclusion of this factor would not change our conclusions by any means.

Now we evaluated the expectation value of the flux of energy as before, but now on a `one-particle state'. For notation convenience, define the operator $\mathfrak A$ and its adjoint which brings the vacuum to the `one-particle state' above: $\mathfrak A^\dagger|0\rangle=|1\rangle$. Simple integration reveals these operators respect the same commutation relations the operators $a$ do.
The identity
\begin{equation}
-\left.\frac{\ud E}{\ud t}\right|_{|1\rangle}=-\int_S\sqrt{-g}\langle0|[[\mathfrak A,J^a],\mathfrak A^\dagger]|0\rangle\ \ud S_a-\left.\frac{\ud E}{\ud t}\right|_{|0\rangle}
\label{ecom}
\end{equation}
follows by noting that, from the definition of vacuum state,
\[\langle0|[[\mathfrak A,J^a],\mathfrak A^\dagger]|0\rangle=\langle0|[\mathfrak A,J^a]\mathfrak A^\dagger|0\rangle=\underbrace{\langle0|\mathfrak AJ^a\mathfrak A^\dagger|0\rangle}_{\langle1|J^a|1\rangle}-\underbrace{\langle0|J^a(\mathfrak A^\dagger\mathfrak A+1)|0\rangle}_{\langle0|J^a|0\rangle},\]
where the commutation relation has been used in the last passage followed again from the definition of vacuum state.

To evaluate the first term of (\ref{ecom}) for the same surface $S$ as before, we first note that all terms in operators $b$ and $b^\dagger$ vanish after commuting with the operators $a$ and $a^\dagger$, then we place the annihilation operators on the right of the creation operators, making use of their commutation relation, and make use of the definition of vacuum state. Integration of the delta functions 
leads to
\begin{equation}
\left.\frac{\ud E}{\ud t}\right|_{|1\rangle}=-\int_S\sqrt{-g}\ud S_a\xi^b\langle1|T^a_b|1\rangle=\left.\frac{\ud E}{\ud t}\right|_{|0\rangle}+\iint\ud\omega_1\ud\omega_2\ \alpha^*(\omega_1)\alpha(\omega_2)I(\varphi(\omega_1),\varphi(\omega_2)),
\label{ecomfb}
\end{equation}
where the notation $I(\eta,\zeta)\equiv \int_S\sqrt{-g}\ud S^a\ \xi^b T_{ab}(\eta,\zeta)$, and $T_{ab}(\eta,\zeta)$ means the expression for the classical energy-momentum tensor substituting $\bar\psi$ by $\bar\eta$ and $\psi$ by $\zeta$, that is $T_{ab}(\zeta,\zeta)$ is the energy momentum tensor for field $\zeta$. Consequently $I(\zeta,\zeta)$ is the energy flux across $S$ associated with $\zeta$.

To describe an approximate important case more explicitly, suppose $\alpha$ to be appreciably different from zero in a very narrow band of frequency compared to $m\Omega$\footnote{More discussion about obtaining explicit results will be present on the next section, about fermions. That techniques apply in present case as well.}. In this case, we can approximate the integrand to evaluate $I$ in just one frequency, leading simply to the classical current $J^a$ evaluated on the mode $\int\ud\omega\ \alpha\varphi$. The Wronskian relations lead immediately to $\frac{1}{\pi}\int\ud\omega\ \omega|\alpha|^2(|R_+|-1)$, which is positive for our choice of $\alpha$, $$\left.\frac{\ud E}{\ud t}\right|_{|1\rangle}-\left.\frac{\ud E}{\ud t}\right|_{|0\rangle}<0,$$ showing the `presence of a particle' with energy and angular momentum in superradiant interval contributes to cause a decrease in black hole's energy. In contrast, if we had built a `one-particle state' of radiation far from the superradiant regime, this calculation shows the sign would be reversed, and the black hole would gain energy or, at least, lose energy more slowly then when compared with the vacuum state. These results seem intuitive.

For angular momentum the calculation is almost identical, except by substituting a factor of $\omega$ in favour of a factor of $m$. The conclusions are, therefore, the same: a `particle' in superradiant regime reinforces angular momentum loss.

As we did when studying the so-called Klein paradox, we can show the existence of particle creation by the black hole in superradiant regime, but we shall not derive this result explicitly here, since we have already derived the variation of relevant physical quantities without invoking this calculation.

\subsection{Fermions}

We can repeat the analysis for fermions, massless neutrinos, for simplicity, whose action is given by
\begin{equation}
S=\int\sqrt{-g}\ud^4x\ \bar\psi\gamma^a\nabla_a\psi,
\label{neutrinoaction}
\end{equation}
where $\psi$ is a `Dirac spinor' as defined on chapter three. It is checked this action actually describes massless fermions on arbitrary (globally hyperbolic) space-times by matching its Euler-Lagrange equations with (\ref{DiracUsual}).

The form of Dirac equation with Dirac spinors makes the task of canonical quantisation, based on Hamiltonian approach much simpler. Indeed, from (\ref{neutrinoaction}) follows the canonical conjugated momentum is $\pi=\ui\sqrt{-g}\bar\psi\gamma^0$. The energy-momentum tensor is given in Dirac formalism by $T_{ab}=\frac{\ui}{2}\left[\bar\psi\gamma_{(a}\nabla_{b)}\psi-(\nabla_{(a}\bar\psi)\gamma_{b)}\psi\right]$. This leads to $T^a_b\xi^b=\frac{\ui}{4}[\bar\psi\gamma^a(\xi^b\nabla_b)\psi+\bar\psi(\xi^b\gamma_b)\nabla^a\psi]+\mathrm{c.c.}$ and similar expression if one replace $\xi^a$ by $\psi^a$.

From the conserved current in the form $j^a=\bar\psi\gamma^a\psi$, the Dirac product defined on chapter three, reads $(\psi_1,\psi_2)_D=\int_\Sigma\sqrt{-g}\ud^4x\ n_a\bar\psi_1\gamma^a\psi_2$, where $n_a$ is the normal of Cauchy surface $\Sigma$. Of course, since this product coincides with the Dirac product defined in terms of usual spinors, it does not depend on the choice of $\Sigma$.

Let $k$ denote the separation constant appearing in the dynamical equation will shall write explicitly below. So, we define the creation/annihilation operators by $a(\omega,m,\lambda,k)=(\varphi(\omega,m,\lambda,k),\Psi(\omega,m,\lambda,k))_D$ for $\kappa(\omega,m,\lambda,k)>0$ and $b^\dagger(\omega,m,\lambda,k)=(\varphi(\omega,m,\lambda,k),\Psi(\omega,m,\lambda,k))_D$ for $\kappa(\omega,m,\lambda,k)<0$. Let $\{\bullet,\bullet\}$ denote the anti-commutator. Then, from canonical anti-commutation relation for fermions $\{\psi,\pi\}$, follows $\{a(\omega,m,\lambda,k),a^\dagger(\omega',m',\lambda',k')\}=\delta(\omega-\omega')\delta_{mm'}\delta_{\lambda\lambda'}\delta_{kk'}$ and similarly for $b$ and $b^\dagger$.  Define once again the vacuum state by $a|0\rangle=b|0\rangle=0$. Already promoting the field to operator, we can expand it in normal modes as
\[\Psi=\sum_m\int_{\kappa>0}\ud\omega\ \sum_{k,\lambda}[a(\omega,m,\lambda,k)\varphi(\omega,m,\lambda,k)+b^\dagger(\omega,m,\lambda,k)\varphi(-\omega,-m,\lambda,k)].\]

Now we may compute the net inflow of energy and angular momentum through a spacelike surface $S$ orthogonal both to $\xi^a$ and $\frac{\partial}{\partial r}$, for convenience respectively by
\begin{equation}
\frac{\ud E}{\ud t}=-\int_S\sqrt{-g}\ud S_a\ \xi^b\langle0|T^a_b|0\rangle\qquad\frac{\ud L}{\ud t}=-\int_S\sqrt{-g}\ud S_a\ \psi^b\langle0|T^a_b|0\rangle,
\label{ELloss}
\end{equation}
where the energy-momentum tensor has been shown in terms of the spinors in (\ref{emtspinor}), but it is convenient to work with the Dirac spinor version, since our expansion in normal modes works with them. In fact, our swap to Dirac spinor formalism in this chapter is intended to give a description of quantisation closer to the ones we are used to. 

Substituting (\ref{volta}) for these spinors, the normal modes are
\[\varphi=\frac{e^{\ui(m\phi-\omega t)}}{(\Delta\sin^2\theta\bar\rho)^\frac{1}{4}}
\begin{bmatrix}
R_{+\frac{1}{2}}S_{+\frac{1}{2}}\\
R_{-\frac{1}{2}}S_{-\frac{1}{2}}\\
LR_{+\frac{1}{2}}S_{+\frac{1}{2}}\\
LR_{-\frac{1}{2}}S_{-\frac{1}{2}}
\end{bmatrix},\]
where $L=\pm1$ gives account of the difference between left and right-handed neutrinos \cite{casals}.

We could resource to (\ref{wywz}) to find the asymptotic behaviour of these functions and the corresponding Wronskian relations, but it is not necessary to do so. We will not proceed this way because, we are interested in a second linearly independent solution, as before in this chapter. We take \cite{unruh,casals}
\[(R_{1/2}^+, R_{-1/2}^+)\to\frac{1}{\sqrt{2\pi}}
\begin{cases}
(A_+e^{\ui\omega r^*},e^{-\ui\omega r^*}) & r^*\to\infty\\
(0,B_+e^{-\ui(\omega-m\Omega)r^*}) & r^*\to-\infty
\end{cases}\]
and
\[(R_{1/2}^-, R_{-1/2}^-)\to\frac{1}{\sqrt{2\pi}}
\begin{cases}
(B_-e^{\ui\omega r^*},0) & r^*\to\infty\\
(e^{\ui(\omega-m\Omega)r^*},A_-e^{-\ui(\omega-m\Omega)r^*}) & r^*\to-\infty
\end{cases},\]
where the coefficients $A_\pm$ and $B_\pm$ may, of course, depend on $\omega$, $m$ and on the separation constant $k$.

Expanding an arbitrary mode in terms of normal modes and substituting on the expression of loss of energy and angular momentum (\ref{ELloss}) above, one gets
\begin{subequations}
\begin{align}
\frac{\ud E}{\ud t}=-\frac{1}{2\pi}\int\ud\omega\ \omega\sum_{k, m}[\kappa(\omega,m,+)(|A_+|^2-1)+\kappa(\omega,m,-)|B_-|^2]\\
\frac{\ud L}{\ud t}=-\frac{1}{2\pi}\int\ud\omega\ \sum_{k, m}m[\kappa(\omega,m,+)(|A_+|^2-1)+\kappa(\omega,m,-)|B_-|^2],
\end{align}
\label{ELlosssub}
\end{subequations}
where we used the expression $$T_{ab}=\frac{\ui}{2}[\bar\psi,(\gamma_{(a}\nabla_{b)}\psi)]+\mathrm{h.c.}$$ for the operator corresponding to the energy-momentum tensor. Analogously to the bosonic case, it is antisymmetric under the interchange $\psi$ and $\bar\psi$ to avoid infinities coming from the evaluation of the anticommutator of fields evaluated at the same events.

The Wronskian relations are found by means we developed on chapter two, for linear, first order, homogeneous system of ODEs. As it can be seen directly from Teukolsky equations for neutrinos radial functions (\ref{teud}a,c) with $\mu=0$, which can be re-written as
\footnote{We recall, for reference, the angular functions are subjected to the same normalisation as they were on chapter three and obey
\[\begin{bmatrix}
\frac{\ud S_{1/2}}{\ud\theta}\\
\frac{\ud S_{-1/2}}{\ud\theta}
\end{bmatrix}
=
\begin{bmatrix}
-\omega a\sin\theta+\frac{m}{\sin\theta} & k\\
-k & \omega a\sin\theta-\frac{m}{\sin\theta}
\end{bmatrix}
\begin{bmatrix}
S_{1/2}\\
S_{-1/2}
\end{bmatrix}\]}
\[\begin{bmatrix}
\frac{\ud R_{1/2}}{\ud r}\\
\frac{\ud R_{-1/2}}{\ud r}
\end{bmatrix}=\underbrace{
\begin{bmatrix}
\frac{\ui}{\Delta}[\omega(r^2+a^2)-ma] & \frac{k}{\sqrt\Delta}\\
\frac{k}{\sqrt\Delta} & -\frac{\ui}{\Delta}[\omega(r^2+a^2)-ma]
\end{bmatrix}}_{\text{traceless}}
\begin{bmatrix}
R_{1/2}\\
R_{-1/2}
\end{bmatrix}\]
clearly satisfies the tracelessness condition for Wronskian conservation, which means Wronskian between any pair of solutions is conserved from (\ref{GAI}). It is immediate also that if $(X,Y)$ is a solution for this system, so is $(Y^*,X^*)$. This fact was already expressed on chapter three when we proved the equations for one component of Dirac equation was the same as for its complex conjugate after the interchange $s\leftrightarrow-s$.

We derive the consequence of conservation of Wronskian between four possible pairs of solutions: $(R_{1/2}^\pm,R_{-1/2}^\pm)$ and $((R_{-1/2}^\pm)^*, (R_{1/2}^\pm)^*)$;  $(R_{1/2}^+,R_{-1/2}^+)$ and $(R^-_{1/2},R^-_{-1/2})$; $(R_{1/2}^+,R_{-1/2}^+)$ and $((R_{-1/2}^-)^*,(R_{1/2}^-)^*)$. Therefore, using these asymptotic modes, one finds, for each independent set of quantum numbers, $|A_\pm|^2+|B_\pm|^2=1$ (meaning once again absence of superradiance\footnote{As in chapter three, the interpretation of these coefficients on spinors are not immediate, it is required to compute energy momentum tensor to link them to the reflection coefficient. We are going to perform this calculation few lines below, to evaluate an energy flux.}), $B_-=B_+$ and $A_+B_-^*=-A_-^*B_+$, respectively. Applying these relations on (\ref{ELlosssub}), we find the integrand vanishes for frequencies outside the interval $0<\omega<m\Omega$, leaving

\[\frac{\ud E}{\ud t}=\frac{2}{\pi}\sum_m\int_0^{m\Omega}\ud\omega\ \omega\sum_k(|A_+|^2-1),\]
\[\frac{\ud L}{\ud t}=\frac{2}{\pi}\sum_m\int_0^{m\Omega}\ud\omega\ m\sum_k(|A_+|^2-1).\]
Thanks to the interval of integration we see this spontaneous emission has the effect of increase in horizon area.

This sheds light into an interesting question. As we saw on section 4.2.4, from (\ref{beke3}) the laws of black hole thermodynamics predict the existence of superradiance for bosonic radiation accompanied by increase in horizon area; similar arguments show absence of superradiance of fermions is accompanied by decrease in horizon area. As we made clear on the previous chapter, this bears no contradiction with the laws itself, since this kind of radiation explicitly violates one of the hypothesis of area theorem. On the other hand, we apparently face problems with the Generalised Second Law of Thermodynamics. For, this area decrease would be accompanied with the decrease of intensity of radiation. If incident radiation has sufficiently high intensity (but low enough in order maintain the problem treated as a perturbation problem, as we did, neglecting back reaction effects), its entropy increases with the logarithm of the intensity, that is a monotonic function, which means the entropy of radiation would also decrease. The Generalised Second Law has been surviving to many challenges since it has been originally proposed by J. Bekenstein.

A wide class of possible quantum violations of the Generalised Second Law, or even the Second Law of Thermodynamics itself has been studied and ruled out by Ford \cite{ford}. The proposed mechanism relies on the possibility of producing negative-energy fluxes in quantum mechanics and a hot body absorbing it, provoking the body to lose energy and entropy.  The author shows examples for which these fluxes $F$ are bounded to occur in a finite time interval $\tau$ and to respect $|F|\lesssim\tau^{-2}$. Although as we showed on chapter four, violation of weak energy condition surely occurs when incident low-frequency fermionic radiation in a spinning black hole, fact that could be interpreted as the cause of such apparent violation of Generalised Second Law, since if weak energy condition is violated, energy density itself can become negative as seen for a particular observer, let alone energy fluxes\footnote{Even a violation of the dominant energy condition can produce negative energy fluxes.}. On the other hand, as we saw on section 3.4, the time-averaged flux of $T^a_b\xi^b$ is negative, which means that the integrated flux does not obey the mentioned inequality.

We are going to seek the answer by considering stimulated emission of the phenomena we have been discussing. The construction of an `one-particle state' as we did,$|1\rangle=\int\ud\omega\ \alpha(\omega)a^\dagger(m,\omega,k)|0\rangle\equiv\mathfrak A^\dagger|0\rangle$ for $\mathrm{supp}\ \alpha\subset(0,m\Omega)$ and $\int\ud\omega|\alpha(\omega)|^2=1$, is a pure state with vanishing entropy, consequently. If we substitute the field in the energy momentum tensor operator for its expansion in normal modes and make use of the identities valid \emph{only for after vacuum expectation value}, since terms with annihilation operators on right and creation operators on left are omitted,
\[a(\omega_1)a(\omega_2)a^\dagger(\omega_3)a^\dagger(\omega_4)=\delta(\omega_2-\omega_3)\delta(\omega_1-\omega_4)-\delta(\omega_2-\omega_4)\delta(\omega_1-\omega_3)\]
\[-a(\omega_1)a^\dagger(\omega_3)a(\omega_2)a^\dagger(\omega_4)=-\delta(\omega_1-\omega_3)\delta(\omega_2-\omega_4)\]
and
\[a(\omega_1)[b(\omega_2)b^\dagger(\omega_3)-b^\dagger(\omega_3)b(\omega_2)]a^\dagger(\omega_4)=\delta(\omega_2-\omega_3)\delta(\omega_1-\omega_4),\]
we obtain
\begin{equation}
\left.\frac{\ud E}{\ud t}\right|_{|1\rangle}=-\int_S\sqrt{-g}\ud S_a\xi^b\langle1|T^a_b|1\rangle=\left.\frac{\ud E}{\ud t}\right|_{|0\rangle}+\iint\ud\omega_1\ud\omega_2\ \alpha^*(\omega_1)\alpha(\omega_2)I(\varphi(\omega_1),\varphi(\omega_2)),
\label{ecomf}
\end{equation}
where the notation is similar to the one appearing in (\ref{ecomfb}), apart from using the energy-momentum tensor for neutrinos, naturally.

An important difference arises here. Both terms in (\ref{ecom}) have the same sign as we saw, which means that the presence of `a particle' contributes for the emission the black hole would spontaneously do. Here, as we could foresee on physical grounds, after solving the classical problem and concluding superradiance to be absent, the effect of `a particle' is to include a term with the opposite sign comparing with the vacuum contribution. Therefore, in order to decide the sign of the final expression we have to evaluate it explicitly, but because fields evaluated at different frequencies (or different values of $m$) obey different system of ODE's, we do not have Wronskian relations to decide the sign of the final expression. Analytic results for the fields are known (Page, \cite{page}) only in a narrow band of frequencies, we ought to resort to numerical calculations to address properly this question. But we have an analytic result if state is monochromatic, apart from an infinity which arises from the fact that such states are not normalisable. To overcome this difficulty one may consider the frequencies as discrete variables, instead of allowing them to have any value. This procedure is equivalent of placing the field in a box with periodic boundary conditions and after letting the size of the box approach infinity. In this case, comparing the integrands of both terms above we conclude there is flux of energy outgoing the black hole and  similarly for angular momentum, consequently $\left.\frac{\ud E}{\ud t}\right|_{|1\rangle}\geqslant\Omega\left.\frac{\ud L}{\ud t}\right|_{|1\rangle}$, obeying the Generalised Second Law.

This result shows the presence of a `particle' in the superradiant regime can at most suppress the loss of energy and angular momentum the black hole would spontaneously would lose and, moreover, satisfies the above inequality. From the first law of black hole thermodynamics, we see that, in contrast to what we predicted classically, the area of horizon will \emph{not} decrease! This reasoning always shows the presence of fermions spontaneously emitted (and carrying their entropy, of course) is at most suppressed, never reversed by the presence of a `particle', so the entropy associated with matter (outside the black hole) will increase, or at most remain the same and will precisely do when the horizon area also remain unchanged.

As a consequence of the anticommutation relations, $(a^\dagger)^2=0$, which means it is impossible to construct a `many particle' state, a manifestation of Pauli's exclusion principle, consequently no other state can threat our conclusions. See the remarkable similarity of this solution of the apparent paradox and the solution for the similar issue in a non-gravitational context on last paragraph of section 4.1.

Therefore, it is fair to say that quantum mechanics may prevent Generalised Second Law violation from the absence of fermionic superradiance. In fact, this is not a significant surprise. The Generalised Second Law itself has already quantum origin, as can be seen from a black hole's entropy predicted by Hawking radiation, where it appears the constant $\hslash$ (see chapter four). It was proposed after Hawking's discovery that black hole evaporation would lead to area decrease, and it was later proved that the entropy of emitted radiation do compensate such decrease.

Also, this result is a contribution in favour of the maintenance of the weak cosmic censorship conjecture, as we were discussing last chapter. Because quantum mechanics prevents area reduction, it also prevents overspinning a black hole, as Hod had suggestively related to Pauli principle.

It is worthwhile to notice the same calculation used here to compute the effect of the `presence of a particle' could be used almost without any change for the section 5.1, to investigate the energy fluxes across a surface of constant $x$ not in the vacuum, but in a occupied state. Our conclusions are similar to the black hole case: the vacuum contribution is more significant than the `particle' contribution for fermions, and contributes to the vacuum emission for bosons.

\backmatter

\begin{savequote}[65mm]
A work is never completed except by some accident such as weariness, satisfaction, the need to deliver, or death: for, in relation to who or what is making it, it can only be one stage in a series of inner transformations.
\qauthor{Paul Valéry (1871-1945)}
\end{savequote}
\chapter{Conclusions}

We dedicated a significant part of this work exploring the rich relation between superradiance and thermodynamics. We saw superradiance can be predicted from the laws of ordinary and black hole thermodynamics to be thermodynamically favourable in the superradiant regime. In both cases when we treat physical laws classically but consider fermionic radiation, which is intrinsically quantum mechanical, the second law of thermodynamics appears to be threatened by the absence of superradiance, and the threat is released when the Pauli exclusion principle is taken into account. These results relating absence of fermionic superradiance and the second law of thermodynamics are believed to be original from this work. It is particularly worthwhile to recall how different in nature ordinary and black hole thermodynamics arguments are and at the same time how similar they are in their physical interpretations. This similarity is seen as a theoretical clue about the existence of a still deeper connection between black holes and thermodynamics as well as for the validity of the Generalised Second Law.

A possible criticism applicable to most of this work is the fact we considered the approximation of space-time as a fixed background and considered fields as a perturbation. We followed this procedure explicitly in perturbations from chapter three, while establishing the connection with black hole thermodynamics and also when we did Quantum Field Theory in curved space-time. Indeed we have few clues of what back reaction effects can interfere in, especially what role they can play on deriving superradiance (or its absence) using black hole thermodynamics. These effects are extremely hard to be taken into account, and we only say our main conclusions will hardly be invalidated from these effects, at least at sufficiently low intensities, since low intensities lead to small value of energy-momentum tensor and via Einstein's equations, low influence on space-time.

Among the most interesting issues concerning superradiance is its deep and evident connection with the spin and statistics, we explored in depth here. We had the opportunity to see several mathematically distinct forms this fact is expressed (regularity on effective potentials, number current conservation, null energy condition violation and so on). Until present, we are in lack of a definitive general explanation on the context of the second chapter of why fermions are not subjected to superradiance of any kind. However, as we have seen thorough this dissertation, superradiant scattering conserves the frequency and other relevant quantum numbers. This is an ultimate consequence of symmetries. Mathematically, they may be expressed in terms of the boundary conditions imposed, for example, the preservation of frequency is required when the boundary conditions do not depend on time: the only way for them to be respected is if all exponential terms $e^{i\omega t}$ and $e^{i\omega' t}$ cancel one another for every $t$, which can only be achieved if $\omega=\omega'$; or, as in the Kerr-Newman black hole problem, where the notion of time may appear not well established, the space-time possesses a timelike isometry and a spacelike isometry associated with a Killing vector whose orbits are closed, that is, an axial symmetry. Therefore we do not expect any scattering problem whose perturbations preserves these symmetries to violate them, implying on conservation of both frequency and the angular momentum density component along the axis of symmetry. So we must understand absence of superradiance for fermions as (once again) a consequence of Pauli exclusion principle on a particle level point of view, which prevents  a second particle to occupy a same outgoing mode, as already suggested by Hawking \cite{Hawking}. Yet, a formalisation, as we did on the last chapter, of this argument is still missing nowadays. Difficulties arise if one tries to put this assertion in rigorous grounds, since Pauli principle appears from canonical anticommutation relations, and therefore, only when one is quantising fermionic fields (see chapter five), whilst superrradiance is ruled out on a classical level (see chapters three, four and five), directly from the field dynamical equations. This issue remains open nowadays.


\begin{thebibliography}{1}

\bibitem{HE}Hawking, S.W.; Ellis, G.F.R.; \emph{The Large Scale Structure of Space-Time}. Cambridge Monographs on Mathematical Physics.(1973).
\bibitem{friedman} John Friedman, private communication, 2012.
\bibitem{mauricioalberto} Richartz, M.; Saa, A.; \emph{Phys. Rev. D}, \textbf{88}, 044008 (2013).
\bibitem{hod} Hod, S.; Hod, O.; \emph{Phys. Rev. D} \textbf{81}, 061502 (2010).
\bibitem{rosa} Rosa, J.G.; \emph{JHEP} 1006, 015 (2010).
\bibitem{unruh81} Unruh, W.G., \emph{Phys. Rev. Lett.} \textbf{21} 1351 (1981).
\bibitem{unruh02} Schützhold, R; Unruh, W.G.; \emph{Phys. Rev. D} \textbf{66}, 044019 (2002).
\bibitem{unruh11} Weinfurtner, S.; Tedford E.W.; Penrice, M.C.J.; Unruh, W.G.; Lawrence, G.A.; \emph{Phys. Rev. Lett.} \textbf{106}, 021302 (2011).
\bibitem{Mauricio}M. Richartz, S. Weinfurtner, A. J. Penner, W. G. Unruh,
Phys. Rev. D{\bf 80}, 124016 (2009) [arXiv:0909.2317].
\bibitem{kwa}Kwa, K.H.; \emph{Basic Theory of Systems of First Order Linear Equations}, notes available on 
http://www.math.osu.edu/\texttildelow kwa.1/
\bibitem{beke} J. D. Bekenstein e M. Schiffer, Phys. Rev. D {\bf 58}, 064014 (1998) [gr-qc/9803033].
\bibitem{arfken} Arfken, G.B.; Webber, H.J.; \emph{Mathematical Methods for Physicists -- International Edition}, Sixth Edition. Elsevier. (2005).
\bibitem{jackson} Jackson, J.D., \emph{Classical Electrodynamics}, Third Edition. Wiley. (1998).
\bibitem{penrose} Penrose, R.; Rindler, W.; \emph{Spinors and Space-time}, Cambridge Monographs on Mathematical Physics. (1986).
\bibitem{streater} Streater, R.F; Wightman, A.S.; \emph{PCT, Spin and Statistics, and All That}, Princeton University Press, Third Printing (1980).
\bibitem{WaldGR}R. M. Wald, \emph{General Relativity}. University of Chicago Press. (1984).
\bibitem{Survey}Hawking, S.W.; Israel, W. (editors);\emph{General Relativity: An Einstein Centenary Survey.} Cambridge University Press. (1979).
\bibitem{chandrasekhar} Chandrasekhar, S.; \emph{The Mathematical Theory of Black Holes}, Oxford University Press. (1983).
\bibitem{landau2} Landau, L.D.; Lifshitz, E.M.; \emph{The Classical Theory of Fields}, Fourth Revised English Edition, Butterworth-Heinemann. (1975).
\bibitem{abramowitz} Abramowitz, M.; Stegun, I.A.; \emph{Handbook of Mathematical Functions with Formulas, Graphs and Mathematical Tables}. Dover. (1964).
\bibitem{epoisson}Poisson, E; \emph{A Relativist's Toolkit: The Mathematics of Black-Holes Mechanics.} Cambridge University Press. (2004).
\bibitem{barata} Barata, J.C.A.; \emph{Curso de Física-Matemática}, available on \\ http://denebola.if.usp.br/$\sim$jbarata/Notas\underline{ }de\underline{ }aula/capitulos.html (2013).
\bibitem{bch} Bardeen, J.M.; Carter, B.; Hawking, S.W.; \emph{Commun. Math. Phys.} \textbf{31}, 161 (1973).
\bibitem{lee} Lee, J.M.; \emph{Introduction to Smooth Manifolds}, Springer-Verlag, (2002).
\bibitem{Nakahara}Nakahara, M.,\emph{Geometry, Topology and Physics.}, Taylor \& Francis. (2003).
\bibitem{beig} Beig, R.; Phys. Lett. \textbf{69A}, 153 (1978).
\bibitem{brewin} Brewin, L. arXiv [gr-qc]: 0609079v1 (2006).
\bibitem{iyerwald} Iyer, V.; Wald, R. M.; \emph{Phys. Rev. D} \textbf{50}, 846-864 (1994).
\bibitem{beke73} Bekenstein, J.D.; \emph{Phys. Rev. D} \textbf{7}, 949 (1973).
\bibitem{ANEC} Wald, R.M.; Yurtsever, U.; \emph{Phys. Rev. D} {\bf 44}, 403 (1991).
\bibitem{hodccc} Hod, S.; \emph{Phys. Lett. B} {\bf 668}, 346, (2008).
\bibitem{Hawking}Hawking, S.W., \emph{Commun. Math. Phys.} \textbf{43}, 199 (1975).
\bibitem{Birrel}Birrel, N.D.; Davies, P.C.W.; \emph{Quantum Fields in Curved Spaces}. Cambridge Monographs on\\ Mathematical Physics. (1984).
\bibitem{Manogue} C. A. Manogue, Annals of Physics {\bf 181}, 261 (1988).
\bibitem{sakurai} Sakurai, J.J.; \emph{Advanced Quantum Mechanics}, Addison-Wesley. (1967).
\bibitem{landau4} Berestetskii, V.B.; Lifshitz, E.M.; Pitaevskii, L.P.; \emph{Quantum Electrodynamics}, Second Edition, Butterworth-Heinemann. (1982).
\bibitem{unruh} Unruh, W. G.; \emph{Phys. Rev. D} \textbf{10}, 3194-3205  (1974).
\bibitem{casals} Casals, M.; Dolan, S.R.; Nolan, B.C.; Ottewill, A.C.; Winstanley, E.; \emph{Phys. Rev. D} {\bf 87}, 064027 (2013).
\bibitem{ford} Ford, L.H.; \emph{Proc. R. Soc. Lond. A} \textbf{364} 227 (1978).
\bibitem{page} Page, D.; \emph{Phys. Rev. D} \textbf{13}, 198 (1976).
\end{thebibliography}
\end{document}